\documentclass[11pt,onecolumn,draft]{IEEEtran} 
\usepackage{cite}
\usepackage{psfig}
\usepackage{subfigure}
\usepackage{graphicx}
\usepackage{amsmath}
\usepackage{bm}
\usepackage{amssymb}

\newcommand{\bb}[1]{{\pmb{#1}}}
\newcommand{\uu}[1]{{\underline{#1}}}

\newcommand{\nn}{\nonumber}

\DeclareMathOperator{\sinc}{sinc}
\DeclareMathOperator{\diag}{diag}
\DeclareMathOperator{\tr}{tr}

\newtheorem{theorem}{Theorem}

\newtheorem{corollary}{Corollary}
\newtheorem{lemma}{Lemma}

\newtheorem{definition}{Definition}
\begin{document}

\title{Asymptotic Spectral Distribution of Crosscorrelation Matrix in Asynchronous CDMA}

\author{Chien-Hwa Hwang\\
Institute of Communications Engineering\\
\& Department of Electrical Engineering,\\
National Tsing Hua University,\\
Hsinchu, Taiwan\\
E-mail: chhwang@ee.nthu.edu.tw}

\footnotetext[1]{This paper was presented in part at IEEE ISIT'05,
Adelaide, Australia, Sept. 4-9, 2005 and IEEE ISIT'07, Nice, France,
June 24-29, 2007.}

\maketitle

\begin{abstract}
Asymptotic spectral distribution (ASD) of the crosscorrelation
matrix is investigated for a random spreading short/long-code
asynchronous direct sequence-code division multiple access (DS-CDMA)
system. The discrete-time decision statistics are obtained as the
output samples of a bank of symbol matched filters of all users. The
crosscorrelation matrix is studied when the number of symbols
transmitted by each user tends to infinity. Two levels of
asynchronism are considered. One is symbol-asynchronous but
chip-synchronous, and the other is chip-asynchronous. The existence
of a nonrandom ASD is proved by moment convergence theorem, where
the focus is on the derivation of asymptotic eigenvalue moments
(AEM) of the crosscorrelation matrix. A combinatorics approach based
on noncrossing partition of set partition theory is adopted for AEM
computation. The spectral efficiency and the minimum mean-square-error (MMSE)
achievable by a linear receiver of asynchronous CDMA are plotted by
AEM using a numerical method.
\end{abstract}

\section{Introduction}

Direct sequence-code division multiple access (DS-CDMA) is one of
the most flexible and commonly proposed multiple access techniques
for wireless communication systems. To gain deeper insights into the
performance of receivers in a CDMA system, much work has been
devoted to the analysis of random spreading CDMA in the large-system
regime, i.e., both the number of  users $K$ and the number of chips
$N$ per symbol approach infinity with their ratio $K/N$ kept as a
finite positive constant $\beta$ \cite{grant98,verdu99,tse99}. Such
asymptotic analysis of random spreading CDMA enables random matrix
theory to enter communication and information theory. In the last
few years, a considerable amount of CDMA research has made
substantial use of results in random matrix theory (see
\cite{tulino2004} and references therein).

In this paper, we focus on the derivation of the asymptotic spectral
distribution (ASD) of the crosscorrelation matrix in asynchronous
CDMA systems. Consider the linear vector memoryless channels of the
form $\pmb y=\pmb H\pmb x+\pmb w$, where $\pmb x$, $\pmb y$ and
$\pmb w$ are the input vector, output vector and additive white
Gaussian noise (AWGN) vector, respectively, and $\pmb H$ denotes the
random channel matrix independent of $\pmb w$. This linear model
encompasses a variety of applications in communications such as
multiuser channels, multi-antenna channels, multipath channels, and,
in particular, asynchronous CDMA channels of our interest in this
research, etc., with $\pmb x$, $\pmb y$ and $\pmb H$ taking
different meanings in each case. Concerned with the linear model, it
is of particular interest to investigate the limiting distribution
of eigenvalues of the random matrix $\pmb H\bb H^\dag$ or $\pmb
H^\dag\bb H$, called the ASD of the random matrix, when the column
and row sizes of $\bb H$ tend to infinity but the ratio of sizes is
fixed as a finite constant. Since ASD is deterministic and
irrelevant to realizations of random parameters, it is convenient to
use the asymptotic limit as an approximation for finite-sized system
design and analysis. Moreover, it is quite often that ASD provides
us with much more insights than an empirical spectral distribution
(ESD) does. Even though ASD is obtained with the large-system
assumption, in practice, the system enjoys large-system properties
for a moderate size of the channel matrix.

Some applications of ASD in communication and information theory are
exemplified below. Take the linear model $\bb y=\bb H\bb x+\bb w$
for illustration. A number of the system performance measures, e.g.
channel capacity and the minimum mean-square-error (MMSE) achievable
by a linear receiver, is determined by the ESD of the matrix $\bb
H\bb H^\dag$. The asymptotic capacity and MMSE obtained by using ASD
as an approximation of ESD can often result in closed-form
expressions \cite{verdu98,verdu99}. It is also shown in
\cite{moshavi96,muller2001,tulino01,honig2001} that empirical
eigenvalue moments (or, more conveniently, AEM) can be used to find
the optimal weights of the reduced-rank MMSE receivers and its
output signal-to-interference-ratio (SIR) in a large system.
Moreover, a functional related to AEM is defined as the free
expectation of random matrices in free probability theory
\cite{voiculescu92}, which has been recently applied to the
asymptotic random matrix analysis.

For synchronous DS-CDMA systems, it is well known that the ASD of
the crosscorrelation matrix follows Mar\u{c}enko-Pastur law
\cite{verdu98}. Also, explicit expressions for the AEM under the
environments of unfaded, frequency-flat fading with single and
multiple antennas, and frequency-selective fading are derived in
\cite{li01,li04}. Actually, most of the research results on random
spreading CDMA making use of random matrix theory are confined to
synchronous systems. Just a few of them investigate asynchronous
systems
\cite{kiran00,zhang01,cottatellucci04,mantravadi02,hwang05,cottatellucci06,hwang07}.
The goal of this work is to find out the ASD of crosscorrelation
matrix in asynchronous CDMA systems given a set of users' relative
delays and an arbitrary chip waveform. As the uplink of a CDMA
system is asynchronous, this work is motivated by the needs to study
the problem of asynchronous transmission that is important but much
less explored in the area of random matrix theory.

Two levels of asynchronism are considered in this paper, i.e.,
symbol-asynchronous but chip-synchronous, and chip-asynchronous. In
the sequel, \textit{chip-synchronous} is used for short to denote
the former, and \textit{symbol-synchronous} represents an ideal
synchronous system. To be more specific, the relative delays among
users are integer multiples of the chip duration in chip-synchronous
CDMA, while they are any real numbers in chip-asynchronous CDMA.

Some previous results on asynchronous CDMA are reviewed. In
\cite{kiran00}, it is shown that the output SIR of the linear MMSE
receiver in chip-synchronous CDMA converges to a deterministic limit
characterized by the solution of a fixed-point equation that depends
on the received power and the relative delay distributions of the
users. When the width of the observation window during detection
tends to infinity, the limiting output SIR converges to that of a
symbol-synchronous system having all identical parameters.
The system model of \cite{kiran00} splits each interferer into two
virtual users, which leads to a crosscorrelation matrix with neither
independent nor identically distributed entries. Results of
\cite{kiran00} are obtained by employing Stieltjes transform for
random matrices of that type. In \cite{mantravadi02}, some
equivalence results about the MMSE receiver output are provided for
CDMA systems with various synchronism levels. In specific, when the
ideal Nyquist sinc pulse is adopted as the chip waveform of a
chip-asynchronous system, the asymptotic SIR at the MMSE detector
output is the same as that in an equivalent chip-synchronous system
for any observation window width; moreover, as the observation
window width increases, the output SIR in chip-asynchronous CDMA
converges further to that in an equivalent symbol-synchronous
system. In \cite{cottatellucci06}, the analysis of linear multiuser
detectors is provided for a symbol quasi-synchronous but
chip-asynchronous system, called symbol-quasi-synchronous for short.
It is demonstrated that, when the bandwidth of the chip waveform is
smaller than $1/(2T_c)$, where $T_c$ is the chip duration, the
performance of symbol-quasi-synchronous and symbol-synchronous
systems coincides independently of the chip waveform and the
distribution of relative delays among users, where the performance
is characterized by the output SIR of a reduced-rank MMSE detector
when a square-root raised cosine pulse is adopted as the chip
waveform. If the bandwidth is larger than the threshold, the former
system outperforms the latter. Actually, when the chip waveform
bandwidth is narrower than the threshold, the inter-chip
interference (ICI) free property is lost \cite{proakis00}, which
leads to a severe degradation in performance.

In this paper, the system model is constructed without the user
splitting executed in \cite{kiran00}. In stead, sufficient
statistics obtained in the same way as \cite{verdu86,lupas90,rupf94}
are employed. We assume the width of the observation window for
symbol detection tends to infinity. The formulas for AEM of the
crosscorrelation matrix are derived using a combinatorics approach.
In specific, we use noncrossing partition in set partition theory as
the solving tool to exploit all nonvanishing terms in the
expressions of AEM. The combinatorics approach has been adopted in
\cite{li01,xiao00,xiao05,jonsson82,yin83,yin86} to compute AEM of
random matrices in symbol-synchronous systems. All of them, either
explicitly or implicitly, make use of graphs to signify noncrossing
partitions. In this work, a graphical representation of
$K$-\textit{graph}, which is able to simultaneously represent a
noncrossing partition and its Kreweras complementation map
\cite{kreweras72}, is adopted. This property of a $K$-graph
facilitates the employment of noncrossing partition and free
probability theory in solving problems of interest.

In some applications of probability theory, it is frequent that the
(infinite) moment sequence of an unknown distribution is available,
and these moments determine a unique distribution. Suppose that the
goal is to calculate the expected value of a certain function $g$ of
the random variable $X$ whose distribution is unknown. One of the
most widely used techniques is based on the Gauss quadrature rule
method \cite{golub69}, where the expected value of $g(X)$ is
expressed as a linear combination of samples of $g(x)$, and moments
of $X$ are used to determine the coefficients in the combination and
the points to be sampled. In this paper, the Gauss quadrature rule
method is employed to compute the spectral efficiency and MMSE of
asynchronous CDMA using the derived AEM.

This paper is organized as follows. In Section II, the
crosscorrelation matrices are given for chip-synchronous and
chip-asynchronous CDMA systems. Some definitions regarding the limit
of of a random matrix are also introduced. In Section III, we derive
AEM and ASD of corsscorrelation matrices in both chip-synchronous
and chip-asynchronous systems. In Section IV, free probability
theory is employed to obtain the spectra of sum and product of
crosscorrelation matrix and a random diagonal matrix. In Section V,
mathematical results demonstrated in this paper are connected to
some known results. Discussions of the spectral efficiency and MMSE
in an asynchronous CDMA system are provided in Section VI. Finally,
this paper is concluded in Section VII.

\section{Crosscorrelation Matrix of Asynchronous CDMA}\label{section:mouse1128}

Consider asynchronous direct sequence-code division multiple access
(DS-CDMA) systems where each user's spreading sequence is chosen
randomly and independently. There are $K$ users in the system, and
the number of chips in a symbol is equal to $N$. We focus on the
uplink of the system and assume the receiver knows the spreading
sequences and relative delays of all users. Systems with two levels
of asynchronism are considered, i.e., symbol-asynchronous but
chip-synchronous, and chip-asynchronous. In the sequel,
\textit{chip-synchronous} is used for short to denote the former,
and \textit{symbol-synchronous} refers to an ideal synchronous
system. To differentiate notations of chip-synchronous and
chip-asynchronous systems, subscripts in text form of
"$\textrm{cs}$" and "$\textrm{ca}$" are used for notations in the
former and the latter systems, respectively.

\subsection{Chip-Synchronous CDMA}

Denote the relative delay of user $k$ as $\tau_k$. For convenience,
users are labelled chronologically by their arrival time, and
$\{\tau_k\}_{k=1}^{K}$ satisfy
\begin{equation}\label{eq:tiger0420}
0=\tau_1\leq\tau_2\leq\cdots\leq\tau_{K}< N T_c,
\end{equation}
where $T_c$ is the chip duration, and all $\tau_k$'s are integer
multiples of $T_c$. Suppose that each user sends a sequence of
symbols with indices from $-M$ to $M$. In the complex baseband
notation, the contribution of user $l$ to the received signal in a
frequency-flat fading channel is
$$
x_l(t)=\sum_{n=-M}^M A_l(n) b_l(n)
\sum_{q=nN}^{(n+1)N-1}c_l^{(q)}\psi(t-q T_c-\tau_l),
$$
where $b_l(n)$ is the $n$-th symbol of user $l$, $A_l(n)$ is the
complex amplitude at the time $b_l(n)$ is received, $\{c_l^{(q)}:
nN\leq q\leq (n+1)N-1\}$ is the spreading sequence assigned to the
$n$-th symbol of user $l$, and $\psi(t)$ is the normalized chip
waveform having the zero ICI condition of
\begin{equation}\label{eq:mouse0320}
\int_{-\infty}^\infty \psi(t)\psi(t-r T_c)\textrm{d}t=\left\{
\begin{array}{ll}
1, & r=0, \\
0, & r\in \mathbb{Z}\setminus\{0\}.
\end{array} \right.
\end{equation}
It is assumed that $\{b_l(n)\}_{n=-M}^M$ is a collection of
independent equiprobable $\pm 1$ random variables. The symbol
streams of different users are independent. The distribution of a
scaled chip $\sqrt{N} c_l^{(q)}$ has zero mean, unit variance and
finite higher order moments. We do not assume a particular
distribution of $c_l^{(q)}$. Two distinct spreading mechanisms are
considered, i.e., short-code and long-code. In a short-code system,
the spreading sequences are randomly chosen for the first symbols,
i.e., $b_k(-M)$ for user $k$, and remain the same for every symbol.
In a long-code system, the spreading sequences are randomly and
independently picked from symbol to symbol. The sequences of
received amplitudes $\{A_k(m)\}_{m=-M}^M$ and $\{A_l(n)\}_{n=-M}^M$
are independent if $k\neq l$.

The complex baseband received signal is given by
\begin{equation}\label{eq:dragon0421}
r(t)=\sum_{l=1}^{K} x_l(t)+w(t),
\end{equation}
where $w(t)$ is the baseband complex Gaussian ambient noise with
independent real and imaginary components. The correlation function
of $w(t)$ is $\textrm{E}\{w(t)w(t+\tau)^*\}=N_0\delta_D(\tau)$ with
$\delta_D(\tau)$ being the Dirac delta function. The symbol matched
filter output of user $k$'s symbol $m$, denoted as $y_k(m)$, is
obtained by correlating $r(t)$ with the signature waveform of user
$k$'s symbol $m$
\begin{eqnarray}
y_k(m)&=&\int_{-\infty}^\infty r(t) \left(\sum_{p=mN}^{(m+1)N-1}
c_k^{(p)}\psi(t-p T_c-\tau_k)\right)
\textrm{d}t\label{eq:tiger1018} \\
&=&\sum_{l=1}^{K}
\sum_{n=-M}^M A_l(n) b_l(n)
\rho_{\textrm{cs}}(m,n;k,l)+v_k(m),\label{eq:cow0421}
\end{eqnarray}
where $v_k(m)$ results from the ambient noise $w(t)$, and
$\rho_{\textrm{cs}}(m,n;k,l)$ is the crosscorrelation of spreading
sequences at user $k$'s $m$-th symbol and user $l$'s $n$-th symbol,
given as
\begin{equation}\label{eq:mouse0712}
\sum_{q=nN}^{(n+1)N-1}\sum_{p=mN}^{(m+1)N-1} c_l^{(q)}
c_k^{(p)}\int_{-\infty}^\infty
\psi(t-qT_c-\tau_l)\psi(t-pT_c-\tau_k)\textrm{d}t.
\end{equation}
Due to the zero ICI condition of (\ref{eq:mouse0320}), the integral
in (\ref{eq:mouse0712}) is nonzero (equal to one) if and only if $p
T_c+\tau_k=q T_c+\tau_l$. Thus, we obtain
\begin{eqnarray}\label{eq:tiger0320}
\rho_{\textrm{cs}}(m,n;k,l)=\sum_{p=mN}^{(m+1)N-1}
\sum_{q=nN}^{(n+1)N-1} c_k^{(p)}c_l^{(q)}\delta(p T_c+\tau_k,q
T_c+\tau_l),
\end{eqnarray}
with $\delta(i,j)$ the Kronecker delta function. Since $0\leq
\tau_k,\tau_l\leq (N-1) T_c$, for a specific symbol index $m$, the
$\delta$ function in (\ref{eq:tiger0320}) is equal to zero if
$n\notin\{m-1,m,m+1\}$. Thus, we rewrite (\ref{eq:cow0421}) as
\begin{equation}\label{eq:mouse0906}
y_k(m)=\sum_{l=1}^{K} \sum_{n=\max\{m-1,-M\}}^{\min\{m+1,M\}} A_l(n)
b_l(n) \rho_\textrm{cs}(m,n;k,l)+v_k(m),\qquad -M\leq m\leq M.
\end{equation}
Define the symbol matched filter output vector at the $m$-th symbol
as
$$
\uu y(m)=[y_1(m),y_2(m),\cdots,y_{K}(m)]^T,
$$
and let the transmitted symbol vector $\uu b(m)$ and the noise
vector $\uu v(m)$ have the same structures as $\uu y(m)$. Moreover,
we define a block matrix $\pmb R_\textrm{cs}$ whose $(k,l)$-th
element of the $(m,n)$-th block, with $-M\leq m,n\leq M$, $1\leq
k,l\leq K$, is equal to $\rho_\textrm{cs}(m,n;k,l)$ in
(\ref{eq:tiger0320}). The square bracket $[\cdot]$ is used to
indicate a specific element of the matrix $\pmb R_\textrm{cs}$.
Specifically, $[\pmb R_\textrm{cs}]_{mn,kl}$ represents the
$(k,l)$-th entry of the $(m,n)$-th block of the block matrix $\pmb
R_\textrm{cs}$. When we just want to point out a specific block,
only the first set of indices is used, i.e., $[\pmb
R_\textrm{cs}]_{mn}$.

Using the notations defined above, we can show from
(\ref{eq:mouse0906}) that
$$
\uu y(m)=\sum_{n=\max\{m-1,-M\}}^{\min\{m+1,M\}}[\pmb
R_\textrm{cs}]_{mn}\uu A(n)\uu b(n)+\uu v(m),
$$
where $\uu A(n)=\diag\{A_1(n),A_2(n),$ $\cdots,A_{K}(n)\}$. Stacking up
$\uu y(m)$'s to yield the symbol matched filter output of the whole
transmission period as
$$
\pmb y=\left[\uu y^T(-M),\uu y^T(-M+1),\cdots,\uu y^T(M)\right]^T,
$$
we obtain the discrete-time signal model
\begin{equation}\label{eq:cow1027}
\pmb y=\pmb R_\textrm{cs}\pmb A\pmb b+\pmb v,
\end{equation}
where $\pmb b$ and $\pmb v$ have the same structures as $\pmb y$,
$\pmb A=\diag\{\uu A(-M),\uu A(-M+1)\cdots,\uu A(M)\}$, and the
block matrix $\pmb R_\textrm{cs}$ has a tri-diagonal structure of
\begin{equation}\label{eq:rabbit1019}
\left[
\begin{array}{ccccccc}
\ddots & \ddots & \ddots & & & & \\
 &  [\pmb R_\textrm{cs}]_{-1\hspace{.2mm}-2} & [\pmb R_\textrm{cs}]_{-1\hspace{.2mm}-1} &
[\pmb R_\textrm{cs}]_{-1\hspace{.2mm}0} &  &  & \\
 &  & [\pmb R_\textrm{cs}]_{0\hspace{.2mm}-1} & [\pmb R_\textrm{cs}]_{0\hspace{.2mm}0} &
[\pmb R_\textrm{cs}]_{0\hspace{.2mm}1} &  & \\
 &  &  &  [\pmb R_\textrm{cs}]_{1\hspace{.2mm}0} & [\pmb R_\textrm{cs}]_{1\hspace{.2mm}1} &
[\pmb R_\textrm{cs}]_{1\hspace{.2mm}2} &  \\
 &  &  &  &  \ddots & \ddots & \ddots
\end{array}
\right].
\end{equation}
Since $\tau_k\leq\tau_l$ for $k<l$, $[\pmb
R_\textrm{cs}]_{m\hspace{1mm}m-1}$ and $[\pmb
R_\textrm{cs}]_{m\hspace{1mm}m+1}$ are strict (zero diagonal) upper-
and lower-triangular matrices, respectively. From the signal model
given in (\ref{eq:cow1027}), $\pmb R_\textrm{cs}$ can be viewed as
the crosscorrelation matrix of chip-synchronous CDMA. It can be
shown that the correlation matrix of the noise vector $\bb v$ in
(\ref{eq:cow1027}) is $\textrm{E}\{\bb v\bb v^\dag\}=N_0 \bb
R_\textrm{cs}$. Let $\bb R_\textrm{cs}=\bb H_\textrm{cs}\bb
H_\textrm{cs}^\dag$ be a decomposition of $\bb R_\textrm{cs}$. We
can perform the whitening process by left-multiplying $\bb y$ in
(\ref{eq:cow1027}) with $\bb H_\textrm{cs}^{-1}$, resulting in
\begin{equation}\label{eq:mouse0213}
\bb z=\bb H_\textrm{cs}^{-1}\bb y=\bb H_\textrm{cs}^\dag\bb A\bb
b+\bb w,
\end{equation}
where the noise vector $\bb w=\bb H_\textrm{cs}^{-1}\bb v$ is white
and has the correlation matrix $N_0\bb I$.

\subsection{Chip-Asynchronous CDMA}

In chip-asynchronous CDMA, the assumption that $\tau_k$'s are
integer multiples of the chip duration $T_c$ no longer exists.
Although the relative delays of users are not integer multiples of
$T_c$, it is assumed that the zero ICI condition still holds. Thus,
the property of zero ICI exists for chips of each particular user.
Similarly to (\ref{eq:cow0421}), the symbol matched filter output
$y_k(m)$ can be expressed as
\begin{equation}
y_k(m)=\sum_{l=1}^{K} \sum_{n=-M}^M A_l(n) b_l(n)
\rho_{\textrm{ca}}(m,n;k,l)+v_k(m),
\end{equation}
where $\rho_{\textrm{ca}}(m,n;k,l)$ is different from
$\rho_{\textrm{cs}}(m,n;k,l)$ in (\ref{eq:cow0421}) since the zero
ICI condition does not hold when the time difference of chip
waveforms is not integer multiples of $T_c$. At this moment, the
crosscorrelation $\rho_{\textrm{ca}}(m,n;k,l)$ is given by
\begin{equation}\label{eq:mouse0323}
\rho_{\textrm{ca}}(m,n;k,l)=\sum_{p=mN}^{(m+1)N-1}\sum_{q=nN}^{(n+1)N-1}
c_k^{(p)}c_l^{(q)} R_\psi((p-q)T_c+\tau_k-\tau_l),
\end{equation}
where
\begin{equation}\label{eq:mouse1027}
R_\psi(x)= \int_{-\infty}^\infty \psi(t)\psi(t-x)\textrm{d}t
\end{equation}
is the autocorrelation function of the chip waveform $\psi(t)$. We
define the block matrix $\pmb R_\textrm{ca}$ whose $(k,l)$-th
component of the $(m,n)$-th block, with $-M\leq m,n\leq M$ and
$1\leq k,l\leq K$, is equal to $\rho_{\textrm{ca}}(m,n;k,l)$ given
in (\ref{eq:mouse0323}). It can be shown that
\begin{equation}\label{eq:mouse0803}
\uu y(m)=\sum_{n=-M}^M [\pmb R_\textrm{ca}]_{mn}\uu A(n)\uu b(n)+\uu
v(m),
\end{equation}
and we obtain the discrete-time signal model
\begin{equation}\label{eq:mouse0219}
\pmb y=\pmb R_\textrm{ca}\pmb A\pmb b+\pmb v,
\end{equation}
where $\pmb R_\textrm{ca}$ is thus seen as the crosscorrelation
matrix of a chip-asynchronous CDMA system. Note that, unlike the
tri-diagonal structure of $\pmb R_\textrm{cs}$ shown in
(\ref{eq:rabbit1019}), the matrix $\pmb R_\textrm{ca}$ generally
does not possess such structure except when the autocorrelation
function $R_\psi(x)$ has a finite span. We can perform whitening on
(\ref{eq:mouse0219}) to yield a linear model
\begin{equation}\label{eq:cow0219}
\pmb z=\pmb H_\textrm{ca}^\dag\pmb A\pmb b+\pmb w,
\end{equation}
where $\bb R_\textrm{ca}=\pmb H_\textrm{ca}\pmb H_\textrm{ca}^\dag$
and $\bb w=\pmb H_\textrm{ca}^{-1}\bb v$ is a white noise vector.

The discrete statistics $\pmb y$ in (\ref{eq:cow1027}) and
(\ref{eq:mouse0219}), and hence $\pmb z$ in (\ref{eq:mouse0213}) and
(\ref{eq:cow0219}), for chip-synchronous and chip-asynchronous
systems, respectively, are sufficient and are obtained in the same
way as \cite{verdu86,lupas90,rupf94}. These sufficient statistics
are the output samples of a bank of filters matched to the symbol
spreading waveforms of all users. An alternative approach to
generating statistics, adopted by
\cite{kiran00,schramm99,mantravadi02,cottatellucci04}, is to pass
the received signal to a chip matched filter and sample the output.
Statistics yielded in this way are sufficient only under symbol- and
chip-synchronous assumptions, and are not sufficient in the
chip-asynchronous case. For a chip-asynchronous system, it is
reported in \cite{mantravadi01} that, when the chip waveform is time
limited to the chip interval, statistics obtained by sampling the
chip matched filtering output at the chip rate leads to significant
degradation in performance. On the other hand, if the output is
sampled at up to the Nyquist rate, the loss in performance is
negligible.

\subsection{Asymptotic Spectral Distribution of Crosscorrelation
Matrix}\label{subsection:cow0831}

The analysis of asynchronous CDMA systems will be conducted in a
large system regime. That is, we assume both the number of users $K$
and the spreading gain $N$ approach infinity with their ratio $K/N$
converging to a non-negative constant $\beta$. To proceed the
analysis, some definitions regarding the limit of a random matrix
\cite{bai99} are introduced. Let $\bb S^{(p)}$ denote a $p\times p$
Hermitian random matrix whose each element is a random variable.
Suppose that $\bb S^{(p)}$ has eigenvalues
$\nu_1\leq\nu_2\leq\cdots\leq\nu_{p}$. Since $\bb S^{(p)}$ is
Hermitian, all $\nu_i$'s are real. The ESD of $\bb S^{(p)}$ is
defined as
\begin{equation}\label{eq:rabbit0904}
F^{(p)}(x)=p^{-1}\#\{i:\nu_i\leq x\},
\end{equation}
where $\#\{\cdots\}$ denotes the number of elements in the indicated
set. A simple fact is the $n$-th moment of $F^{(p)}(x)$ can be
represented as
\begin{equation}\label{eq:tiger0120}
m_n^{(p)}=\int_{-\infty}^\infty x^n \textrm{d}
F^{(p)}(x)=p^{-1}\tr((\bb S^{(p)})^n),
\end{equation}
where $\textrm{tr}$ is the trace operator, and the second equality
holds because $\tr((\bb S^{(p)})^n)=\sum_{i=1}^p \nu_i^n$. If
$F^{(p)}(x)$ converges to a nonrandom function $F(x)$ as
$p\rightarrow\infty$, then we say that the sequence $\{\bb
S^{(p)}:p=1,2,\cdots\}$ has an ASD $F(x)$. To show that $F^{(p)}(x)$
tends to a limit, the moment convergence theorem \cite{frechet31}
can be employed. To be specific, the theorem is stated here in a
form convenient for this paper.

\begin{theorem}\label{theorem:rabbit0121}
[\textit{Moment Convergence Theorem}] Let $\{F^{(p)}(x):
p=1,2,\cdots\}$ be a sequence of distribution for which the moments
$$
m_n^{(p)}=\int_{-\infty}^\infty x^n\textrm{d} F^{(p)}(x)
$$
exist for all $n=0,1,2,\cdots$. Furthermore, let $F(x)$ be a
distribution function for which the moments
\begin{equation}\label{eq:dragon0121}
m_n=\int_{-\infty}^\infty x^n\textrm{d} F(x)
\end{equation}
exist for all $n=0,1,2,\cdots$. If
\begin{equation}\label{eq:snake0121}
\lim_{p\rightarrow\infty} m_n^{(p)}=m_n
\end{equation}
for all $n=0,1,2,\cdots$ in some sense, and if $F(x)$ is uniquely
determined by the sequence of moments $m_0,m_1,m_2,\cdots$, then
$$
\lim_{p\rightarrow\infty}F^{(p)}(x)=F(x),
$$
and the convergence holds in the same sense as that of
(\ref{eq:snake0121}). \hfill{$\blacksquare$}
\end{theorem}

The moment convergence theorem has a long history. The details can
be found in \cite{takacs91}. In applying this theorem to show the
existence of the ASD $F(x)$, we should determine the asymptotic
moment sequence $\{m_n\}$ in (\ref{eq:dragon0121}) and prove that a
unique distribution is determined by $\{m_n\}$. In
\cite{carleman22}, Carleman gave a sufficient condition
$\sum_{n=1}^\infty m_{2n}^{-1/(2n)}=\infty$, called Carleman's
criterion, for the uniqueness of a distribution given a moment
sequence $\{m_n\}$.

Concerned with the linear models of (\ref{eq:mouse0213}) and
(\ref{eq:cow0219}), it is of interest to consider the ESD of the
random matrix $\bb H^\dag_\textrm{x}\bb A(\bb H^\dag_\textrm{x}\bb
A)^\dag$, $\textrm{x}\in\{\textrm{cs},\textrm{ca}\}$
[\citenum{tulino2004}, Chapter 1]. Represent matrices $\bb
H_\textrm{x}$, $\bb A$, and $\bb R_\textrm{x}$ by $\bb
H_\textrm{x}^{(K)}$, $\bb A^{(K)}$, and $\bb R_\textrm{x}^{(K)}$,
respectively, when the user size of the system is $K$. In order to
use Theorem~\ref{theorem:rabbit0121} to find the ASD of the random
matrix ${\bb H_\textrm{x}^{(K)}}^\dag\bb A^{(K)}{\bb
A^{(K)}}^\dag\bb H_\textrm{x}^{(K)}$, it is required to find the
limits of the empirical moments
\begin{eqnarray}
&&(2M+1)^{-1}K^{-1}\textrm{tr}(({\bb H_\textrm{x}^{(K)}}^\dag\bb
A^{(K)}{\bb A^{(K)}}^\dag\bb H_\textrm{x}^{(K)})^n),\nn\\
&=&(2M+1)^{-1}K^{-1}\textrm{tr}(({\bb A^{(K)}}^\dag\bb
R_\textrm{x}^{(K)}\bb A^{(K)})^n),\quad n\geq 0\label{eq:tiger0219}.
\end{eqnarray}
In an unfaded channel, i.e., $|A_k(m)|^2$'s are equal for all $k$
and $m$, the matrix $\bb A^{(K)}$ is a scaled identity matrix. Thus,
the quantity
\begin{equation}\label{eq:rabbit0219}
(2M+1)^{-1}K^{-1}\textrm{tr}((\bb
R_\textrm{x}^{(K)})^n)
\end{equation}
is of interest in (\ref{eq:tiger0219}).

\section{ASD of Crosscorrelation Matrix in Asynchronous CDMA}\label{section:III}

The goal of this section is to show that the ESD's of ${\bb
A^{(K)}}^\dag\bb R_\textrm{cs}^{(K)}\bb A^{(K)}$ and ${\bb
A^{(K)}}^\dag\bb R_\textrm{ca}^{(K)}\bb A^{(K)}$ converge to
nonrandom limits when $K,N\rightarrow\infty$ and
$K/N\rightarrow\beta$. We consider chip-synchronous and
chip-asynchronous CDMA in Sections \ref{subsection:mouse0121} and
\ref{subsection:cow0121}, respectively.

\subsection{Chip-Synchronous CDMA}\label{subsection:mouse0121}

We will use Theorem~\ref{theorem:rabbit0121} to prove the result
stated in the previous paragraph. We first consider the case of
unfaded channel, and then we extend to a frequency-flat fading
channel. The proof starts from showing the existence of
\begin{equation}\label{eq:mouse0120}
\mu(\bb
R_\textrm{cs}^n)=\mathop{\lim_{M,K,N\rightarrow\infty}}_{K/N\rightarrow\beta}
(2M+1)^{-1}K^{-1}\textrm{E}\left\{\tr((\bb
R_\textrm{cs}^{(K)})^n)\right\},\quad n\geq 0,
\end{equation}
where the functional $\mu(\cdot)$ is a limiting normalized expected
trace of the matrix in the argument, and the limit
$M\rightarrow\infty$ is placed because we investigate the system
behavior when the width of the observation window for symbol
detection tends to infinity. We take the relative delays $\tau_k$'s
as deterministic quantities, and the expectation of
(\ref{eq:mouse0120}) is with respect to (w.r.t.) the random
spreading sequences. We have the following lemma.

\begin{lemma}\label{lemma:rabbit0120}
In both short-code and long-code chip-synchronous CDMA systems, for
any relative delays $\{\tau_k\}_{k=1}^K$ and any chip waveform
$\psi(t)$ satisfying (\ref{eq:mouse0320}), $\mu(\bb
R_\textrm{cs}^n)$ exists and is given by
\begin{equation}\label{eq:dragon0124}
\mu(\bb R_\textrm{cs}^n)=\frac{1}{n}\sum_{j=1}^n {n\choose
j}{n\choose j-1}\beta^{j-1},\quad n\geq 0.
\end{equation}
\end{lemma}
\begin{proof}
See Appendix~\ref{appendix:dragon0120}.
\end{proof}

In Appendix~\ref{appendix:dragon0120}, we prove this lemma with the
aid of techniques from noncrossing partition. Results of noncrossing
partitions necessary for the proof are summarized in
Appendix~\ref{appendix:noncrossing}. The same tool has been employed
in [\citenum{li01},\citenum{xiao00}] for a symbol-synchronous
system. Note that, in the proof of Lemma~\ref{lemma:rabbit0120}, the
spreading sequences are only assumed to be independent across users.
For a particular user, we do not assume that the sequence is
independent across symbols. Thus, the proof is applicable to both
short-code and long-code systems. Moreover, the relative delays
$\{\tau_k\}_{k=1}^K$ are treated as deterministic constants, and we
do not adopt a particular chip waveform except for the zero ICI
condition. Thus, $\mu(\bb R_\textrm{cs}^n)$ does not depend on the
asynchronous delays and the chosen chip waveform.

\begin{lemma}\label{lemma:cow0123}
The $n$-th moment of the ESD of $\bb R_\textrm{cs}^{(K)}$ converges
a.s. to $\mu(\bb R_\textrm{cs}^n)$ when $M,K,N\rightarrow\infty$ and
$K/N\rightarrow\beta$. Moreover, the moment sequence $\{\mu(\bb
R_\textrm{cs}^n):n\geq 1\}$ satisfies the Carleman's criterion
$\sum_{n=1}^\infty \mu(\bb R_\textrm{cs}^{2n})^{-1/(2n)}=\infty$.
\end{lemma}
\begin{proof}
See Appendix~\ref{appendix:dragon0831}.
\end{proof}

Since the $n$-th moment of the ESD of $\bb R_\textrm{cs}^{(K)}$
converges to $\mu(\bb R_\textrm{cs}^n)$, we refer to $\mu(\bb
R_\textrm{cs}^n)$ as the $n$-th AEM of $\bb R_\textrm{cs}$.
It is seen that $\mu(\bb R_\textrm{cs}^n)$ given in
(\ref{eq:dragon0124}) is equal to the $n$-th moment of the
Mar\u{c}enko-Pastur distribution \cite{marcenko67} with ratio index
$\beta$, having density
$$
f_\beta(x)=\left(1-\frac{1}{\beta}\right)^+
\delta(x)+\frac{\sqrt{(x-a)^+(b-x)^+}}{2\pi \beta x},
$$
where $(z)^+=\max\{0,z\}$, $a=(1-\sqrt{\beta})^2$ and
$b=(1+\sqrt{\beta})^2$. As Lemma~\ref{lemma:cow0123} shows the
$n$-th empirical moment of $\bb R_\textrm{cs}^{(K)}$ converges to
$\mu(\bb R_\textrm{cs}^n)$ for $n\geq 0$ and the moment sequence
$\{\mu(\bb R_\textrm{cs}^n)\}$ satisfies Carleman's condition, the
following theorem holds straightforwardly due to
Theorem~\ref{theorem:rabbit0121}.

\begin{theorem}\label{theorem:snake0831}
In both short-code and long-code chip-synchronous CDMA systems, for
any relative delays $\{\tau_k\}_{k=1}^K$ of users and any arbitrary
chip waveform $\psi(t)$ satisfying the ICI free condition, the ESD
of the crosscorrelation matrix converges a.s. to the
Mar\u{c}enko-Pastur distribution with ratio index $\beta$ when
$M,K,N\rightarrow\infty$ and
$K/N\rightarrow\beta$.\hfill{$\blacksquare$}
\end{theorem}

It is known that, in symbol-synchronous CDMA, the ASD of the
crosscorrelation matrix at $K,N\rightarrow\infty$ and
$K/N\rightarrow\beta$ is the Mar\u{c}enko-Pastur law
[\citenum{verdu98}, Proposition 2.1]. Thus, an equivalence result
about symbol-synchronous and chip-synchronous CDMA in terms of the
ASD's of crosscorrelation matrices can be established as follows.
Under an unfaded channel\footnote{In an unfaded channel, the matrix
$\bb A$ in (\ref{eq:cow1027}) governing the amplitude of the
received signal is a scaled identity matrix. Thus, it is the matrix
$\bb R_\textrm{cs}$ that determines the performance of the system.},
the ASD of the crosscorrelation matrix in a chip-synchronous system
converges, as $M$ increases, to the ASD of the crosscorrelation
matrix in a symbol-synchronous system with the same $K/N$ ratio.

To consider a more realistic scenario that the signal is subject to
a fading channel, we define a quantity analogous to $\mu(\bb
R_\textrm{cs}^n)$ given in (\ref{eq:mouse0120}), expressed as
\begin{equation}\label{eq:sheep0216}
\mu((\bb A^\dag\bb R_\textrm{cs}\bb
A)^n)=\mathop{\lim_{M,K,N\rightarrow\infty}}_{K/N\rightarrow\beta}
(2M+1)^{-1}K^{-1}\textrm{E}\left\{\tr(({\bb A^{(K)}}^\dag\bb
R_\textrm{cs}^{(K)}\bb A^{(K)})^n)\right\},\quad n\geq 0.
\end{equation}
We will show below that $(2M+1)^{-1}K^{-1}\mbox{tr}(({\bb
A^{(K)}}^\dag\bb R_\textrm{cs}^{(K)}\bb A^{(K)})^n)$ converges to
its limiting mean, i.e., $\mu((\bb A^\dag\bb R_\textrm{cs}\bb
A)^n)$.

\begin{lemma}\label{corollary:cow0807}
Let ${\cal P}^{(n)}$ denote the $n$-th moment of the random variable
governing the asymptotic empirical distribution of the square
magnitudes of received amplitudes $\{|A_k(m)|^2: -\infty<m<\infty,
k=1,2,\cdots,K\}$. When $M,K,N\rightarrow\infty$ with
$K/N\rightarrow\beta$, the $n$-th moment of the ESD of ${\pmb
A^{(K)}}^\dag\pmb R_\textrm{cs}^{(K)}\pmb A^{(K)}$ converges to the
nonrandom limit $\mu((\pmb A^\dag\pmb R_\textrm{cs}\pmb A)^n)$,
given by
\begin{equation}\label{eq:dragon0728}
\mu((\pmb A^\dag\pmb R_\textrm{cs}\pmb A)^n)=\sum_{j=1}^n
\beta^{j-1}\mathop{\sum_{c_1+c_2+\cdots+c_j=n}}_{c_1\geq
c_2\geq\cdots\geq c_j\geq
1}\dfrac{n(n-1)\cdots(n-j+2)}{f(c_1,c_2,\cdots,c_j)}\prod_{r=1}^j
{\cal P}^{(c_r)},\quad n\geq 0,
\end{equation}
where $f(c_1,c_2,\cdots,c_j)$ is defined in (\ref{eq:mouse0128}) of
Appendix~\ref{appendix:noncrossing}.
\end{lemma}
\begin{proof}
See Appendix~\ref{appendix:cow0128}.
\end{proof}

We call $\mu((\pmb A^\dag\pmb R_\textrm{cs}\pmb A)^n)$ the $n$-th
AEM of the random matrix $\pmb A^\dag\pmb R_\textrm{cs}\pmb A$. The
convergence of ESD of chip-synchronous CDMA in a fading channel is
stated in the following theorem.

\begin{theorem}\label{theorem:mouse0321}
In a chip-synchronous system, if the moment sequence $\{{\cal
P}^{(n)}\}$ holds for the Carleman's criterion, then for any
relative delays $\{\tau_k\}_{k=1}^K$ and arbitrary ICI free chip
waveform $\psi(t)$, the ESD of $\{{\bb A^{(K)}}^\dag\bb
R_\textrm{cs}^{(K)}\bb A^{(K)}: K=1,2,\cdots\}$ converges to a
nonrandom limit whose $n$-th moment is equal to $\mu((\pmb
A^\dag\pmb R_\textrm{cs}\pmb A)^n)$ when $M,K,N\rightarrow\infty$
and $K/N\rightarrow\beta$.
\end{theorem}
\begin{proof}
By similar arguments as in \cite{yin83}, it can be shown that
$\sum_{n=1}^\infty \left({\cal P}^{(2n)}\right)^{-1/(2n)}=\infty$ is
a sufficient condition for $\sum_{n=1}^\infty \mu((\bb A^\dag\bb
R_\textrm{cs}\bb A)^{2n})^{-1/(2n)}=\infty$, which implies
$\{\mu((\pmb A^\dag\pmb R_\textrm{cs}\pmb A)^n)\}$ determines a
unique distribution. Thus, with Lemma~\ref{corollary:cow0807}, this
theorem follows directly from Theorem~\ref{theorem:rabbit0121}.
\end{proof}

It is shown in \cite{li01} that the counterpart of $\mu((\bb
A^\dag\bb R_\textrm{cs}\bb A)^n)$ in a symbol-synchronous system has
the same expression as (\ref{eq:dragon0728}). Thus, in a fading
channel, the ASD of the chip-synchronous system for large $M$ is
identical to that of a symbol-synchronous system, and the
equivalence result of symbol-synchronous and chip-synchronous
systems presented above for the unfaded channel can be generalized
to the case of fading channel. Actually, the equivalence of the two
systems can be discovered in an easier way. When all $\tau_k$'s are
set to zero, a chip-synchronous system becomes symbol-synchronous.
As Theorems~\ref{theorem:snake0831} and \ref{theorem:mouse0321} hold
for any realizations of relative delays, it is immediate to see the
equivalence of chip-synchronous and symbol-synchronous systems.

A related result has been demonstrated previously in \cite{kiran00}.
By assuming the density of the relative delay distribution symmetric
about $NT_c/2$, it is shown in \cite{kiran00} that, as
$M\rightarrow\infty$, a lower bound of the output SIR of the linear
MMSE receiver for chip-synchronous CDMA attains that of the same
receiver in a symbol-synchronous system. It is known that, given the
linear model of a received signal, the MMSE achievable by a linear
receiver, and hence the maximum output SIR, is dictated by the
empirical distribution of the covariance matrix of the random
channel matrix. It follows that our equivalence result on the ASD's
of the crosscorrelation matrices of chip-synchronous and
symbol-synchronous systems assures the equivalence of MMSE receiver
output SIR in the two systems. Thus, the equivalence result we
establish above holds in a more general sense, and neither an
assumption about the distribution of relative delays nor a bound is
employed.

\subsection{Chip-Asynchronous CDMA}\label{subsection:cow0121}

In computing the moments $\mu(\bb R_\textrm{ca}^n)$, defined as
(\ref{eq:mouse0120}) with $\bb R_\textrm{cs}$ replaced by $\bb
R_\textrm{ca}$, the relative delays $\tau_k$'s are regarded as
either deterministic constants or random variables depending on the
bandwidth of chip waveform $\psi(t)$. To be specific, it is known
that, to satisfy the ICI free condition, the minimum bandwidth of
$\psi(t)$ is $1/(2T_c)$ \cite{proakis00}, which corresponds to the
ideal Nyquist sinc pulse. When $\psi(t)$ has a bandwidth of
$1/(2T_c)$, $\tau_k$'s are treated as deterministic constants in the
calculation of $\mu(\bb R_\textrm{ca}^n)$; when the bandwidth of
$\psi(t)$ is larger than $1/(2T_c)$, $\tau_k$'s are taken as
independent and identically distributed (i.i.d.) random variables
whose density function possesses certain symmetry. The reason for
this setting is due to the property of chip waveform presented below
in Lemma~\ref{proposition:mouse0515}. Thus, when the sinc pulse is
employed, $\tau_k$'s are deterministic and the expectation taken in
$\mu(\bb R_\textrm{ca})$ is w.r.t. random spreading sequences. If
other chip waveforms are used, resulting in a bandwidth lager than
$1/(2T_c)$, the expectation in $\mu(\bb R_\textrm{ca})$ is w.r.t.
both spreading sequences and users' relative delays.

\begin{lemma}\label{proposition:mouse0515}
Denote the Fourier transform of a real pulse $\psi(t)$ by
$$
\Psi(\Omega)=\int_{-\infty}^\infty \psi(t)e^{-j \Omega
t}\textrm{d}t.
$$
Let
$$
R_\psi(x)=\int_{-\infty}^\infty \psi(t)\psi(t-x)\textrm{d}t
$$
be the autocorrelation function of $\psi(t)$. Define
\begin{eqnarray}
&&\Xi_\psi(\{n_i\}_{i=0}^{m-1};\{\eta_i\}_{i=0}^{m-1})\label{eq:mouse0222}\\
&=&R_\psi((n_0-n_1)T_c+\eta_0-\eta_1)R_\psi((n_1-n_2)T_c+\eta_1-\eta_2)
\cdots R_\psi((n_{m-1}-n_0)T_c+\eta_{m-1}-\eta_0).\nn
\end{eqnarray}
We have the following results about
$\Xi_\psi(\{n_i\}_{i=0}^{m-1};\{\eta_i\}_{i=0}^{m-1})$.
\begin{enumerate}
\item For any $n_0\in
\mathbb{Z}$ and $\{\eta_j\}_{j=0}^{m-1}\in \mathbb{R}^m$, we have
\begin{eqnarray}
&&
\sum_{\{n_1,\cdots,n_{m-1}\}\in[-\infty,\infty]^{m-1}}\Xi_\psi(\{n_i\}_{i=0}^{m-1};\{\eta_i\}_{i=0}^{m-1})
\label{eq:tiger0826}\\
&=& \frac{1}{2\pi T_c^{m-1}}\int_{-\pi/T_c}^{\pi/T_c}
\left|\Psi\left(\Omega\right)\right|^{2m}\textrm{d}\Omega,\qquad
m=1,2,\cdots,\label{eq:horse0804}
\end{eqnarray}
if the bandwidth of $\psi(t)$ is less than $1/(2T_c)$, i.e.,
$\Psi(\Omega)=0$ for $\Omega>\pi/T_c$.
\item For any $n_0\in \mathbb{Z}$,
$\eta_0\in\mathbb{R}$, and i.i.d. random variables
$\{\eta_i\}_{i=1}^{m-1}$ satisfying $\textrm{E}\left\{\cos(2\pi
k\eta_i/T_c)\right\}=0$ for any nonzero integer $k$, we have
\begin{eqnarray}
&&
\sum_{\{n_1,\cdots,n_{m-1}\}\in[-\infty,\infty]^{m-1}}\textrm{E}\left\{\Xi_\psi(\{n_i\}_{i=0}^{m-1};
\{\eta_i\}_{i=0}^{m-1})\right\}
\label{eq:chicken0804}\\
&=& \frac{1}{2\pi T_c^{m-1}}\int_{-\infty}^\infty
\left|\Psi\left(\Omega\right)\right|^{2m}\textrm{d} \Omega,\qquad
m=1,2,\cdots,\label{eq:sheep0804}
\end{eqnarray}
if the bandwidth of $\psi(t)$ is greater than $1/(2T_c)$.
\end{enumerate}
\end{lemma}
\begin{proof}
(Outline) This lemma can be proved by applying Parseval's theorem
repeatedly for each summation variable $n_1,n_2,\cdots,n_{m-1}$ in
(\ref{eq:tiger0826}) and (\ref{eq:chicken0804}). Since the arguments
of $R_\psi(\cdot)$'s are cyclic, i.e., in the forms of
$(n_0-n_1)T_c+\eta_0-\eta_1,
(n_1-n_2)T_c+\eta_1-\eta_2,\cdots,(n_{m-1}-n_0)T_c+\eta_{m-1}-\eta_0$,
the complex exponentials due to Fourier transforms of time-shifted
autocorrelation functions cancel each other. For the detail of the
proof, see Appendix~\ref{appendix:mouse0826}.
\end{proof}

Convergence of the ESD of $\{\bb R_\textrm{ca}^{(K)}:K=1,2,\cdots\}$
to a nonrandom limit when $M,K,N\rightarrow\infty$ and
$K/N\rightarrow\beta$ is proved below.
We define ${\cal W}_\psi^{(m)}$ as the quantity given in
(\ref{eq:horse0804}) and (\ref{eq:sheep0804}), i.e.,
$$
{\cal W}_\psi^{(m)}=\frac{1}{2\pi T_c^{m-1}}\int_{-\infty}^\infty
\left|\Psi\left(\Omega\right)\right|^{2m}\textrm{d} \Omega,\qquad
m=1,2,\cdots.
$$

\begin{lemma}\label{theorem:cow0902}
Consider a chip-asynchronous system whose quantity ${\cal
W}_\psi^{(m)}$ corresponding to the chip waveform exists for all
$m\geq 1$. When the sinc pulse is employed as the chip waveform, the
relative delays $\tau_k$'s are treated as deterministic; while if
the bandwidth of the chip waveform is larger than $1/(2T_c)$, then
$\tau_k$'s are viewed as i.i.d. random variables having
$\textrm{E}\{\cos(2\pi n\tau_k/T_c)\}=0$ for any nonzero integer
$n$. For both short-code and long-code systems, when
$M,K,N\rightarrow\infty$ with $K/N\rightarrow\beta$, $\mu(\pmb
R_\textrm{ca}^n)$ exists and is given by
\begin{equation}\label{eq:tiger0728}
\mu(\pmb R_\textrm{ca}^n)=\sum_{j=1}^n
\beta^{j-1}\mathop{\sum_{b_1+b_2+\cdots+b_{n-j+1}=n}}_{b_1\geq
b_2\geq\cdots\geq b_{n-j+1}\geq
1}\dfrac{n(n-1)\cdots(j+1)}{f(b_1,b_2,\cdots,b_{n-j+1})}
\prod_{r=1}^{n-j+1}{\cal W}_\psi^{(b_r)},\quad n\geq 0.
\end{equation}
\end{lemma}
\begin{proof}
See Appendix~\ref{appendix:dragon0120}.
\end{proof}
In the proof, when the bandwidth of $\psi(t)$ is greater than
$1/(2T_c)$, the formula of $\mu(\bb R_\textrm{ca}^n)$ is obtained by
means of the chip waveform property in part 2) of
Lemma~\ref{proposition:mouse0515}, which holds when distribution of
$\tau_k$'s has $\textrm{E}\left\{\cos(2\pi n\tau_k/T_c)\right\}=0$
for any nonzero integer $n$. A special case for this zero
expectation is the uniform distribution in the interval $[0,rT_c)$,
$r\in \mathbb{N}$, which encompasses the symbol quasi-synchronous
but chip-asynchronous system considered in \cite{cottatellucci06}.
Thus, as the equivalence in AEM leads to an equivalence in ASD,
Lemma~\ref{theorem:cow0902} provides with a proof for the conjecture
proposed in \cite{cottatellucci06} that the symbol quasi-synchronous
but chip-asynchronous system has the same performance as a
chip-asynchronous system.

\begin{theorem}\label{theorem:cow0217}
Suppose that the chip waveform $\psi(t)$ has a finite bandwidth
denoted by $\textit{BW}$. If the sequence $\{{\cal
W}_\psi^{(n)}:n\geq 1\}$ corresponding to $\psi(t)$ satisfies
$\sum_{n=1}^\infty \left({\cal
W}_\psi^{(2n)}/2\textit{BW}\right)^{-1/(2n)}=\infty$, then the ESD
of $\{\bb R_\textrm{ca}^{(K)}: K=1,2,\cdots\}$ converges a.s. to a
nonrandom limit whose $n$-th moment is equal to $\mu(\bb
R_\textrm{ca}^n)$ when $M,K,N\rightarrow\infty$ and
$K/N\rightarrow\beta$.
\end{theorem}
\begin{proof}
We rewrite ${\cal W}_\psi^{(n)}$ in (\ref{eq:horse0804}) and
(\ref{eq:sheep0804}) as
$$
{\cal W}_\psi^{(n)}=\int_S\hspace{1mm}
\Biggl|\dfrac{\Psi\Bigl(\frac{2\pi}{T_c}f\Bigr)}{\sqrt{T_c}}\Biggr|^{2n}\textrm{d}f,
$$
where $f\in S$ if $\Psi(2\pi f/T_c)\neq 0$. The measure of $S$ is
equal to $2\textit{BW}$. It is clear that $|\Psi(2\pi
f/T_c)/\sqrt{T_c}|^2$ belongs to the space of integrable functions,
and the set $S$ is a measurable subset of real numbers with the
Lebesgue measure. By a generalization of H\"{o}lder's inequality
\cite{finner92}, we have
$$
{\cal W}_\psi^{(k)}\leq (2\textit{BW})^{1-k/n}\left({\cal
W}_\psi^{(n)}\right)^{k/n},\quad 1\leq k<n.
$$
Thus, we have the product of ${\cal W}_\psi^{(b_r)}$'s in
(\ref{eq:tiger0728}) upper-bounded by
\begin{equation}\label{eq:tiger0131}
\prod_{r=1}^{n-j+1}{\cal W}_\psi^{(b_r)}\leq \prod_{r=1}^{n-j+1}
(2\textit{BW})^{1-b_r/n}\left({\cal
W}_\psi^{(n)}\right)^{b_r/n}=(2\textit{BW})^{n-j}{\cal
W}_\psi^{(n)}.
\end{equation}
We use similar arguments as in \cite{yin83} to show $\{\mu(\bb
R_\textrm{ca}^n)\}$ satisfies the Carleman's criterion. That is, we
can bound $\mu(\bb R_\textrm{ca}^n)$ by
\begin{eqnarray}
\mu(\bb R_\textrm{ca}^n)&\leq& (2\textit{BW})^{n-1}{\cal
W}_\psi^{(n)}\sum_{j=1}^n
\left(\frac{\beta}{2\textit{BW}}\right)^{j-1}\mathop{\sum_{b_1+b_2+\cdots+b_{n-j+1}=n}}_{b_1\geq
b_2\geq\cdots\geq b_{n-j+1}\geq
1}\dfrac{n(n-1)\cdots(j+1)}{f(b_1,b_2,\cdots,b_{n-j+1})}\nn\\
&=&(2\textit{BW})^{n-1}{\cal W}_\psi^{(n)}\sum_{j=1}^n
\left(\frac{\beta}{2\textit{BW}}\right)^{j-1}\dfrac{1}{n}{n\choose
j}{n\choose j-1}\nn\\
&\leq& (2\textit{BW})^{n-1}{\cal
W}_\psi^{(n)}\left(1+\frac{\beta}{2\textit{BW}}\right)^{2n}.\label{eq:mouse0217}
\end{eqnarray}
So,
$$
\sum_{n=1}^\infty \mu(\bb R_\textrm{ca}^{2n})^{-1/(2n)}\geq
(2\textit{BW})^{-1}\left(1+\frac{\beta}{2\textit{BW}}\right)^{-2}
\sum_{n=1}^\infty \left({\cal
W}_\psi^{(2n)}/2\textit{BW}\right)^{-1/(2n)}=\infty.
$$
It follows tha the moment sequence $\{\mu(\pmb R_\textrm{ca}^n)\}$
determines a unique distribution. Besides, pursuing the same lines
of the proof for Lemma~\ref{lemma:cow0123} presented in
Appendix~\ref{appendix:dragon0831}, we can show the $n$-th moment of
the ESD of $\bb R_\textrm{ca}^{(K)}$ converges a.s. to $\mu(\bb
R_\textrm{ca}^n)$ when $M,K,N\rightarrow\infty$ and
$K/N\rightarrow\beta$. Thus, this theorem follows directly from
Theorem~\ref{theorem:rabbit0121}.
\end{proof}

We now consider the situation that the signal is subject to a
frequency-flat fading channel. Define a quantity $\mu((\pmb
A^\dag\pmb R_\textrm{ca}\pmb A)^n)$ analogous to $\mu((\pmb
A^\dag\pmb R_\textrm{cs}\pmb A)^n)$ of (\ref{eq:sheep0216}) by
replacing $\bb R_\textrm{cs}$ therein with $\bb R_\textrm{ca}$. We
give the following theorem.

\begin{theorem}
When $M,K,N\rightarrow\infty$ with $K/N\rightarrow\beta$, the ESD of
$\{{\bb A^{(K)}}^\dag \bb R_\textrm{ca}^{(K)}\bb A^{(K)}:
K=1,2,\cdots\}$ converges to a nonrandom limit whose $n$-th moment
is given by
\begin{eqnarray}\label{eq:mouse0523}
&&\mu((\pmb A^\dag\pmb R_\textrm{ca}\pmb A)^n)=\sum_{j=1}^n
\beta^{j-1}\mathop{\sum_{b_1+b_2+\cdots+b_{n-j+1}=n}}_{b_1\geq
b_2\geq\cdots\geq b_{n-j+1}\geq
1}\nn\\
&&\times\mathop{\sum_{c_1+c_2+\cdots+c_j=n}}_{c_1\geq
c_2\geq\cdots\geq c_j\geq
1}\dfrac{n(n-j)!(j-1)!}{f(b_1,b_2,\cdots,b_{n-j+1})
f(c_1,c_2,\cdots,c_j)}\prod_{t=1}^{n-j+1}{\cal
W}_\psi^{(b_t)}\prod_{r=1}^{j}{\cal P}^{(c_r)},\quad n\geq 1,
\end{eqnarray}
if the sequences $\{{\cal P}^{(n)}:n\geq 1\}$ and $\{{\cal
W}_\psi^{(n)}:n\geq 1\}$ satisfy $\sum_{n=1}^\infty \left({\cal
P}^{(2n)}{\cal W}_\psi^{(2n)}/2\textit{BW}\right)^{-1/(2n)}=\infty$.
\end{theorem}

\begin{proof}
First, we prove the $n$-th AEM of ${\pmb A}^\dag\pmb
R_\textrm{ca}\pmb A$, i.e., $\mu((\pmb A^\dag\pmb R_\textrm{ca}\pmb
A)^n)$, is given as (\ref{eq:mouse0523}). The proof follows the
lines of Lemma~\ref{corollary:cow0807}'s proof given in
Appendix~\ref{appendix:cow0128}. It can be shown that $\mu((\pmb
A^\dag\pmb R_\textrm{ca}\pmb A)^n)$ is expressed as (cf.
(\ref{eq:mouse0807}))
\begin{eqnarray}
&&
\mathop{\lim_{K,N,M\rightarrow\infty}}_{K/N\rightarrow\beta}K^{-1}
\sum_{j=1}^n\mathop{\sum_{b_1+b_2+\cdots+b_{n-j+1}=n}}_{b_1\geq
b_2\geq\cdots b_{n-j+1}\geq
1}\mathop{\sum_{c_1+c_2+\cdots+c_j=n}}_{c_1\geq c_2\geq\cdots
c_j\geq 1} \dfrac{n(n-j)!(j-1)!}
{f(b_1,b_2,\cdots,b_{n-j+1})f(c_1,c_2,\cdots,c_j)}
\nn\\
&&\times  \prod_{s=0}^{j-1}(K-s)\cdot
N^{-j+1}\prod_{t=1}^{n-j+1}{\cal W}_\psi^{(b_t)} \cdot\prod_{r=1}^j
{\cal P}^{(c_r)},
\end{eqnarray}
which is equal to (\ref{eq:mouse0523}).

Secondly, we would show $\sum_{n=1}^\infty \left({\cal
P}^{(2n)}{\cal W}_\psi^{(2n)}/2\textit{BW}\right)^{-1/(2n)}=\infty$
is a sufficient condition that the sequence $\{\mu((\pmb A^\dag\pmb
R_\textrm{ca}\pmb A)^n)\}$ determines a unique distribution. By a
generalization of H\"{o}lder's inequality \cite{finner92}, we have
${\cal P}^{(k)}\leq \left({\cal P}^{(n)}\right)^{k/n}$ for $1\leq
k<n$. Consequently, the product of ${\cal P}^{(c_r)}$'s in
(\ref{eq:mouse0523}) is bounded as
\begin{equation}
\prod_{r=1}^j {\cal P}^{(c_r)}\leq \left({\cal
P}^{(n)}\right)^{(c_1+c_2+\cdots+c_j)/n}={\cal P}^{(n)}.
\end{equation}
Incorporating the inequality of (\ref{eq:tiger0131}), we can
upper-bound $\mu((\bb A^\dag \bb R_\textrm{ca}\bb A)^n)$ by
\begin{eqnarray}
\mu((\bb A^\dag \bb R_\textrm{ca}\bb A)^n)&\leq&
(2\textit{BW})^{n-1}{\cal W}_\psi^{(n)}{\cal P}^{(n)}\sum_{j=1}^n
\left(\frac{\beta}{2\textit{BW}}\right)^{j-1}\nn\\
&&\times\mathop{\sum_{b_1+b_2+\cdots+b_{n-j+1}=n}}_{b_1\geq
b_2\geq\cdots b_{n-j+1}\geq
1}\mathop{\sum_{c_1+c_2+\cdots+c_j=n}}_{c_1\geq c_2\geq\cdots
c_j\geq 1} \dfrac{n(n-j)!(j-1)!}
{f(b_1,b_2,\cdots,b_{n-j+1})f(c_1,c_2,\cdots,c_j)}.\nn
\end{eqnarray}
Proceeding in a similar way as the proof of
Theorem~\ref{theorem:cow0217}, we are able to demonstrate that the
condition $\sum_{n=1}^\infty \left({\cal P}^{(2n)}{\cal
W}_\psi^{(2n)}/2\textit{BW}\right)^{-1/(2n)}=\infty$ is sufficient
for $\sum_{n=1}^\infty \mu((\bb A^\dag\bb R_\textrm{ca}\bb
A)^{2n})^{-1/(2n)}=\infty$, which gaurantees that $\{\mu((\bb
A^\dag\bb R_\textrm{ca}\bb A)^{n})\}$ determines a unique
distribution.
\end{proof}

We use the following corollary to establish the equivalence result
of systems with three synchronism levels when $M\rightarrow\infty$
and the sinc chip waveform is employed.

\begin{corollary}\label{corollary:mouse0902}
If $M\rightarrow\infty$ and the ideal Nyquist sinc chip waveform
\begin{equation}\label{eq:dragon0216}
\psi^\star(t)=\dfrac{1}{\sqrt{T_c}}\sinc\left(\dfrac{t}{T_c}\right)
\end{equation}
is used, the ESD of $\{\bb R_\textrm{ca}^{(K)}: K=1,2,\cdots\}$
converges to the Mar\u{c}enko-Pastur law with ratio index $\beta$.
Under the same premise, the ESD's of $\{{\bb A^{(K)}}^\dag\bb
R_\textrm{ca}^{(K)}\bb A^{(K)}: K=1,2,\cdots\}$ and $\{{\bb
A^{(K)}}^\dag\bb R_\textrm{cs}^{(K)}\bb A^{(K)}: K=1,2,\cdots\}$
converge to the same limiting distribution, provided that
$\sum_{n=1}^\infty \left({\cal
P}^{(2n)}T_c\right)^{-1/(2n)}=\infty$.
\end{corollary}

\begin{proof}
The Fourier transform of $\psi^\star(t)$ is
$$
\Psi^\star(\Omega)=\sqrt{T_c}\textrm{rect}
\left(\dfrac{T_c\Omega}{2\pi}\right),
$$
where $\textrm{rect}(x)=1$ for $-1/2\leq x\leq 1/2$ and equal to $0$
otherwise. By part 1) of Lemma~\ref{proposition:mouse0515}, ${\cal
W}_{\psi^\star}^{(m)}=1$ for all $m\in \mathbb{N}$. Due to
(\ref{eq:mouse0804}), the formula of $\mu(\pmb R_\textrm{ca}^n)$ in
(\ref{eq:tiger0728}) is equal to (\ref{eq:dragon0124}), which is the
$n$-th moment of the Mar\u{c}enko-Pastur distribution. By the moment
convergence theorem, the first part of this corollary follows. The
proof of the second part is straightforward, where the equality of
(\ref{eq:mouse0218}) is helpful.
\end{proof}

It is demonstrated in \cite{mantravadi02} that, when the sinc chip
waveform is used and $M\rightarrow\infty$, the asymptotic SIR at the
linear MMSE detector output is the same for all of the three
synchronism levels\footnote{The equivalence results shown in
\cite{mantravadi02} holds in a more general sense. That is, for any
finite $M$, the output SIR of the MMSE detector in the
chip-asynchronous system converges in mean-square sense to the SIR
for the chip-synchronous system.}. This equivalence result on output SIR can be seen as a direct
consequence of the equivalence of ASD demonstrated by
Theorem~\ref{theorem:snake0831} and
Corollary~\ref{corollary:mouse0902}. It is shown in \cite{guo2002}
that the linear MMSE receiver belongs to the family of linear
receivers that can be arbitrarily well approximated by polynomials
receivers\footnote{Although the result is presented in
\cite{guo2002} for symbol-synchronous CDMA, the proof (Lemma 5 of
\cite{guo2002}) can be extended to asynchronous systems in a
straightforward manner.}, i.e., in the form of
$$
f(\bb A^\dag\bb R\bb A)=a_0 \bb I+a_1 \bb A^\dag\bb R\bb
A+\cdots+a_n (\bb A^\dag\bb R\bb A)^n,
$$
with $\bb R$ standing for the crosscorrelation matrix in the system.
In general, the accuracy of the approximation is in proportional to
the order $n$ of the polynomial. Both the coefficients $a_i$'s and
the receiver output SIR can be determined by the AEM of $\bb
A^\dag\bb R\bb A$ \cite{moshavi96,muller2001,tulino01,honig2001}. As
AEM are equivalent in systems of three synchronism levels under the
indicated circumstances, both the coefficients of the three
polynomial receivers approximating linear MMSE receivers and their
output SIR are identical. It is readily seen that the equivalence
result is true not only for the linear MMSE receiver but also for
all receivers in the family defined in \cite{guo2002}, which proves
the conjecture proposed in \cite{mantravadi02}.

Up to now, the chip waveform is assumed to be ICI free for both
chip-synchronous and chip-asynchronous systems. This ICI free
condition requires that the chip waveform has a bandwidth no less
than $1/(2T_c)$ \cite{proakis00}. Here we extend the equivalence
results to the circumstance where the bandwidth of $\psi(t)$ is less
than $1/(2T_c)$ so that zero ICI condition does not exist. At this
moment, the crosscorrelation $\rho_\textrm{cs}(m,n;k,l)$ given in
(\ref{eq:tiger0320}) is no longer correct. Instead, it has the same
form as that of a chip-asynchronous system given in
(\ref{eq:mouse0323}). The crosscorrelation in a symbol-synchronous
system has the same expression as well by letting $\tau_k=\tau_l=0$.
Setting $\eta_i$'s in part 1) of Lemma~\ref{proposition:mouse0515}
as the relative delays among users, it is shown by the lemma that
$\sum_{\{n_1,\cdots,n_{m-1}\}\in[-\infty,\infty]^{m-1}}\Xi_\psi(\{n_i\}_{i=0}^{m-1};\{\tau_i\}_{i=0}^{m-1})$
does not depend on realizations of relative delays. That is, this
expression yields the same value in systems of three synchronism
levels. Tracing Appendix~\ref{appendix:dragon0120} for the proof of
Lemma~\ref{theorem:cow0902}, we find out AEM formulas $\mu(\bb
R_\textrm{cs}^n)$ and $\mu((\bb A^\dag\bb R_\textrm{cs}\bb A)^n)$
have the same expressions as their counterparts in chip-asynchronous
system, given by (\ref{eq:tiger0728}) and (\ref{eq:mouse0523}),
respectively.
Consequently, symbol-synchronous, chip-synchronous and
chip-asynchronous systems have the same ASD when the chip waveform
bandwidth is less than $1/(2T_c)$. Along with the equivalence result
concerning the sinc chip waveform in
Corollary~\ref{corollary:mouse0902}, the above discussion leads to
the following corollary.
\begin{corollary}\label{corollary:horse0216}
Suppose that $M\rightarrow\infty$ and a chip waveform with bandwidth
no greater than $1/(2T_c)$ is adopted. In either the unfaded or
fading channel, systems with three levels of synchronism have the
same ASD. $\hfill{\small\blacksquare}$
\end{corollary}

\section{More Results by Free Probability Theory}\label{section:freeprob}

In this section, we use free probability theory to obtain more
results about the asymptotic convergence of eigenvalues of
crosscorrelation matrices in asynchronous CDMA. Free probability is
a discipline founded by Voiculescu \cite{voiculescu83} in 1980s that
studies non-commutative random variables. Random matrices are
non-commutative objects whose large-dimension asymptotes provide the
major applications of the free probability theory. For convenience,
the definition of asymptotic freeness of two random matrices by
Voiculescu \cite{voiculescu91} is given below.

\begin{definition}
\cite{voiculescu91} The Hermitian random matrices $\pmb B$ and $\pmb
C$ are asymptotically free if, for all polynomials $p_j(\cdot)$ and
$q_j(\cdot)$, $1\leq j\leq n$, such that $\mu(p_j(\pmb
B))=\mu(q_j(\pmb C))=0$, we have
$$
\mu(p_1(\pmb B)q_1(\pmb C)\cdots p_n(\pmb B)q_n(\pmb C))=0.
$$
\begin{flushright}
$\blacksquare$
\end{flushright}
\end{definition}
In this definition, the functional $\mu(\cdot)$ is used. As we have
shown in (\ref{eq:mouse0120}), $\mu(\cdot)$ is a limiting normalized
expected trace of the matrix in the argument. Let $\bb B$ be sized
by $b\times b$, and we have a polynomial $p(x)=\sum_{i=0}^n a_i
x^i$. Then
$$
\mu(p(\bb B))=\lim_{b\rightarrow\infty}b^{-1}\sum_{i=0}^n
a_i\textrm{E}\{\textrm{tr}(\bb B^i)\}.
$$
Asymptotic freeness is related to the spectra of algebra of random
matrices $\pmb B$ and $\pmb C$ when their sizes tend to infinity. In
our context, the random matrices $\pmb R_\textrm{cs}$ and $\pmb
R_\textrm{ca}$ have column and row sizes equal to $(2M+1)K$
controlled by two parameters $M$ and $K$. Since the asymptotes of
$\pmb R_\textrm{cs}$ and $\pmb R_\textrm{ca}$ are studied when the
size of observation window $2M+1$ is large, we let both $M$ and $K$
approach infinity.

In the following theorem, we show that $\pmb R_\textrm{x}$,
$\textrm{x}\in\{\textrm{cs},\textrm{ca}\}$, is asymptotically free
with a diagonal random matrix $\bb D$ whose statistical description
is detailed in the theorem. This asymptotic freeness property enable
us to find the free cumulants of $\bb R_\textrm{x}$ and AEM's of
matrix sum $\pmb R_\textrm{x}+\pmb D$ and matrix product $\pmb
R_\textrm{x}\pmb D$.

\begin{theorem}\label{theorem:mouse0810}
Suppose that $\uu D(m)=\diag\{d_1(m),d_2(m),\cdots,d_{K}(m)\}$ and
$\pmb D=\diag\{\uu D(-M),$ $\uu D(-M+1),\cdots,\uu D(M)\}$, where
$d_k(m)$'s are random variables having bounded moments, and
$d_k(m_1)$ and $d_l(m_2)$ are independent if $k\neq l$. Also, $\pmb
R_\textrm{x}$, $\textrm{x}\in\{\textrm{cs,ca}\}$, and $\pmb D$ are
independent. Then $\pmb R_\textrm{cs}$ and $\pmb D$ are
asymptotically free as $M,K,N\rightarrow\infty$ with
$K/N\rightarrow\beta$. Moreover, if any of the following two
conditions holds:
\begin{enumerate}
\item The random variables $d_k(m)$'s are non-negative, and ${\cal W}_\psi^{(m)}$ exists for all $m\geq
1$,

\item For any $n_0\in \mathbb{Z}$ and $\{\eta_j\}_{j=0}^{m-1}\in
\mathbb{R}^m$, we have
$$
\sum_{n_1,\cdots,n_{m-1}\in[-\infty,\infty]}\left|\Xi_\psi(\{n_i\}_{i=0}^{m-1};\{\eta_i\}_{i=0}^{m-1})\right|
=O(1),\qquad m=1,2,\cdots,
$$
\end{enumerate}
then $\pmb R_\textrm{ca}$ and $\pmb D$ are asymptotically free.
\end{theorem}

\begin{proof}
See Appendix~\ref{appendix:rabbit0826}.
\end{proof}
Before we proceed, some results of free probability theory about
random matrices (see, for example, \cite{hiai2000}) are summarized
in the following theorem.
\begin{theorem}\label{theorem:sheep0811}
\cite{hiai2000} Let $\pmb B$ and $\pmb C$ be asymptotically free
random matrices. The $n$-th AEM of the sum $\pmb B+\pmb C$ and
product $\pmb B\pmb C$ can be given by
\begin{equation}\label{eq:rabbit0227}
\mu((\pmb B+\pmb C)^n)=\sum_{\varpi}\prod_{V\in\varpi}
\left(c_{|V|}(\pmb B)+ c_{|V|}(\pmb C)\right),
\end{equation}
and
\begin{equation}\label{eq:dragon0227}
\mu((\pmb B\pmb C)^n)=\sum_{\varpi}\prod_{V\in\varpi} c_{|V|}(\pmb
B)\prod_{U\in \mathit{KC}(\varpi)} \mu(\pmb C^{|U|}),
\end{equation}
where each summation is over all noncrossing partitions $\varpi$ of
a totally ordered $n$-element set, $V\in\varpi$ means $V$ is a class
of $\varpi$, $|V|$ denotes the cardinality of $V$, $c_k(\pmb B)$ is
the $k$-th free cumulant of $\pmb B$, and $\mathit{KC}(\varpi)$ is
the Kreweras complementation map of $\varpi$. Moreover, the
relations between the asymptotic moment and free cumulant sequences
are
\begin{eqnarray}
\mu(\pmb B^n)&=&
\sum_{\varpi} \prod_{V\in\varpi}c_{|V|}(\pmb B),\label{eq:tiger0811}\\
c_n(\pmb B)&=&\sum_{\varpi} \prod_{V\in\varpi} \mu(\pmb
B^{|V|})\prod_{U\in \mathit{KC}(\varpi)} {\cal
S}_{|U|},\label{eq:rabbit0811}
\end{eqnarray}
where
$$
{\cal S}_k=(-1)^{k-1} \frac{1}{k}{2k-2\choose k-1}.
$$
\begin{flushright}
$\blacksquare$
\end{flushright}
\end{theorem}
With the aid of Theorem~\ref{theorem:sheep0811}, we consider free
cumulants of $\pmb R_\textrm{x}$ and $\bb A^\dag\pmb R_\textrm{x}\bb
A$ for $\textrm{x}\in\{\textrm{cs},\textrm{ca}\}$. Rewrite
(\ref{eq:tiger0728}) as
\begin{equation}\label{eq:cow0811}
\mu(\pmb R_\textrm{ca}^n)=\sum_{j=1}^n
\mathop{\sum_{b_1+b_2+\cdots+b_j=n}}_{b_1\geq b_2\geq\cdots\geq
b_j\geq 1}\dfrac{n(n-1)\cdots(n-j+2)}{f(b_1,b_2,\cdots,b_j)}
\prod_{r=1}^j{\cal W}_\psi^{(b_r)}\beta^{b_r-1}.
\end{equation}
Let us interpret the summation variable $j$ in (\ref{eq:cow0811}) as
the number of classes of a noncrossing partition $\varpi$ of an
$n$-element ordered set, and $b_r$ is the size of the $r$-th class
of $\varpi$. From (\ref{eq:tiger0811}), it is readily seen that the
$n$-th free cumulant of $\pmb R_\textrm{ca}$ is
$$
c_n(\pmb R_\textrm{ca})={\cal W}_\psi^{(n)}\beta^{n-1}.
$$
Similarly, we obtain the $n$-th free cumulant of $\pmb
R_\textrm{cs}$ as
$$
c_n(\pmb R_\textrm{cs})=\beta^{n-1}.
$$
Regarding the free cumulants of $\bb A^\dag\bb R_\textrm{cs}\bb A$
and $\bb A^\dag\bb R_\textrm{ca}\bb A$, they are difficult to be
identified directly from (\ref{eq:tiger0811}). Instead, we rewrite
(\ref{eq:rabbit0811}) in a more detailed way as
\begin{eqnarray}
c_n(\pmb B)&=&\sum_{j=1}^n
\mathop{\sum_{b_1+b_2+\cdots+b_j=n}}_{b_1\geq b_2\geq\cdots\geq
b_j\geq 1} \mathop{\sum_{c_1+c_2+\cdots+c_{n-j+1}=n}}_{c_1\geq
c_2\geq\cdots\geq c_{n-j+1}\geq 1}
\dfrac{n(n-j)!(j-1)!}{f(b_1,b_2,\cdots,b_j)
f(c_1,c_2,\cdots,c_{n-j+1})}\nn\\
&&\times\prod_{r=1}^{j}\mu(\pmb B^{b_r}) \prod_{t=1}^{n-j+1} {\cal
S}_{c_t}.\label{eq:tiger0227}
\end{eqnarray}
As AEM's are available for both $\bb A^\dag\bb R_\textrm{cs}\bb A$
and $\bb A^\dag\bb R_\textrm{ca}\bb A$ in (\ref{eq:dragon0728}) and
(\ref{eq:mouse0523}), respectively, their free cumulants can be
computed from (\ref{eq:tiger0227}).

Let $\bb D$ be a $(2M+1)K\times (2M+1)K$ diagonal random matrix with
the statistical properties stated in
Theorem~\ref{theorem:mouse0810}. Since $\bb D$ and $\bb
R_\textrm{x}$, $\textrm{x}\in\{\textrm{cs},\textrm{ca}\}$ are
asymptotically free, (\ref{eq:rabbit0227}) and (\ref{eq:dragon0227})
hold. Suppose that either the AEM or free cumulants of $\bb D$ are
available. We have the $n$-th AEM of $\pmb R_\textrm{ca}+\pmb D$ and
$\pmb R_\textrm{ca}\pmb D$ given as
\begin{equation}\label{eq:snake0811}
\mu((\pmb R_\textrm{ca}+\pmb D)^n)=\sum_{j=1}^n
\mathop{\sum_{b_1+b_2+\cdots+b_j=n}}_{b_1\geq b_2\geq\cdots\geq
b_j\geq 1}\dfrac{n(n-1)\cdots(n-j+2)}{f(b_1,b_2,\cdots,b_j)}
\prod_{r=1}^j\left({\cal W}_\psi^{(b_r)}\beta^{b_r-1}+c_{b_r}(\pmb
D)\right),
\end{equation}
and
\begin{eqnarray}\label{eq:horse0811}
\mu((\pmb R_\textrm{ca}\pmb D)^n)&=&\sum_{j=1}^n
\mathop{\sum_{b_1+b_2+\cdots+b_j=n}}_{b_1\geq b_2\geq\cdots\geq
b_j\geq 1} \mathop{\sum_{c_1+c_2+\cdots+c_{n-j+1}=n}}_{c_1\geq
c_2\geq\cdots\geq c_{n-j+1}\geq 1}
\dfrac{n(n-j)!(j-1)!}{f(b_1,b_2,\cdots,b_j)
f(c_1,c_2,\cdots,c_{n-j+1})}\nn\\
&&\times\prod_{r=1}^{j}{\cal W}_\psi^{(b_r)}\beta^{b_r-1}
\prod_{t=1}^{n-j+1}\mu(\pmb D^{c_t}).
\end{eqnarray}
By setting $\pmb D=\pmb A\pmb A^\dag$, we have $\mu(\pmb D^k)={\cal
P}^{(k)}$, where ${\cal P}^{(k)}$ is defined in
Lemma~\ref{corollary:cow0807}. In this way, (\ref{eq:horse0811})
becomes (\ref{eq:mouse0523}).

The AEM $\mu((\pmb R_\textrm{cs}+\pmb D)^n)$ and $\mu((\pmb
R_\textrm{cs}\pmb D)^n)$ can be obtained from (\ref{eq:snake0811})
and (\ref{eq:horse0811}), respectively, by setting all ${\cal
W}_\psi^{(k)}$'s equal to one. In this way, $\mu((\pmb
R_\textrm{cs}\pmb D)^n)$ has a simpler form of
$$
\mu((\pmb R_\textrm{cs}\pmb D)^n)=\sum_{j=1}^n
\beta^{n-j}\mathop{\sum_{c_1+c_2+\cdots+c_{n-j+1}=n}}_{c_1\geq
c_2\geq\cdots\geq c_{n-j+1}\geq
1}\frac{n(n-1)\cdots(j+1)}{f(c_1,c_2,\cdots,c_{n-j+1})}
\prod_{r=1}^{n-j+1}\mu(\pmb D^{c_r}).
$$

\section{Connections with Known Results in Symbol-Synchronous CDMA}

We relate the results of this paper with those in \cite{yin83},
which find applications in symbol-synchronous CDMA. Consider a
symbol-synchronous CDMA system. Define $C=[\uu c_1 \hspace{1mm}\uu
c_2\hspace{1mm} \cdots\hspace{1mm} \uu c_{K}]$ where $\uu c_k$ is
the $N\times 1$ random spreading sequence vector of user $k$. Let
$S$ be an $N\times N$ symmetric random matrix independent of $C$
with compactly supported asymptotic averaged empirical eigenvalue
distribution. It is shown in \cite{yin83} that the $n$-th AEM of
\begin{equation}\label{eq:mouse0305}
C^T S C=\left(
\begin{array}{cccc}
\uu c_1^T S\uu c_1 & \uu c_1^T S\uu c_2 & \cdots & \uu c_1^T S\uu
c_K\\
\vdots & \vdots &  & \vdots \\
\uu c_K^T S\uu c_1 & \uu c_K^T S\uu c_2 & \cdots & \uu c_K^T S\uu
c_K
\end{array}
\right)
\end{equation}
is given by
\begin{eqnarray}\label{eq:monkey0811}
\mu((C^T S C)^n)=\sum_{j=1}^n \beta^{j-1}
\mathop{\sum_{b_1+b_2+\cdots+b_{n-j+1}=n}}_{b_1\geq
b_2\geq\cdots\geq b_{n-j+1}\geq
1}\frac{n(n-1)\cdots(j+1)}{f(b_1,b_2,\cdots,b_{n-j+1})}
\prod_{r=1}^{n-j+1}\mu(S^{b_r}).
\end{eqnarray}

We now establish the relationship of $\mu(\bb R_\textrm{cs}^n)$,
$\mu(\bb R_\textrm{ca}^n)$ and $\mu((C^T S C)^n)$. Denote the
spreading sequence vector of user $k$'s $m$-th symbol as $\uu
c_k(m)=[c_k^{(mN)}\hspace{1mm}c_k^{(mN+1)}\hspace{1mm}\cdots$
$c_k^{((m+1)N-1)}]^T$, and we define
\begin{eqnarray}
&&\uu C(m)=\diag\{\uu c_1(m),\uu c_2(m),\cdots,\uu c_{K}(m)\},
\label{eq:cow0529}\\
&&\pmb C=\diag\{\uu C(-M),\uu C(-M+1),\cdots,\uu
C(M)\}.\label{eq:mouse0529}
\end{eqnarray}
Let $\pmb\Delta$ be a block matrix whose $(m,n)$-th block, $-M\leq
m,n\leq M$, is denoted by $\pmb\Delta(m,n)$. Each $\pmb\Delta(m,n)$
is also a block matrix with the $(k,l)$-th block, $1\leq k,l\leq K$,
represented by $\pmb\Delta(m,n;k,l)$. The matrix
$\pmb\Delta(m,n;k,l)$ is an $N\times N$ matrix whose $(p,q)$-th
entry, $0\leq p,q\leq N-1$, is equal to
$\delta((mN+p)T_c+\tau_k,(nN+q)T_c+\tau_l)$. Then, the $(m,n)$-th
block's $(k,l)$-th element of $\bb R_\textrm{cs}$ can be expressed
as
\begin{equation}\label{eq:tiger0305}
[\bb R_\textrm{cs}]_{mn,kl}=\uu c_k(m)^T\pmb\Delta(m,n;k,l)\uu
c_l(n),
\end{equation}
and the crosscorrelation matrix $\bb R_\textrm{cs}$ can be
decomposed as
$$
\bb R_\textrm{cs}=\bb C^T\pmb\Delta\bb C.
$$
Similarly, we have
\begin{equation}\label{eq:rabbit0305}
[\bb R_\textrm{ca}]_{mn,kl}=\uu c_k(m)^T\pmb\Omega(m,n;k,l)\uu
c_l(n)\qquad\mbox{and}\qquad \bb R_\textrm{ca}=\bb C^T\pmb\Omega\bb
C,
\end{equation}
where matrix $\pmb\Omega$ has the same structure as $\pmb\Delta$
with the $(p,q)$-th component of $\pmb\Omega(m,n;k,l)$ equal to
$R_\psi(((m-n)N+(p-q))T_c+\tau_k-\tau_l)$. Rewrite $\mu(\bb
R_\textrm{cs}^n)$ in (\ref{eq:dragon0124}) as
\begin{equation}\label{eq:cow0305}
\mu((\bb C^T\pmb\Delta\bb C)^n)=\sum_{j=1}^n \beta^{j-1}
\mathop{\sum_{b_1+b_2+\cdots+b_{n-j+1}=n}}_{b_1\geq
b_2\geq\cdots\geq b_{n-j+1}\geq
1}\frac{n(n-1)\cdots(j+1)}{f(b_1,b_2,\cdots,b_{n-j+1})}
\prod_{r=1}^{n-j+1}1.
\end{equation}
We find that $\mu((\bb C^T\pmb\Delta\bb C)^n)$ given in
(\ref{eq:cow0305}) and $\mu((\bb C^T\pmb\Omega\bb C)^n)$ given in
(\ref{eq:tiger0728}) show remarkable similarity as $\mu((C^T S
C)^n)$ in (\ref{eq:monkey0811}). However, even though AEM's of
matrices $C^T S C$, $\bb C^T\pmb\Delta\bb C$ and $\bb
C^T\pmb\Omega\bb C$ have the same form, they have distinct
structures. As seen in (\ref{eq:mouse0305}), elements in the matrix
$C^T S C$ are quadratic forms $\uu c_k^T S\uu c_l$ of a common
matrix $S$. Whereas, in $\bb C^T\pmb\Delta\bb C$ and $\bb
C^T\pmb\Omega\bb C$, the entries are expressed as
(\ref{eq:tiger0305}) and (\ref{eq:rabbit0305}), respectively, with
the matrices $\pmb\Delta(m,n;k,l)$ and $\pmb\Omega(m,n;k,l)$ varying
for each component.

In the following, another expression of $\bb R_\textrm{cs}$ will be
presented. Let $\uu u_k(m)$, $1\leq k\leq K$ and $-M\leq m\leq M$,
be an $N$-dimensional column vector whose $\sqrt{N}$ times scaled
entries are i.i.d. random variables with zero-mean, unit variance,
and bounded higher order moments. Besides, $\uu u_{k}(m)$ and $\uu
u_{l}(n)$ are independent when either $k\neq l$ or $m\neq n$. Given
a set of integers $\{\gamma_k:1\leq k\leq K\}\in[0,N-1]^K$, define
the $(2M+2)N$-dimensional vector $\widetilde{\uu u}_k(m)$
$$
\widetilde{\uu
u}_k(m)=[\underbrace{0,0,\cdots,0}_{(M+m)N+\gamma_k\textrm{ times}},
\uu u_k(m)^T, \underbrace{0,0,\cdots,0}_{(M-m+1)N-\gamma_k\textrm{
times}}]^T.
$$
We also define a $(2M+2)N\times K$ matrix $\uu U(m)$, given by
\begin{equation}\label{eq:snake0305}
\uu U(m)=[\widetilde{\uu u}_1(m),\widetilde{\uu u}_2(m),\cdots,
\widetilde{\uu u}_{K}(m)],
\end{equation}
and a $(2M+2)N\times (2M+1)K$ matrix
\begin{equation}\label{eq:dog0811}
\pmb U=[\uu U(-M),\uu U(-M+1),\cdots,\uu U(M)].
\end{equation}
We have the following theorem, whose proof demonstrates that $\bb
R_\textrm{cs}$ can be expressed as $\bb U^T\bb U$ with certain
choices of $\{\gamma_k\}$ and $\uu u_k(m)$'s.
\begin{theorem}
For any $\{\gamma_k\}_{k=1}^K\in[0,N-1]^K$, the ESD of the random
matrix $\bb U^T\bb U$ converges a.s. to the Mar\u{c}enko-Pastur
distribution with ratio index $\beta$ when $M,K,N\rightarrow\infty$
and $K/N\rightarrow\beta$. Moreover, let $\bb D$ be a diagonal
random matrix as stated in Theorem~\ref{theorem:mouse0810}. Then the
$n$-th free cumulant of $\bb U\bb D\bb U^T$ is equal to $\mu(\bb
D^n)\beta$.
\end{theorem}
\begin{proof}
Setting $\uu u_k(m)$ as the spreading sequence vector of the $m$-th
symbol of user $k$ and $\gamma_k:=\tau_k/T_c$ in a chip-synchronous
system, we have $\uu U^T(m)\uu U(n)=[\pmb R_\textrm{cs}]_{mn}$ and
$\bb U^T\bb U=\bb R_\textrm{cs}$. Thus, the first part of this
theorem is a direct consequence of Theorem~\ref{theorem:snake0831}.

For the second part, we have
$\mu((\pmb U\bb D\pmb U^T)^n)=\beta\cdot\mu((\bb R_\textrm{cs}\bb
D)^n)$, written as
$$
\sum_{j=1}^n \mathop{\sum_{c_1+c_2+\cdots+c_j=n}}_{c_1\geq
c_2\geq\cdots\geq c_j\geq
1}\dfrac{n(n-1)\cdots(n-j+2)}{f(c_1,c_2,\cdots,c_j)}\prod_{r=1}^j
\mu(\bb D^{c_r})\beta.
$$
By the moment-free cumulant formula of (\ref{eq:tiger0811}), the
$n$-th free cumulant of $\pmb U\bb D\pmb U^T$ is $\mu(\bb
D^n)\beta$.
\end{proof}

Let us particularly use $\uu U(m)_\star$ and $\bb U_\star$ to denote
the matrices $\uu U(m)$ and $\bb U$, respectively, when $\tau_k=0$
for all $k$'s. Clearly, $\bb U_\star^T\bb U_\star$ is a block
diagonal matrix with each block $\uu U(m)_\star^T\uu U(m)_\star$ a
crosscorrelation matrix in symbol-synchronous CDMA. It has been
derived in \cite{li01} that the $n$-th free cumulant of $\uu
U(m)_\star\uu D(m)\uu U(m)^T_\star$ is $\mu(\uu D(m)^n)\beta$, which
has the same form as its counterpart in chip-synchronous CDMA.

\section{Spectral Efficiency and MMSE of Asynchronous CDMA}

In some applications of probability, it is frequent that the
(infinite) moment sequence of an unknown distribution $F$ is
available, and these moments determine a unique distribution.
Suppose that the final aim is to calculate the expected value of
function $g(X)$ of the random variable $X$ whose distribution $F$ is
unknown. One of the most widely used techniques for evaluating
$\textrm{E}\{g(X)\}$ is based on the Gauss quadrature rule method
\cite{golub69}, where $2Q+1$ moments $\{m_n\}_{n=0}^{2Q}$ of
distribution $F$ are used to determine a $Q$-point quadrature rule
$\{w_q,x_q\}_{q=1}^Q$ such that
$$
\textrm{E}\{g(X)\}=\int_{-\infty}^\infty g(x)\textrm{d}F(x)\approx \sum_{q=1}^Q w_q g(x_q),
$$
and the approximation error becomes negligible when $Q$ is large.
However, this method often suffers from serious numerical problems
due to finite precision of a computing instrument. Fortunately, by
using the \textit{modified moments} technique \cite{sack71} which
requires only regular moments $\{m_n\}$, the algorithm becomes
exceptionally stable especially when the density of the distribution
$F$ has a finite interval. In case that the interval is infinite,
the algorithm does not completely remove the ill-conditioning
[\citenum{press92}, Section 4.5]. Some remedies can be found in the
above reference.

In this section, the Gauss quadrature rule method with modified
moments technique is employed to compute the spectral efficiency and
MMSE of asynchronous CDMA using AEM derived in previous sections.
Square-root raised cosine (SRRC) pulses with various roll-off
factors $\alpha$, denoted by SRRC-$\alpha$, are adopted as chip
waveforms. Since the employed method is numerically based and cannot
refrain from computing errors, we are careful in drawing conclusions
from the numerical results. Cares are taken to avoid making wrong
claims caused by numerical errors. For example, that the spectral
efficiency curve of system A is above the curve of system B may come
from different amounts of numerical errors on spectral efficiency
curves of the two systems.
In the sequel, $\bb R_{\textrm{ca},\alpha}$ (or $\bb
R_{\textrm{cs},\alpha}$) is used to represent the crosscorrelation
matrix corresponding to the SRRC-$\alpha$ pulse. Given a random
matrix $\bb M$, we use $\lambda_{\bb M}$ to denote the limiting
random variable governing eigenvalues of $\bb M$ when the matrix
size tends to infinity.

Assume the channel is unfaded and the per-symbol signal-to-noise
ratio $\textsf{\footnotesize SNR}$ is common to all users. We
consider chip-asynchronous systems. The spectral efficiency of the
optimum receiver is given as \cite{verdu99}
\begin{equation}\label{eq:cow0329}
{\cal C}^{\textsf{opt}}(\alpha,\beta,\textsf{\footnotesize SNR})=
\frac{\beta}{1+\alpha}\textrm{E}\left\{\log_2
(1+\textsf{\footnotesize SNR}\cdot\lambda_{\bb
R_{\textrm{ca},\alpha}})\right\},
\end{equation}
where the spectral efficiency is scaled by a factor
$(1+\alpha)^{-1}$ because the nonideal signaling scheme of
SRRC-$\alpha$ pulse has each complex dimension occupies $(1+\alpha)$
seconds $\times$ hertz. On the other hand, since the limiting
distribution of the linear MMSE receiver output is Gaussian, the
spectral efficiency of the receiver is asymptotically equal to the
spectral efficiency of a single-user channel with signal-to-noise
ratio equal to the output SIR of the MMSE receiver \cite{verdu99}.
It is known that the MMSE receiver has the output SIR given as
\cite{verdu98}
$$
\left[\overline{\tr}\left(\bb I+ \textsf{\footnotesize SNR}\bb
R_{\textrm{ca},\alpha}\right)^{-1}\right]^{-1}-1,
$$
whose limit is lower-bounded by 
$$
\textrm{E}\left\{\frac{1}{1+\textsf{\footnotesize
SNR}\cdot\lambda_{\bb R_{\textrm{ca},\alpha}}}\right\}^{-1}-1,
$$
and $\overline{\tr}$ denotes the normalized trace. Thus,
\begin{equation}\label{eq:tiger0329}
{\cal C}^{\textsf{mmse}}(\alpha,\beta,\textsf{\footnotesize SNR})
\geq
-\frac{\beta}{1+\alpha}\log_2\textrm{E}\left\{\frac{1}{1+\textsf{\footnotesize
SNR}\cdot\lambda_{\bb R_{\textrm{ca},\alpha}}}\right\}.
\end{equation}
When $\alpha=0$, the equality holds\footnote{The MMSE spectral
efficiency can be obtained as ${\cal
C}^{\textsf{mmse}}(\alpha,\beta,\textsf{\scriptsize SNR})=\beta
\log_2(1+\textsf{\scriptsize SNR}\eta(\textsf{\scriptsize
SNR}))/(1+\alpha)$, where $\eta(\textsf{\scriptsize SNR})$ is the
asymptotic multiuser efficiency of the linear MMSE receiver
\cite{shamai01}. However, $\eta(\textsf{\scriptsize SNR})$ is not
known to the author for nonzero $\alpha$.}. To compare systems with
chip waveforms of different roll-off factors, the spectral
efficiency must be given as a function of the energy-per-bit
relative to one-sided noise spectral level $E_b/N_0$. It can be
shown that a system achieving ${\cal
C}^{\textsf{opt}}(\alpha,\beta,\textsf{\footnotesize SNR})$ has an
energy per bit per noise level equal to \cite{verdu99}
$$
\frac{E_b}{N_0}=\frac{\beta\textsf{\footnotesize
SNR}}{(1+\alpha){\cal
C}^{\textsf{opt}}(\alpha,\beta,\textsf{\footnotesize SNR})},
$$
and the same relation holds for the spectral efficiency of the  MMSE
receiver ${\cal
C}^{\textsf{mmse}}(\alpha,\beta,\textsf{\footnotesize SNR})$ and
$E_b/N_0$.

\begin{figure*}[t]
\begin{center}
\begin{tabular}{c}
\psfig{figure=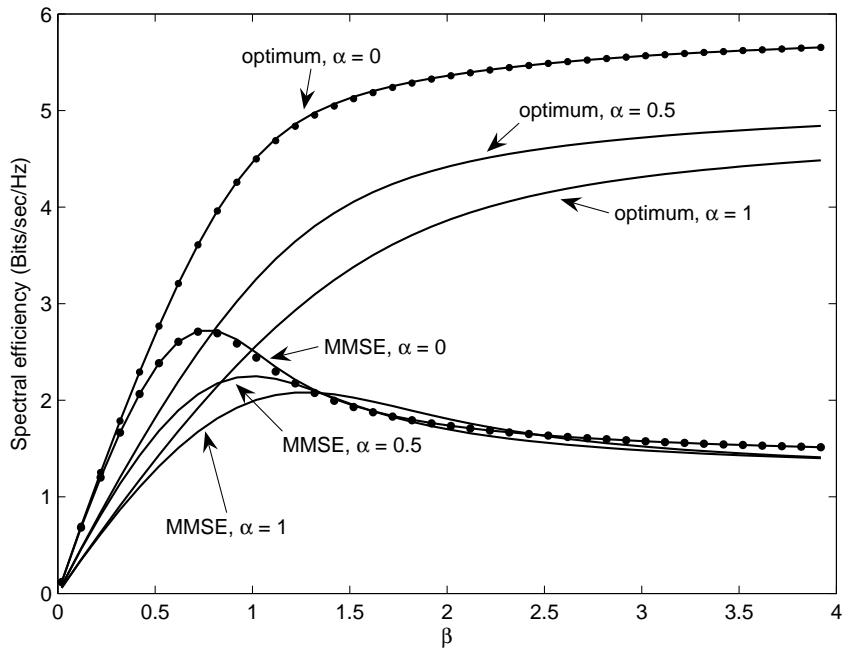,height=9cm}
\end{tabular}
\end{center}
\caption{Large system spectral efficiencies of a chip-asynchronous
CDMA system for $E_b/N_0=10$ dB and unfaded channel.}
\label{fig:efficiency1}
\end{figure*}

Fig.~\ref{fig:efficiency1} shows the spectral efficiencies versus
$\beta$ in a chip-asynchronous system for the optimum and the linear
MMSE receivers, where $E_b/N_0$ is fixed as $10$ dB and an unfaded
channel is assumed. Spectral efficiencies corresponding to SRRC
pulses with different roll-off factors are depicted, and curves in
the figure are obtained from (\ref{eq:cow0329}) and
(\ref{eq:tiger0329}) using a 10-point quadrature rule. The black
dots on the figure are obtained from the analytical results
\begin{eqnarray}
{\cal C}^{\textsf{opt}}(\alpha=0,\beta,\textsf{\footnotesize
SNR})&=& \beta\log_2\left(1+\textsf{\footnotesize
SNR}-\frac{1}{4}{\cal F}(\textsf{\footnotesize
SNR},\beta)\right)+\log_2\left(1+\textsf{\footnotesize
SNR}\beta-\frac{1}{4}{\cal F}(\textsf{\footnotesize
SNR},\beta)\right)\nn\\
&&-\frac{\log_2 e}{4\textsf{\footnotesize SNR}}{\cal
F}(\textsf{\footnotesize SNR},\beta)\nn
\end{eqnarray}
and
\begin{eqnarray}
{\cal C}^{\textsf{mmse}}(\alpha=0,\beta,\textsf{\footnotesize
SNR})=\beta\log_2\left(1+\textsf{\footnotesize SNR}-\frac{1}{4}{\cal
F}(\textsf{\footnotesize SNR},\beta)\right)\nn
\end{eqnarray}
when $\alpha=0$, where
$$
{\cal
F}(x,z)=\left(\sqrt{x(1+\sqrt{z})^2+1}-\sqrt{x(1-\sqrt{z})^2+1}\right)^2.
$$
These results are derived in \cite{verdu99} for a symbol-synchronous
system. However, by Corollary~\ref{corollary:mouse0902}, they are
applicable to a chip-asynchronous system with $\alpha=0$ as well. It
is seen that, when $\beta$ is around 1, there is visible discrepancy
between results of the analytical formula and the Gauss quadrature
method on the spectral efficiency of the MMSE receiver. This is
because the Mar\u{c}enko-Pastur distribution, i.e., the ASD
corresponding to $\alpha=0$, tends to be infinite-interval when
$\beta$ is close to 1, and the Gauss quadrature method is less
accurate when the density function has an infinite interval.

The discussions in the following two paragraphs apply to
chip-asynchronous systems. For the optimum receiver, given any
$\beta$, the spectral efficiency corresponding to $\alpha=0$ is
obviously greater than that of $\alpha=0.5$ and then of $\alpha=1$.
The spectral efficiency grows as $\beta$ increases. When $\beta$ is
small, the ratios of spectral efficiencies of $\alpha=0$, $0.5$ and
$1$ are roughly equal to the ratios of inverses of their bandwidths,
i.e., ratios of $(1+\alpha)^{-1}$, meaning the maximum bit rates
that can be transmitted arbitrarily reliably are the same for
various SRRC-$\alpha$ pulses\footnote{When a chip waveform with
roll-off factor $\alpha$ is chosen, the maximum bit rates that can
be transmitted arbitrarily reliably is equal to the spectral
efficiency times $(1+\alpha)/T_c$.}, although the consumed
bandwidths are different. As $\beta$ gradually increases, the ratios
of spectral efficiencies ($\alpha=0$ to $0.5$ and to $1$) become
smaller and smaller, suggesting that, when a chip waveform with a
larger excess bandwidth is chosen, the the maximum reliable
transmission rate can be increased.

For the linear MMSE receiver, the spectral efficiency is maximized
by a certain $\beta$ depending on $\alpha$. When $\beta$ is small,
it is obvious that chip waveforms with smaller values of $\alpha$
have larger MMSE spectral efficiencies. Nonetheless, as $\beta$ is
greater than around $1.2$, the favor of smaller $\alpha$ in spectral
efficiencies disappears. For low $\beta$, the linear MMSE receiver
achieves near-optimum spectral efficiency. Otherwise, great gains in
efficiency can be realized by nonlinear receivers. When $\beta$ is
small, the MMSE receiver with $\alpha=0$ is superior to the optimum
receiver having $\alpha=0.5$ in terms of spectral efficiency; so is
the MMSE receiver with $\alpha=0.5$ to the optimum receiver having
$\alpha=1$. Comparing curves of two receivers, we comment when more
bandwidths are consumed due to the choice of an SRRC pulse with
higher $\alpha$, the return in channel capacity (maximum reliable
data rate) is larger in the linear MMSE receiver than in the optimum
receiver. For example, when $\beta=2$, twice bandwidth of an SRRC
pulse with $\alpha=1$ than with $\alpha=0$ results in approximately
twice reliable data rate of $\alpha=1$ than $\alpha=0$ in the linear
MMSE receiver. However, for the optimum receiver, the ratio of the
data rates between $\alpha=1$ and $\alpha=0$ is around $1.5$ for the
same $\beta$. Even for values of $\beta$ in a practical system, the
higher return of the MMSE receiver in capacity is still true.
Another interesting observation is that the MMSE spectral efficiency
curve of SRRC pulse with $\alpha=1$ is above the curve of the sinc
pulse in the region around $\beta\in[1.5,2]$. However, as the
difference of the two curves is small, we should be careful in
making comments. In that region, the curve of the sinc pulse is
exact (matching with the analytical result); the curve of SRRC with
$\alpha=1$ is an approximation with two opposite forces
counter-acting on each other. On the one hand, (\ref{eq:tiger0329})
is a lower bound so the curve underestimates the true spectral
efficiency; on the other hand, the numerical method tends to be
optimistic, which yields an overestimate. We surmise that the first
factor dominates; adding the result that the curve of sinc pulse is
below the other, we conjecture: for some $\beta$, the sinc pulse is
not optimal in terms of the MMSE spectral efficiency under the
indicated environments.

\begin{figure*}[t]
\begin{center}
\begin{tabular}{c}
\psfig{figure=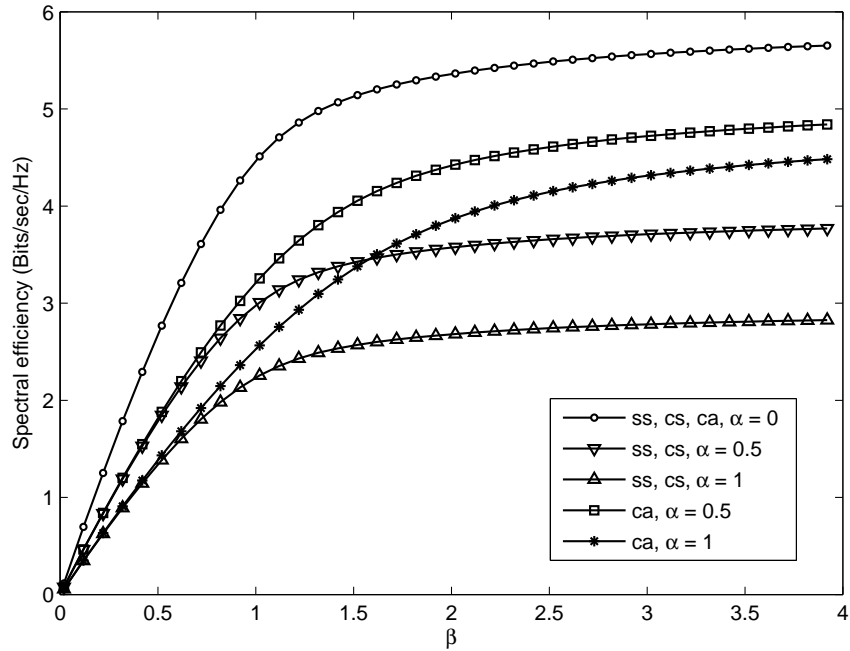,height=9cm}
\end{tabular}
\end{center}
\caption{Large system optimum spectral efficiencies of
symbol-synchronous, chip-synchronous, and chip-asynchronous CDMA
systems when $E_b/N_0=10$~dB and the channel is unfaded, where
"$\textrm{ss}$", "$\textrm{cs}$" and "$\textrm{ca}$" refer to
symbol-synchronous, chip-synchronous, and chip-asynchronous,
respectively.} \label{fig:efficiency3}
\end{figure*}

Fig.~\ref{fig:efficiency3} shows the optimum spectral efficiencies
as a function of $\beta$ for symbol-synchronous, chip-synchronous
and chip-asynchronous systems with various chip waveforms when
$E_b/N_0=10$ dB and the channel is unfaded. The three curves marked
by circles, squares and stars have appeared in
Fig.~\ref{fig:efficiency1}; those two marked by triangles (down and
up) are obtained from (\ref{eq:cow0329}), where $\lambda_{\bb
R_{\textrm{ca},\alpha}}$ in the equation is replaced with
$\lambda_{\bb R_{\textrm{cs},\alpha}}$. In both symbol- and
chip-synchronous systems, the spectral efficiency corresponding to a
particular $\alpha^*$ is equal to the spectral efficiency
corresponding to $\alpha=0$ divided by $1+\alpha^*$. This is because
$\lambda_{\bb R_{\textrm{cs},\alpha}}$ has the same distribution
regardless of $\alpha$.

When $\alpha=0$, three systems have the same optimum spectral
efficiency. This is a direct consequence of
Corollary~\ref{corollary:horse0216}. Given any $\alpha$, when
$\beta$ is low, the differences of the three systems in optimum
spectral efficiency are negligible; as $\beta$ increases, the
chip-asynchronous system is superior to the other two for nonzero
$\alpha$, and the optimum spectral efficiency differences are
proportional to $\beta$. On the other hand, given any $\beta$, as
$\alpha$ increases, the chip-asynchronous system has a larger
spectral efficiency than the other two, and the difference grows
with $\alpha$. Similar comments can be made from
Fig.~\ref{fig:efficiency1} for the MMSE spectral efficiencies of the
three systems. We also observe, while choosing chip waveforms with
larger bandwidths may result in the increase of channel capacity in
a chip-asynchronous system, the statement is not true for symbol-
and chip-synchronous systems. This can be interpreted as follows. It
is the ASD that determines the performance measures of a system such
as the channel capacity, the MMSE achievable by a linear receiver,
and so on. Regretfully, the ASD of symbol- and chip-synchronous
systems does not depend on the chosen chip waveform; hence the
increase of bandwidth due to the replacement of a chip waveform
merely decreases the spectral efficiency and does not help in
boosting the capacity.

\begin{figure*}[t]
\begin{center}
\begin{tabular}{c}
\psfig{figure=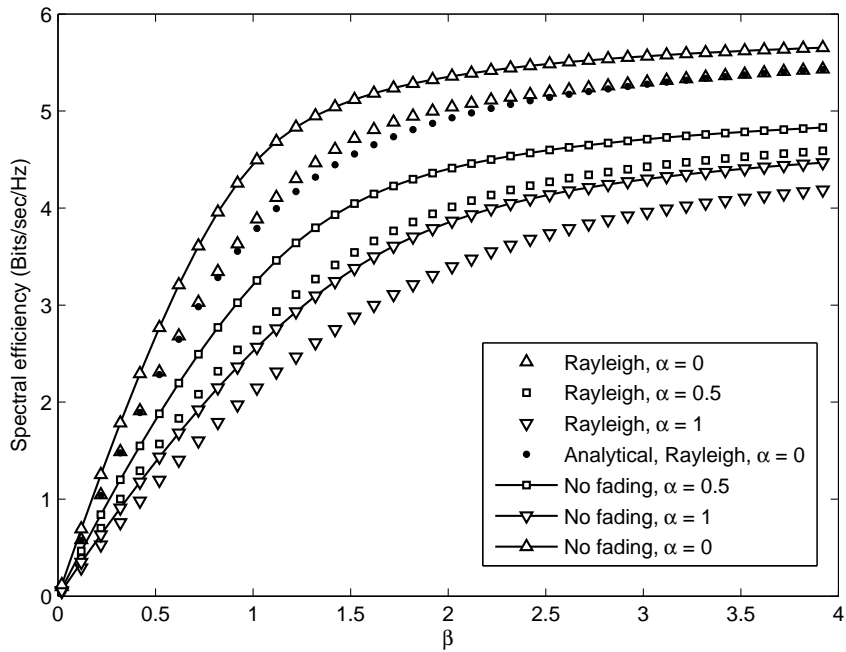,height=9cm}
\end{tabular}
\end{center}
\caption{Large system optimum spectral efficiencies of a
chip-asynchronous system for unfaded and Rayleigh fading channels
when $E_b/N_0=10$~dB.} \label{fig:efficiency2}
\end{figure*}

Consider a channel subject to frequency-flat fading. The square
magnitude of the received signal $|A_k(m)|^2$ is governed by
$\textsf{\footnotesize SNR}$ common to all users and a normalized
random variable $\overline{A}_k(m)$ having
$\textrm{E}\{|\overline{A}_k(m)|^2\}=1$. Thus, the amplitude matrix
$\bb A$ has $\bb A\bb A^\dag=\textsf{\footnotesize SNR}\overline{\bb
A}\overline{\bb A}^\dag$, where $\overline{\bb A}$ has the same
structure as the diagonal amplitude matrix $\bb A$, and
$\overline{A}_k(m)$ is located at the $(k,k)$-th entry in the
$(m,m)$-th block of $\overline{\bb A}$. The spectral efficiencies of
the optimum receiver is given by (\ref{eq:cow0329}) with
$\lambda_{\bb R_{\textrm{ca},\alpha}}$ replaced as
$\lambda_{\overline{\bb A}^\dag\bb
R_{\textrm{ca},\alpha}\overline{\bb A}}$. Although we can also
modify (\ref{eq:tiger0329}) to yield a lower bound for the MMSE
spectral efficiency under fading; however, according to our
experiments, the bound is loose. Fig.~\ref{fig:efficiency2} compares
optimum spectral efficiencies in a chip-asynchronous system with and
without fading for a fixed $E_b/N_0$ equal to $10$~dB. The fading
channel is assumed to be Rayleigh. To generate curves of the fading
channel, a 15-point quadrature rule is used. The black dots shown in
the figure correspond to the analytical result obtained in
\cite{shamai01} for $\alpha=0$ and a Rayleigh fading channel.
Perceptible discrepancy between analytical and numerical results
appear in the region of $\beta\in[1,2]$. We comment that fading in a
chip-asynchronous system leads to a degradation in optimum spectral
efficiency, which is consistent with a mathematical result
demonstrated in \cite{shamai01}.

\begin{figure*}[t]
\begin{center}
\begin{tabular}{c}
\psfig{figure=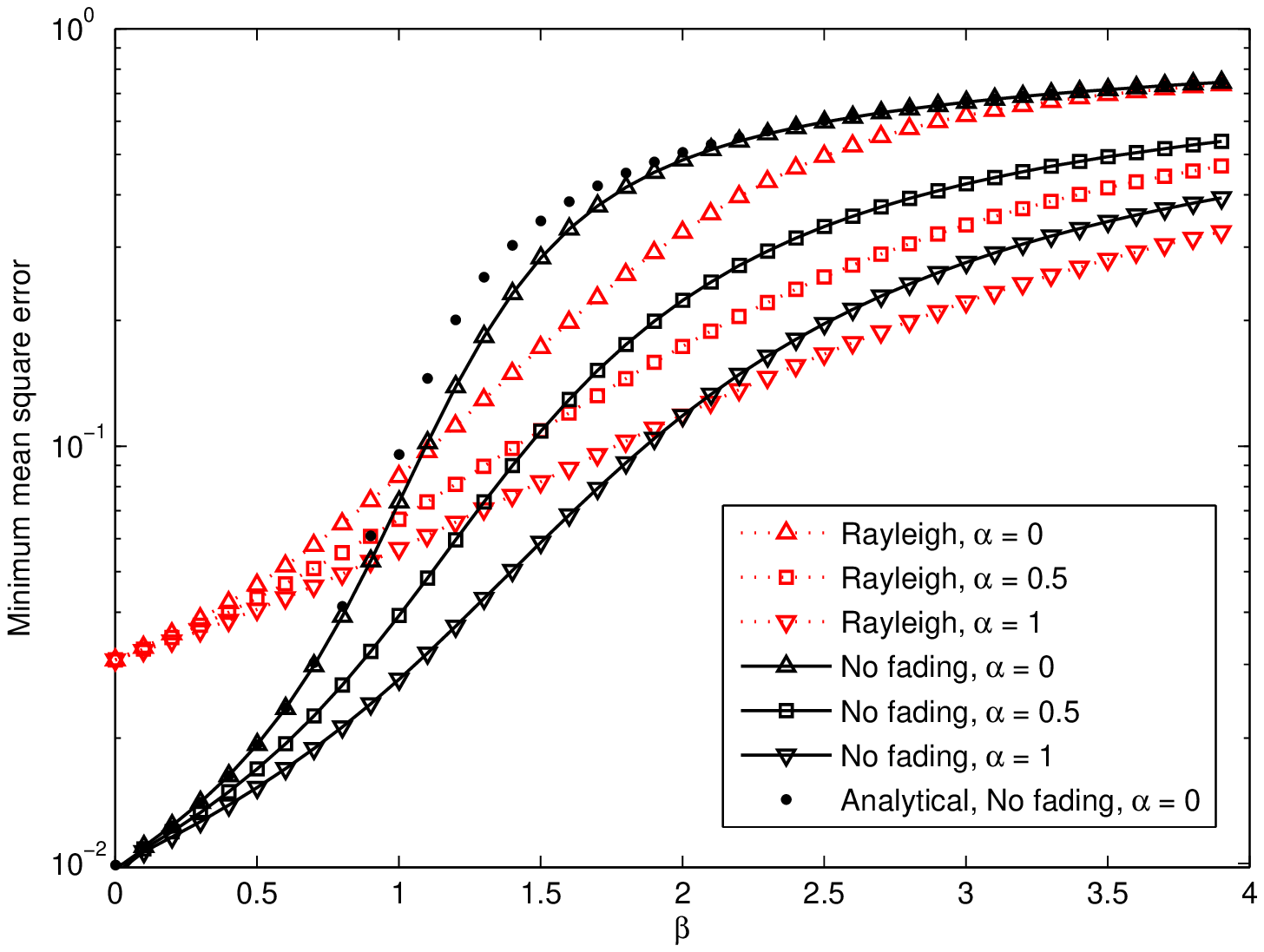,height=9cm}
\end{tabular}
\end{center}
\caption{Large system MMSE versus $\beta$ of a chip-asynchronous
system for unfaded and Rayleigh fading channels when
$\textsf{\footnotesize SNR}=20$~dB.} \label{fig:mmse1}
\end{figure*}

In the presence of fading, the arithmetic mean over the users of the
mean-square-error achieved by the linear MMSE receiver is given as
\cite{verdu98}
\begin{equation}\label{eq:snake0329}
{\cal E}(\alpha,\beta,\textsf{\footnotesize
SNR})=\textrm{E}\left\{\frac{1}{1+ \textsf{\footnotesize
SNR}\cdot\lambda_{\overline{\bb A}^\dag\bb
R_{\textrm{ca},\alpha}\overline{\bb A}}}\right\}.
\end{equation}
For the unfaded case, ${\cal E}(\alpha,\beta,\textsf{\footnotesize
SNR})$ is obtained by setting $\overline{\bb A}$ in
(\ref{eq:snake0329}) as the identity matrix. Fig.~\ref{fig:mmse1}
compares the MMSE achievable by a linear receiver for unfaded and
Rayleigh fading channels when $\textsf{\footnotesize SNR}=20$~dB. We
use 10- and 15-point quadrature rules for unfaded and fading
channels, respectively. Black dots on the figure correspond to the
analytical result of
$$
1-\frac{{\cal F}(\textsf{\footnotesize
SNR},\beta)}{4\beta\textsf{\footnotesize SNR}}
$$
for $\alpha=0$ in the absence of fading \cite{verdu99}. According to
our tests, the discrepancy between the analytical and numerical
results grows with $\textsf{\footnotesize SNR}$. The difference
becomes almost unnoticeable when $\textsf{\footnotesize SNR}<15$~dB.
For low $\beta$, the MMSE in an unfaded channel is lower than in a
Rayleigh fading channel. In the latter case, as $\beta\to 0$, the
analytical result is equal to $\textrm{E}\{(1+\textsf{\footnotesize
SNR}\cdot X)^{-1}\}=0.04079$, where $X$ has the exponential density
$e^{-x}$, $x\geq 0$; the numerical result is equal to $0.03075$ at
$\beta\to 0$.

We observe that, regardless of fading or not, the MMSE is inversely
proportional to $\alpha$ in a chip-asynchronous system. This is
consistent with the conclusion drawn previously that choosing chip
waveforms with larger excess bandwidths leads to a higher capacity.
Nevertheless, for symbol- and chip-synchronous systems, regardless
of $\alpha$, the MMSE are unchanged and correspond to the curve of
$\alpha=0$. Interestingly, we can see that fading decreases the MMSE
in the region of high $\beta$. The explanation is similar to that
for fading increasing the spectral efficiency at high $\beta$ made
in \cite{shamai01}. That is, due to fading, a certain portion of
interferers are low-powered; thus, the number of "effective"
interferers seen by the receiver is reduced. This interference
population control of fading compensates for its harmful effect on
the desired user. It is also observed, as $\alpha$ increases, the
receiver needs a larger $\beta$ to have this phenomenon begin to
operate, and this phenomenon is less obvious for larger $\alpha$.

\section{Conclusion}

In this paper, the ASD of crosscorrelation matrices in random
spreading chip-synchronous and chip-asynchronous CDMA systems are
investigated with a particular emphasis on the derivation of AEM.
Noncrossing partition and the graphical representation of $K$-graph
are the key tools in AEM computation.
We assume an infinite observation window width, known spreading
sequences and relative delays to the receiver, and an
arbitrary chip waveform. We consider both unfaded and frequency-flat fading channels. The
spreading sequences are only assumed to be independent across users.
For a particular user, we do not assume that the sequence is
independent across symbols. Thus, results shown in this paper are
applicable for both short-code and long-code systems.

In the following, results of this paper are summarized. For
chip-synchronous CDMA systems, the explicit expressions for AEM of
the crosscorrelation matrix are given when the users' relative
delays are deterministic constants. We show that AEM do not depend
on the realizations of asynchronous delays and the shape of chip
waveform, as long as the zero ICI condition holds. It is also shown
that the AEM formulas are identical to those of symbol-synchronous
CDMA. In an unfaded channel, as the AEM satisfy the Carleman's
criterion and the a.s. convergence test, it is concluded that the
ASD in a chip-synchronous system converges a.s. to
Mar\u{c}enko-Pastur law with ratio index $\beta$. In the case of
flat fading, the a.s. convergence of ESD to a nonrandom ASD is
established provided that a constraint on the empirical moments of
the fading coefficients is satisfied.

For chip-asynchronous CDMA systems, the convergence of ESD to an ASD
in a.s. sense is proved for general constraints on the chip waveform
and, for a fading channel, the empirical moments of the signal
received power. It is shown that, in contrast to chip-synchronous
CDMA, AEM in a chip-asynchronous system are dependent on the shape
of chip waveform. On the other hand, the relation of AEM and users'
relative delays depends on the bandwidth of the chosen chip
waveform. Specifically, it is mentioned that, for the zero ICI
property to hold, the chip waveform has a bandwidth at least equal
to $1/(2T_c)$, which corresponds to the sinc pulse. When the
bandwidth of the chip waveform is $1/(2T_c)$, AEM do not depend on
the realizations of relative delays. On the contrary, if the
bandwidth is wider than the threshold, AEM do depend on the
asynchronous delays; nonetheless, different relative delays
realizations may result in the same AEM. Suppose that relative
delays are modeled as i.i.d. random variables $\tau_k$'s. Let $G_1$
and $G_2$ be two distinct distributions of relative delays, and both
of them possess the symmetry property of $\textrm{E}\left\{\cos(2\pi
n \tau_k/T_c)\right\}=0$ for nonzero integer $n$. Then, for the same
chip waveform, AEM's averaged over realizations of relative delays
with distributions $G_1$ and $G_2$ are equal. The distribution
symmetry condition given above encompasses the symbol
quasi-synchronous but chip-asynchronous system considered in
\cite{cottatellucci06}. By moment convergence theorem, the
equivalence of AEM leads to an equivalence of ASD provided that the
uniqueness of limiting distribution is true. It follows that our
result proves the conjecture given there that relative delays
ranging uniformly within the chip duration and within the symbol
duration yield the same performance. When the sinc chip waveform is
adopted, no matter fading or not, the AEM of chip-asynchronous CDMA
are shown to be equal to those of chip-synchronous CDMA and hence
those of a symbol-synchronous system. This explains the equivalence
result of \cite{mantravadi02} that the output SIR of the linear MMSE
receiver converges to those of chip- and symbol-synchronous systems
when $M$ is large. Since every receiver in the family constructed in
\cite{guo2002} can be arbitrarily well approximated by a polynomial
receiver, and both the polynomial coefficients and the polynomial
receiver's output SIR are determined by AEM, we can also prove the
conjecture in \cite{mantravadi02} that the equivalence result in the
output SIR of the three systems holds for all receivers in that
family. We also study the situation that the chip waveform bandwidth
is less than $1/(2T_c)$ such that zero ICI condition is lost. It is
shown that, without the zero ICI property, the AEM formulas in
symbol- and chip-synchronous systems bear the same forms as those in
a chip-asynchronous system. Thus, when systems of three synchronism
levels have the same parameters except for the delays of the users,
their AEM's are all the same; consequently, these three systems have
the same ASD.

With the help of free probability theory, free cumulants of
crosscorrelation matrices are also derived for both chip-synchronous
and chip-asynchronous systems. It is also proved that the
crosscorrelation matrix is asymptotically free with a random
diagonal matrix having a general constraint. Based on the asymptotic
freeness property, AEM's for the sum and the product of the
crosscorrelation matrix and a random diagonal matrix are derived
accordingly.

Mathematical results obtained in this paper are connected to those
that are widely used by researchers who apply random matrix theory
to communication problems.

At last, some application cases are provided. The Gauss quadrature
rule method is adopted to depict the spectral efficiencies of the
optimum and linear MMSE detectors and the MMSE achievable by a
linear receiver in asynchronous CDMA. Performance in the measures of
the spectral efficiency, channel capacity, and MMSE are compared for
various chip waveforms, two types of receivers, and different
asynchronism levels.

\appendices

\section{Noncrossing Partition}\label{appendix:noncrossing}

The proofs of Lemmas~\ref{lemma:rabbit0120} and \ref{theorem:cow0902} require results from
\textit{noncrossing partition} of set partition theory.
Our treatment here for noncrossing partition is very brief; for more
details, please consult \cite{speicher06}.

\begin{definition}
(\textit{Noncrossing Partition}
[\citenum{kreweras72},\citenum{speicher06}])\label{def:3} Let $S$ be
a finite totally ordered set.
\begin{enumerate}
\item We call $\varpi=\{B_1,\cdots,B_j\}$ a partition of the set $S$ if
and only if $B_1,\cdots,B_j$ are pairwise disjoint, non-empty
subsets of $S$ such that $B_1\cup\cdots\cup B_j=S$. We call
$B_1,\cdots,B_j$ the classes of $\varpi$. The classes
$B_1,\cdots,B_j$ are ordered according to the minimum element in
each block. That is, the minimum element in $B_k$ is smaller than
that of $B_l$ if $k<l$.

\item The collection of all partitions of $S$ can be viewed as a
partially ordered set (poset) in which the partitions are ordered by
\textit{refinement}: if $\varpi,\sigma$ are two partitions of $S$,
we have $\varpi\leq \sigma$ if each block of $\varpi$ is contained
in a block of $\sigma$. For example, when $S=\{1,2,3,4,5,6,7\}$, we
have
$\{\{1\},\{2,5\},\{3,4\},\{6\},\{7\}\}<\{\{1,3,4\},\{2,5\},\{6,7\}\}$.

\item A partition of the set $S$ is called crossing if there
exist $p_1<q_1<p_2<q_2$ in $S$ such that $p_1$ and $p_2$ belong to
one class and $q_1$ and $q_2$ to another. If a partition is not
crossing, then it is called noncrossing. \hfill{$\blacksquare$}
\end{enumerate}
\end{definition}
The set of all noncrossing partitions of $S$ is denoted by
$\mathit{NC}(S)$. In the special case $S=\{1,\cdots,n\}$, we denote
this by $\mathit{NC}(n)$.

\begin{definition}
(\textit{Kreweras Complementation Map}
[\citenum{kreweras72},\citenum{speicher06}]) Consider elements
$\overline{1},\overline{2},\cdots,\overline{n}$ and interlace them
with $1,2,\cdots,n$ in the alternating way of $1,\overline{1},
2,\overline{2},\cdots,n,\overline{n}$. Let $\varpi\in
\mathit{NC}(n)$. Then its Kreweras complementation map
$\mathit{KC}(\varpi):\mathit{NC}(n)\rightarrow \mathit{NC}(n)\in
\mathit{NC}(\{\overline{1},\overline{2},\cdots,\overline{n}\})$ is
defined as the biggest element among those $\sigma\in
\mathit{NC}(\{\overline{1},\overline{2},\cdots,\overline{n}\})$ such
that the union $\varpi\cup\sigma$ of the two noncrossing partitions
belongs to $\mathit{NC}
(\{1,\overline{1},2,\overline{2},\cdots,n,\overline{n}\})$.
$\hfill{\small \blacksquare}$
\end{definition}
It can be shown that, if $\varpi$ contains $j$ classes, then the
number of classes in $\mathit{KC}(\varpi)$ is $n-j+1$.
\begin{figure*}[t]
\begin{center}
\begin{tabular}{c}
\psfig{figure=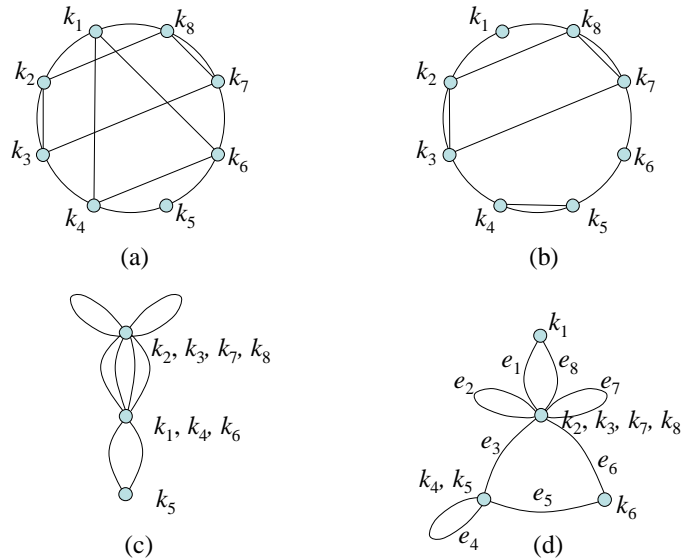,height=8cm}
\end{tabular}
\end{center}
\caption{(a) The partition
$\{\{k_1,k_4,k_6\}\{k_2,k_3,k_7,k_8\}\{k_5\}\}$ of the totally
ordered set $\{k_1,k_2,\cdots,k_8\}$, (b) the partition
$\{\{k_1\}\{k_2,k_3,k_7,k_8\}\{k_4,k_5\}\{k_6\}\}$, (c) the
$K$-graph for the partition represented in (a), and (d) the
$K$-graph for the partition represented in (b).} \label{fig:7}
\end{figure*}

A partition can be represented graphically. For example,
Figs.~\ref{fig:7}(a) and \ref{fig:7}(b) show two partitions of
$\{k_1,k_2,\cdots,k_8\}$, where elements in the same class are
joined \textit{successively} by chords. A noncrossing partition is
such that the chords intersect only at elements $k_1,\cdots,k_{n}$.
For instance, Fig.~\ref{fig:7}(b) is a noncrossing partition, while
Fig.~\ref{fig:7}(a) is not. In the following, we define a
representation, called $K$-\textit{graph}, for any partition of a
totally ordered set. The $K$-graph defined below is similar to the
$W$-graph of \cite{bai99} used to establish the convergence of
moments of a Wigner matrix. We discover several pleasant properties
of $K$-graphs that will be useful in proving the lemmas. They are
enumerated right after the definition of $K$-graph.

\begin{definition}\label{definition:mouse0810}
($K$-graph) The $K$-graph corresponding to a partition
$\varpi=\{B_1,\cdots,B_j\}$ of a totally ordered set
$\{k_1,k_2,\cdots,k_n\}$ is denoted by a graph $G=({\cal V},{\cal
E})$. The vertex set is ${\cal V}=\{v_1,v_2,\cdots,v_j\}$, and the
edge set is ${\cal E}=\{e_1,e_2,\cdots,e_n\}$, where the edge $e_r$
connects vertices $v_s$ and $v_t$ if $k_r$ and $k_{r+1}$ are
partitioned into classes $B_s$ and $B_t$, respectively (with
$n+1:=1$). $\hfill{\small \blacksquare}$
\end{definition}

\textit{Remark}: The $K$-graph for a partition $\varpi$ of
$\{k_1,k_2,\cdots,k_n\}$ can be interpreted in a more visually
convenient way as follows. Let $k_i$'s be arranged orderly (either
clockwise or counter-clockwise) as vertices of an $n$-vertex cycle,
and let edge $e_r$ connect vertices $k_r$ and $k_{r+1}$. The
$K$-graph of $\varpi$ can be obtained by merging vertices that are
partitioned in the same class of $\varpi$ into one. When vertices
are merged, edges originally incident on these vertices become
incident on the merged one. Mergence of two adjacent vertices
results in a self-loop cycle. \hfill{\small $\blacksquare$}

For example, Figs.~\ref{fig:7}(c) and \ref{fig:7}(d) present the
$K$-graphs for the partitions of Figs.~\ref{fig:7}(a) and
\ref{fig:7}(b), respectively.

\textit{Properties of $K$-graphs}: Given a partition $\varpi$ of a
totally ordered $n$-element set ${\cal K}$ and its corresponding
$K$-graph $G=({\cal V},{\cal E})$. We have the following properties.
\begin{enumerate}
\item There is a bijective correspondence between
classes of $\varpi$ and vertices of $G$.

\item $G$ is connected. If and only if $\varpi$ is noncrossing, $G$ is a concatenation of cycles with any two of
them connected by at most one vertex. Moreover, if $\varpi$ is
noncrossing and has $j$ classes, then there are $n-j+1$ cycles in
$G$. For example, Fig.~\ref{fig:7}(d) is composed of $8-4+1=5$
cycles. Any pair of these five cycles are connected by at most one
of the two vertices labelled with $k_2,k_3,k_7,k_8$ and $k_4,k_5$.

\item Consider a partition $\sigma$ of the ordered edge set ${\cal E}=\{e_1,e_2,\cdots,e_n\}$ of $G$
by letting edges in the same cycle of $G$
being in the same class. If $\varpi$ is noncrossing, then $\sigma$
is noncrossing as well. Moreover, $\varpi$ and $\sigma$ are Kreweras
complementation maps of each other. For example, Fig.~\ref{fig:7}(d)
corresponds to $\sigma=\{\{e_1,e_8\},\{e_2\},\{e_3,e_5,e_6\},$
$\{e_4\},\{e_7\}\}$, which is a noncrossing partition of
$\{e_1,e_2,\cdots,e_8\}$. It is seen that
$\{\{1,8\},\{2\},\{3,5,6\},\{4\},\{7\}\}$ is the Kreweras
complementation map of $\{\{1\},\{2,3,7,8\},$ $\{4,5\},\{6\}\}$, and
vice versa.\hfill{\small $\blacksquare$}
\end{enumerate}
From properties 1) and 3), if $\varpi$ is noncrossing, then the
corresponding $K$-graph can represent both $\varpi$ and its Kreweras
complementation map simultaneously. That is, $\varpi$ and
$\mathit{KC}(\varpi)$ can be identified by the vertex set and edge
set of the $K$-graph, respectively.

Some results about noncrossing partition are in order. The number of
noncrossing partitions that partition $n$ elements into $j$ classes
is the Narayana number, given by
\begin{equation}\label{eq:mouse0519}
\dfrac{1}{n}{n\choose j}{n\choose j-1}.
\end{equation}
Moreover, if the $j$ classes have sizes $c_1,c_2,\cdots,c_j$ with
$c_1\geq c_2\geq\cdots\geq c_j\geq 1$ (but not specifying which
class gets which size), the number of noncrossing partitions is
\cite{kreweras72}
\begin{equation}\label{eq:cow0519}
\dfrac{n(n-1)\cdots(n-j+2)}{f(c_1,c_2,\cdots,c_j)},
\end{equation}
where
\begin{equation}\label{eq:mouse0128}
f(c_1,c_2,\cdots,c_j)=\prod_{k\geq 1} n_k !
\end{equation}
with $n_k$ being the number of elements in $(c_1,c_2,\cdots,c_j)$
that are equal to $k$. It is clear to see
\begin{equation}\label{eq:mouse0804}
\mathop{\sum_{c_1+c_2+\cdots+c_j=n}}_{c_1\geq c_2\geq\cdots\geq
c_j\geq 1}\dfrac{n(n-1)\cdots(n-j+2)}{f(c_1,c_2,\cdots,c_j)}=
\dfrac{1}{n}{n\choose j}{n\choose j-1}.
\end{equation}
The number of noncrossing partitions $\varpi$ of an $n$-element set
meeting conditions of
\begin{enumerate}
\item[i)] $\varpi$ has $j$ classes with sizes in non-ascending order of
$(c_1,c_2,\cdots,c_j)$, and

\item[ii)] the classes of $\mathit{KC}(\varpi)$ have sizes in non-ascending order
of $(b_1,b_2,\cdots,b_{n-j+1})$,
\end{enumerate}
is equal to [\citenum{biane96},\citenum{li04}]
\begin{equation}\label{eq:cow0804}
\dfrac{n(n-j)!(j-1)!}
{f(b_1,b_2,\cdots,b_{n-j+1})f(c_1,c_2,\cdots,c_j)}.
\end{equation}

\section{Proofs of Lemma~\ref{lemma:rabbit0120} and Lemma~\ref{theorem:cow0902}}\label{appendix:dragon0120}

With the results of noncrossing partition in Appendix~\ref{appendix:noncrossing}, we now proceed to
prove Lemma~\ref{lemma:rabbit0120} and Lemma~\ref{theorem:cow0902}.
Consider $\mu(\bb R_\textrm{cs}^n)$ in (\ref{eq:mouse0120}) and
$\mu(\bb R_\textrm{ca}^n)$ by replacing $\bb R_\textrm{cs}$ in the
equation with $\bb R_\textrm{ca}$.
They can be rewritten as
\begin{equation}\label{eq:cow0120}
\mu(\bb
R_\textrm{x}^n)=\mathop{\lim_{K,N\rightarrow\infty}}_{K/N\rightarrow\beta}
K^{-1}\left[\lim_{M\rightarrow\infty}(2M+1)^{-1}\textrm{E}\left\{\tr((\bb
R_\textrm{x}^{(K)})^n)\right\}\right],\quad\textrm{x}\in\{\textrm{cs},\textrm{ca}\}.
\end{equation}
For notational convenience, the superscript $^{(K)}$ of $\bb
R_\textrm{x}^{(K)}$ will be omitted below when no ambiguity occurs.
By (\ref{eq:tiger0320}) and (\ref{eq:mouse0323}), we have
\begin{eqnarray}
&&[\bb R_\textrm{cs}]_{m_r m_{r+1},k_r
k_{r+1}}\hspace{2mm}=\hspace{2mm}\rho_{\textrm{cs}}(m_r,
m_{r+1};k_r,
k_{r+1})\nn\\
&=&\sum_{p_r=m_r
N}^{(m_r+1)N-1}\sum_{q_{r+1}=m_{r+1}N}^{(m_{r+1}+1)N-1}
c_{k_r}^{(p_r)} c_{k_{r+1}}^{(q_{r+1})}\delta(p_r
T_c+\tau_{k_r},q_{r+1}T_c+\tau_{k_{r+1}}),\nn
\end{eqnarray}
and
\begin{eqnarray}
&&[\bb R_\textrm{ca}]_{m_r m_{r+1},k_r
k_{r+1}}\hspace{2mm}=\hspace{2mm}\rho_{\textrm{ca}}(m_r,
m_{r+1};k_r,
k_{r+1})\nn\\
&=&\sum_{p_r=m_r
N}^{(m_r+1)N-1}\sum_{q_{r+1}=m_{r+1}N}^{(m_{r+1}+1)N-1}
c_{k_r}^{(p_r)} c_{k_{r+1}}^{(q_{r+1})}
R_\psi((p_r-q_{r+1})T_c+\tau_{k_r}-\tau_{k_{r+1}}),\nn
\end{eqnarray}
for $1\leq r\leq n$ with $m_{n+1}:=m_1$ and $k_{n+1}:=k_1$. By
expanding matrix multiplications of $\bb R_\textrm{x}^n$, the term
inside of square brackets of (\ref{eq:cow0120}) can be expressed as
\begin{eqnarray}
&&\lim_{M\rightarrow\infty} (2M+1)^{-1}\textrm{E}\left\{\tr(\pmb
R_\textrm{x}^n)\right\}\nn\\
&=&\lim_{M\rightarrow\infty}(2M+1)^{-1}\sum_{{\cal K}\in{\cal X}}
\sum_{{\cal M}\in {\cal Y}}\textrm{E}\{ [\pmb R_\textrm{x}]_{m_1
m_2,k_1 k_2} [\pmb R_\textrm{x}]_{m_2 m_3,k_2 k_3} \cdots [\pmb
R_\textrm{x}]_{m_{n}m_1, k_{n} k_1} \}\label{eq:mouse0315},
\end{eqnarray}
where ${\cal K}=\{k_1,\cdots,k_{n}\}$, ${\cal
X}=\underbrace{[1,K]\times\cdots\times[1,K]}_{n \textrm{
times}}=[1,K]^n$, ${\cal M}=\{m_1,\cdots,m_{n}\}$ and ${\cal
Y}=[-M,M]^n$. Equation (\ref{eq:mouse0315}) is equal to
\begin{eqnarray}
&&\lim_{M\rightarrow\infty}(2M+1)^{-1}\sum_{{\cal K}\in{\cal
X}}\sum_{{\cal M}\in {\cal Y}} \sum_{{\cal P}_1\in {\cal Z}_1}\cdots
\sum_{{\cal P}_{n}\in{\cal Z}_{n}}
\textrm{E}\left\{\left(c_{k_1}^{(p_1)} c_{k_2}^{(q_2)}\right)
\left(c_{k_2}^{(p_2)} c_{k_3}^{(q_3)}\right) \cdots
\left(c_{k_{n}}^{(p_{n})} c_{k_1}^{(q_1)}\right) \right\}\nn\\
&&\times \delta(p_1 T_c+\tau_{k_1},q_2 T_c+\tau_{k_2}) \delta(p_2
T_c+\tau_{k_2},q_3 T_c+\tau_{k_3})\cdots \delta(p_{n}
T_c+\tau_{k_{n}},q_1 T_c+\tau_{k_1}),\label{eq:mouse0131}
\end{eqnarray}
when $\textrm{x}=\textrm{cs}$, and
\begin{eqnarray}
&&\lim_{M\rightarrow\infty}(2M+1)^{-1}\sum_{{\cal K}\in{\cal
X}}\sum_{{\cal M}\in {\cal Y}} \sum_{{\cal P}_1\in {\cal Z}_1}\cdots
\sum_{{\cal P}_{n}\in{\cal Z}_{n}}
\textrm{E}\left\{\left(c_{k_1}^{(p_1)} c_{k_2}^{(q_2)}\right)
\left(c_{k_2}^{(p_2)} c_{k_3}^{(q_3)}\right) \cdots
\left(c_{k_{n}}^{(p_{n})} c_{k_1}^{(q_1)}\right) \right\}\label{eq:cow0131}\\
&&\times \textrm{E}\{R_\psi((p_1-q_2) T_c+\tau_{k_1}-\tau_{k_2})
R_\psi((p_2-q_3) T_c+\tau_{k_2}-\tau_{k_3})\cdots R_\psi((p_{n}-q_1)
T_c+\tau_{k_{n}}-\tau_{k_1})\},\nn
\end{eqnarray}
when $\textrm{x}=\textrm{ca}$, where ${\cal P}_r=\{p_r,q_r\}$ and
${\cal Z}_r=[m_r N,(m_r+1)N-1]^2$ for $1\leq r\leq n$.
Owing to the tri-diagonal structure of $\pmb R_{\textrm{cs}}$ shown
in (\ref{eq:rabbit1019}),
there are constraints $|m_r-m_{r+1}|\leq 1$ 
for $1\leq r\leq n$ when (\ref{eq:mouse0131}) is considered.
Moreover, as stated in the beginning of
Section~\ref{subsection:cow0121}, the relative delays $\tau_k$'s are
viewed as deterministic when the chip waveform is the ideal sinc
pulse and viewed as random when otherwise, the expectation in the
second line of (\ref{eq:cow0131}) can be discarded when the sinc
pulse is adopted.

Computations of (\ref{eq:mouse0131}) and (\ref{eq:cow0131}) can be
executed by considering the equivalence patterns of elements in
${\cal K}=\{k_1,k_2,\cdots, k_{n}\}$. As equivalence relation and
partition are essentially equivalent, the computations of
(\ref{eq:mouse0131}) and (\ref{eq:cow0131}) can be carried out with
the aid of set partition theory, where $\cal K$ is a totally ordered
set with ordering $k_1\succ k_2\succ\cdots\succ k_n$, and $k_r$ and
$k_s$ are partitioned in the same class if and only if they take the
same integer in $[1,K]$. Note that the ordering $k_1\succ
k_2\succ\cdots\succ k_n$ is just an arrangement of objects $k_i$'s
as an ordered set. It is different from the ordering of the values
taken by summation variables $k_1,k_2,\cdots,k_n$ in $[1,K]$.

In the following, the summation $\sum_{{\cal K}\in{\cal X}}$ in
(\ref{eq:mouse0131}) and (\ref{eq:cow0131}) is decomposed into
several ones using properties stated in
Appendix~\ref{appendix:noncrossing}. Let ${\cal X}_j\subseteq{\cal
X}=[1,K]^n$ such that each element $x_j=(x_j(1),\cdots,x_j(n))$ in
${\cal X}_j$ corresponds to a $j$-class noncrossing partition of an
$n$-element ordered set. We mean $x_j$ corresponds to a partition by
that $x_j(s)=x_j(t)$ if and only if the $s$- and $t$-th elements are
partitioned in the same class in that partition. Moreover, let
${\cal X}_j(b_1,b_2,\cdots,b_{n-j+1})$, with $b_1\geq
b_2\geq\cdots\geq b_{n-j+1}\geq 1$, stand for the union of $x_j$'s
whose corresponding noncrossing partitions have Kreweras
complementation maps with class sizes $(b_1,b_2,\cdots,b_{n-j+1})$
(but not specifying which class gets which size). Since the Kreweras
complementation map of a noncrossing partition is noncrossing as
well, by (\ref{eq:cow0519}), the number of elements in ${\cal
X}_j(b_1,b_2,\cdots,b_{n-j+1})$ is given by
$$
\#{\cal X}_j(b_1,b_2,\cdots,b_{n-j+1})=\dfrac{n(n-1)\cdots(j+1)}
{f(b_1,b_2,\cdots,b_{n-j+1})}\cdot K(K-1)\cdots(K-j+1).
$$
The above equation is interpreted as follows. The number of
noncrossing partitions associated with ${\cal
X}_j(b_1,b_2,\cdots,b_{n-j+1})$ is
$n(n-1)\cdots(j+1)/f(b_1,b_2,\cdots,b_{n-j+1})$, and each of these
noncrossing partitions has $j$ classes, with each class specified by
a distinct integer in $[1,K]$. Moreover, let ${\cal
X}_{cro}\subseteq{\cal X}$, where each element in ${\cal X}_{cro}$
corresponds to a crossing partition of an $n$-element ordered set.
With these settings, the summation $\sum_{{\cal K}\in{\cal X}}$ in
(\ref{eq:mouse0131}) and (\ref{eq:cow0131}) can be decomposed as
\begin{equation}\label{eq:mouse0603}
\sum_{{\cal K}\in{\cal
X}}\hspace{3mm}\equiv\hspace{3mm}\sum_{j=1}^n\hspace{2mm}
\mathop{\sum_{b_1+b_2+\cdots+b_{n-j+1}=n}}_{b_1\geq
b_2\geq\cdots\geq b_{n-j+1}\geq 1}\hspace{2mm}\sum_{{\cal K}\in
{\cal X}_j(b_1,b_2,\cdots,b_{n-j+1})}+\sum_{{\cal K}\in {\cal
X}_{cro}}.
\end{equation}

\begin{figure*}[t]
\begin{center}
\begin{tabular}{c}
\psfig{figure=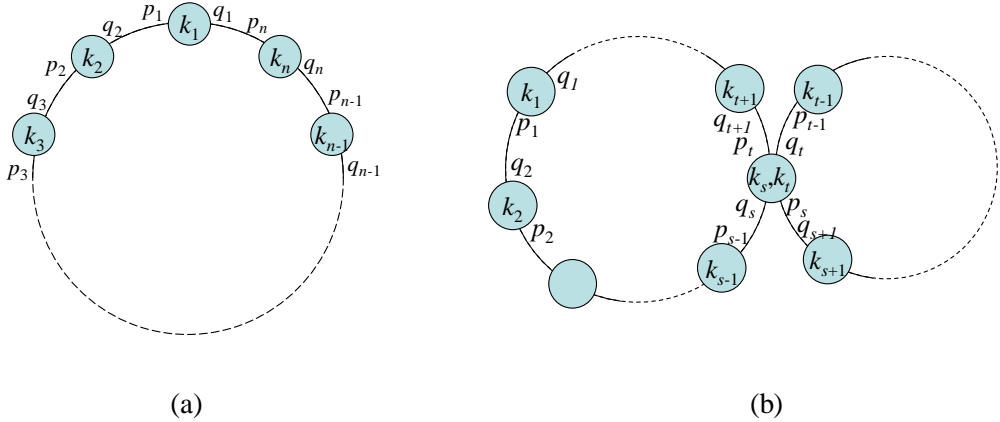,height=6cm}
\end{tabular}
\end{center}
\caption{The $K$-graphs of (a) $n$-class partition
$\{\{k_1\},\{k_2\},\cdots,\{k_{n}\}\}$, and (b) $(n-1)$-class
partition $\{\{k_1\},\{k_2\},\cdots,\{k_s,k_t\},\cdots,\{k_{n}\}\}$
with $s<t$.} \label{fig:mouse1114}
\end{figure*}

Now, we consider $K$-graphs corresponding to elements of ${\cal
X}_j(b_1,b_2,\cdots,b_{n-j+1})$ and of ${\cal X}_{cro}$ in
(\ref{eq:mouse0603}).
To embed the summation variables $p_r$'s and $q_r$'s of
(\ref{eq:mouse0131}) and (\ref{eq:cow0131}) into a $K$-graph, in the
$n$-vertex cycle composed of vertices $k_1,k_2,\cdots,k_n$,
two ends of the edge connecting $k_r$ and $k_{r+1}$ are labelled
with $p_r$ and $q_{r+1}$, with the former and latter touching $k_r$
and $k_{r+1}$, respectively. We call these $p_r$'s and $q_r$'s as
\textit{edge variables}. Fig.~\ref{fig:mouse1114}(a) shows the graph
with edge variables labelled. The edge connecting vertices $k_r$ and
$k_{r+1}$ stands for
$$
c_{k_r}^{(p_r)}c_{k_{r+1}}^{(q_{r+1})}\cdot\delta(p_r
T_c+\tau_{k_r},q_{r+1}T_c+\tau_{k_{r+1}})
$$
and
$$
c_{k_r}^{(p_r)}c_{k_{r+1}}^{(q_{r+1})}\cdot R_\psi((p_r-q_{r+1})
T_c+\tau_{k_r}-\tau_{k_{r+1}})
$$
in (\ref{eq:mouse0131}) and (\ref{eq:cow0131}), respectively, and
the summands in the equations are yielded by multiplying terms
associated with all of the edges together.
Vertices $k_r$'s and edge variables $p_r$'s, $q_r$'s in
Fig.~\ref{fig:mouse1114}(a) are all summation variables of
(\ref{eq:mouse0131}) and (\ref{eq:cow0131}). Another set of
summation variables $m_r$'s are implicitly embedded in $p_r$'s and
$q_r$'s by that the ranges of $p_r$ and $q_r$ are both
$[m_rN,(m_r+1)N-1]$. 
We would like to inspect under what equivalence relation of $p_r$'s
and $q_r$'s would the contributions to (\ref{eq:mouse0131}) and
(\ref{eq:cow0131}) become nonvanishing in the large-system limit.

\subsection{Contributions of Noncrossing Partitions to $\mu(\bb
R_\textrm{cs}^n)$ and $\mu(\bb
R_\textrm{ca}^n)$}\label{sub_appendix:mouse0124}

To compute the contribution of noncrossing partitions of ${\cal
K}=\{k_1,k_2,\cdots,k_n\}$ to $\mu(\bb R_\textrm{cs}^n)$ and
$\mu(\bb R_\textrm{ca}^n)$, we replace $\sum_{{\cal K}\in{\cal X}}$
in (\ref{eq:mouse0131}) and (\ref{eq:cow0131}) with the first term
at the right-hand-side of (\ref{eq:mouse0603}) and then use
(\ref{eq:cow0120}). Given non-ascending ordered natural numbers
$(b_1,b_2,\cdots,b_{n-j+1})$ such that $b_1+b_2+\cdots+b_{n-j+1}=n$.
The contribution of each element of ${\cal
X}_j(b_1,b_2,\cdots,b_{n-j+1})$ to (\ref{eq:mouse0131}) and
(\ref{eq:cow0131}) is to be evaluated. First, we consider $j=n$.
There is only one $K$-graph, shown in Fig.~\ref{fig:mouse1114}(a).
Since all of $k_1,k_2,\cdots,k_n$ are distinct, the expectations of
spreading sequences in (\ref{eq:mouse0131}) and (\ref{eq:cow0131})
are nonzero (equal to $N^{-n}$) if and only if $p_r=q_r$ for $1\leq
r\leq n$. Note that, since $c_{k_r}^{(p_r)}$ and $c_{k_r}^{(q_r)}$
are chips in the same transmitted symbol, i.e., $m_r N\leq
p_r,q_r\leq (m_r+1)N-1$, the necessary and sufficient condition
$p_r=q_r$ stated above holds for both short-code and long-code
systems. The contribution of each element of ${\cal X}_n(n)$ to
(\ref{eq:mouse0131}) becomes
\begin{eqnarray}\label{eq:cow1115}
&& N^{-n} \lim_{M\rightarrow\infty} (2M+1)^{-1}\sum_{{\cal M}\in
{\cal Y}}\sum_{p_1\in {\cal Z}'_1}\sum_{p_2\in {\cal Z}'_2}\cdots
\sum_{p_{n}\in{\cal Z}'_{n}} \delta(p_1 T_c+\tau_{k_1},p_2
T_c+\tau_{k_2})\nn\\
&&\times\delta(p_2 T_c+\tau_{k_2},p_3 T_c+\tau_{k_3})\cdots
\delta(p_{n} T_c+\tau_{k_{n}},p_1 T_c+\tau_{k_1}),
\end{eqnarray}
where ${\cal Z}'_r=[m_r,(m_r+1)N-1]$ and $|m_r-m_{r+1}|\leq 1$. The
product of $\delta(\cdot,\cdot)$'s is nonzero and equal to one if
and only if all $p_r T_c+\tau_{k_r}$, $1\leq r\leq n$, are equal.
Since we have $p_1\in [m_1 N,(m_1+1)N-1]$ for each $m_1\in[-M,M]$,
it is not difficult to see, in (\ref{eq:cow1115}), the term behind
$N^{-n}$ is equal to $N$. Consequently, (\ref{eq:cow1115}) is equal
to $N^{-n}\cdot N=N^{-n+1}$. On the other hand, the contribution of
each element of ${\cal X}_n(n)$ to (\ref{eq:cow0131}) is
\begin{eqnarray}\label{eq:cow0213}
&& N^{-n} \lim_{M\rightarrow\infty} (2M+1)^{-1}\sum_{{\cal M}\in
{\cal Y}}\sum_{p_1\in {\cal Z}'_1}\sum_{p_2\in {\cal Z}'_2}\cdots
\sum_{p_{n}\in{\cal Z}'_{n}} \textrm{E}\left\{R_\psi((p_1-p_2)
T_c+\tau_{k_1}-\tau_{k_2})\right.\nn\\
&&\times \left. R_\psi((p_2-p_3) T_c+\tau_{k_2}-\tau_{k_3})\cdots
R_\psi((p_{n}-p_1) T_c+\tau_{k_{n}}-\tau_{k_1})\right\}.
\end{eqnarray}
By Lemma~\ref{proposition:mouse0515}, it is readily seen that, in
(\ref{eq:cow0213}), the term behind $N^{-n}$ is equal to $N{\cal
W}_\psi^{(n)}$. Consequently, (\ref{eq:cow0213}) is equal to
$N^{-n}\cdot N{\cal W}_\psi^{(n)}=N^{-n+1}{\cal W}_\psi^{(n)}$. Note
that the equivalence condition of $p_r=q_r$ for $1\leq r\leq n$
makes \textit{edge variables touching the same vertex in the
K-graph, i.e., Fig.~\ref{fig:mouse1114}(a), to be equal}.

Next, we consider $j=n-1$. Following the statement in the remark of
Definition~\ref{definition:mouse0810}, we understand that any
$K$-graph corresponding to ${\cal X}_{n-1}(b_1,b_2)$ can be obtained
from the $K$-graph of ${\cal X}_n(n)$ (denoted by $G_n$) by merging
two vertices of $G_n$ into one. Suppose that vertices $k_s$ and
$k_t$ of $G_n$ are merged and $s<t$, meaning that $k_s=k_t$ in
(\ref{eq:mouse0131}) and (\ref{eq:cow0131}). Then, the original
$n$-vertex cycle is decomposed into two cycles with numbers of edges
equal to $t-s$ and $n-t+s$ as shown in Fig.~\ref{fig:mouse1114}(b).
This $K$-graph corresponds to elements of ${\cal
X}_{n-1}(\max(t-s,n-t+s),\min(t-s,n-t+s))$. With a slight abuse of
notational simplification, we let $e=t-s$ and write ${\cal
X}_{n-1}(\max(t-s,n-t+s),\min(t-s,n-t+s))$ as ${\cal
X}_{n-1}(e,n-e)$ afterwards, although $e$ may be smaller than $n-e$.

\begin{figure*}[t]
\begin{center}
\begin{tabular}{c}
\psfig{figure=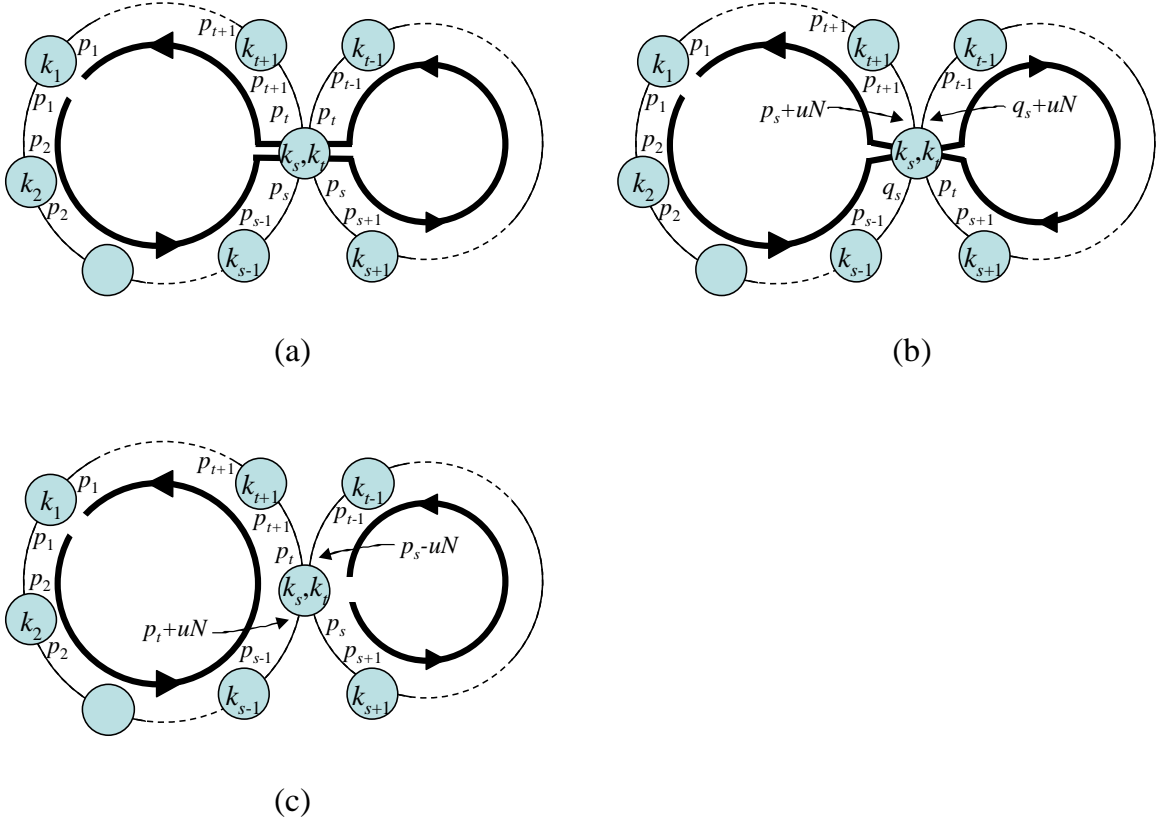,height=11.5cm}
\end{tabular}
\end{center}
\caption{The directed flows represented by thick lines in (a), (b)
and (c) denote the product of $\delta$ functions corresponding to
the equivalence patterns of i) $q_s=p_s$ and $q_t=p_t$, ii)
$p_t=p_s+uN$ and $q_t=q_s+uN$ and iii) $q_s=p_t+uN$ and
$q_t=p_s-uN$, respectively.} \label{fig:rabbit1115}
\end{figure*}

We are going to demonstrate that, concerned with the four edge
variables $p_s,q_s,p_t,q_t$ of the merged vertex in
Fig.~\ref{fig:mouse1114}(b), it is sufficient to consider $p_t=q_s$
and $p_s=q_t$. Since $k_i$'s are all distinct except for $k_s=k_t$,
to yield a nonzero expectation of
$\prod_{r=1}^{n}c_{k_r}^{(q_r)}c_{k_r}^{(p_r)}$ in
(\ref{eq:mouse0131}) and (\ref{eq:cow0131}), it is required that
$p_r=q_r$ for $r\in\{1,2,\cdots, n\} \setminus\{s,t\}$ and
$c_{k_s}^{(p_s)},c_{k_s}^{(q_s)},c_{k_s}^{(p_t)},c_{k_s}^{(q_t)}$
are in pairs, i.e., anyone of the following three conditions
\begin{enumerate}
\item[i)] $q_s=p_s$ and $q_t=p_t$,
\item[ii)] $p_t=p_s+uN$ and $q_t=q_s+uN$ with any integer $u$,
\item[iii)] $q_s=p_t+uN$ and $q_t=p_s-uN$ with any integer $u$,
\end{enumerate}
for a short-code system, and
\begin{enumerate}
\item[iv)] $q_s=p_s$ and $q_t=p_t$,
\item[v)] $p_t=p_s$ and $q_t=q_s$,
\item[vi)] $q_s=p_t$ and $q_t=p_s$,
\end{enumerate}
for a long-code system. The reasons for same sign of two $uN$ in
condition ii) and opposite signs in condition iii) is because
$m_sN\leq p_s,q_s\leq (m_s+1)N-1$ and $m_t N\leq p_t,q_t\leq
(m_t+1)N-1$.
As conditions iv)--vi) are special cases of i)--iii) with $u=0$, we
will use the latter to demonstrate our goal, i.e., it is sufficient
to consider only the equivalence relation of case vi).

With conditions i)--iii) and $p_r=q_r$ for $r\in\{1,2,\cdots, n\}
\setminus\{s,t\}$, the contribution of each element of ${\cal
X}_{n-1}(e,n-e)$ to (\ref{eq:mouse0131}) and (\ref{eq:cow0131})
becomes
\begin{eqnarray}
&& N^{-n+2} \lim_{M\rightarrow\infty} (2M+1)^{-1}\sum_{{\cal M}\in
{\cal Y}}\sum_{p_1\in {\cal Z}'_1}\sum_{p_2\in {\cal Z}'_2}\cdots
\sum_{p_{n}\in{\cal Z}'_{n}}
\textrm{E}\left\{c_{k_s}^{(p_s)}c_{k_s}^{(q_s)}c_{k_s}^{(p_t)}c_{k_s}^{(q_t)}\right\}
\nn\\
&&\times\delta(p_1 T_c+\tau_{k_1},p_2 T_c+\tau_{k_2})\delta(p_2
T_c+\tau_{k_2},p_3 T_c+\tau_{k_3})\cdots \delta(p_{n}
T_c+\tau_{k_{n}},p_1 T_c+\tau_{k_1})|_{k_s=k_t},\label{eq:mouse0318}
\end{eqnarray}
and
\begin{eqnarray}
&& N^{-n+2} \lim_{M\rightarrow\infty} (2M+1)^{-1}\sum_{{\cal M}\in
{\cal Y}}\sum_{p_1\in {\cal Z}'_1}\sum_{p_2\in {\cal Z}'_2}\cdots
\sum_{p_{n}\in{\cal Z}'_{n}}
\textrm{E}\left\{c_{k_s}^{(p_s)}c_{k_s}^{(q_s)}c_{k_s}^{(p_t)}c_{k_s}^{(q_t)}\right\}\nn\\
&&\times \textrm{E}\left\{R_\psi((p_1-p_2)
T_c+\tau_{k_1}-\tau_{k_2}) R_\psi((p_2-p_3)
T_c+\tau_{k_2}-\tau_{k_3})\cdots \right.\nn\\
&&\left.\hspace{1cm}R_\psi((p_{n}-p_1)
T_c+\tau_{k_{n}}-\tau_{k_1})\right\}|_{k_s=k_t},\label{eq:cow0318}
\end{eqnarray}
respectively. A careful inspection reveals that the product of
$\delta$ functions in (\ref{eq:mouse0318}) becomes
\begin{eqnarray}
&&\mathop{\prod_{(\gamma,\epsilon)\in {\cal I}_1}}_{k_t=k_s}
\delta(p_\gamma T_c+\tau_{k_\gamma},p_\epsilon
T_c+\tau_{k_\epsilon}),\quad {\cal
I}_1=\{(1,2),(2,3),\cdots,(n-1,n),(n,1)\}\label{eq:mouse0825_1}
\end{eqnarray}
for condition i), where we let $q_s:=p_s$ and $q_t:=p_t$,
\begin{eqnarray}
&&\delta(p_{s-1}T_c+\tau_{k_{s-1}},q_sT_c+\tau_{k_s})\delta((q_s+uN)T_c+\tau_{k_s},p_{t-1}T_c+\tau_{k_{t-1}})\nn\\
&&\hspace{1cm}\times\delta((p_s+uN)T_c+\tau_{k_s},p_{t+1}T_c+\tau_{k_{t+1}})
\prod_{(\gamma,\epsilon)\in {\cal I}_2} \delta(p_\gamma
T_c+\tau_{k_\gamma},p_\epsilon T_c+\tau_{k_\epsilon}),\nn\\
&&{\cal
I}_2=\{(1,2),\cdots,(s-2,s-1),(t+1,t+2),\cdots,(n,1)\}\nn\\
&&\hspace{2cm}\cup\{(s,s+1),(s+1,s+2),\cdots,(t-2,t-1)\}\label{eq:horse0122}
\end{eqnarray}
for condition ii), where we let $p_t:=p_s+uN$ and $q_t:=q_s+uN$, and
\begin{eqnarray}
&& \delta(p_{s-1}T_c+\tau_{k_{s-1}},(p_t+uN)T_c+\tau_{k_s})
\delta(p_{t-1}T_c+\tau_{k_{t-1}},(p_s-uN)T_c+\tau_{k_s})\nn\\
&&\hspace{1cm}\times\mathop{\prod_{(\gamma,\epsilon)\in{\cal
I}_3}}_{k_t=k_s}\delta(p_\gamma T_c+\tau_{k_\gamma},p_\epsilon
T_c+\tau_{k_\epsilon}) \prod_{(\eta,\zeta)\in{\cal
I}_4}\delta(p_\eta
T_c+\tau_{k_\eta},p_\zeta T_c+\tau_{k_\zeta}),\nn\\
&&{\cal I}_3=\{(1,2),\cdots,(s-2,s-1),(t,t+1),\cdots,(n,1)
\},\nn\\
&&{\cal I}_4=\{(s,s+1),\cdots,(t-2,t-1)\}\label{eq:sheep0122}
\end{eqnarray}
for condition iii), where $q_s:=p_t+uN$ and $q_t:=p_s-uN$. The
product of $R_\psi$ functions in (\ref{eq:cow0318}) can be obtained
from the above equations by replacing each $\delta(a,b)$ with
$R_\psi(a-b)$. According to
(\ref{eq:mouse0825_1})--(\ref{eq:sheep0122}), conditions i)--iii)
are graphically represented by Figs.~\ref{fig:rabbit1115}(a)--(c),
respectively. For instance, in Fig.~\ref{fig:rabbit1115}(a), except
for the merged vertex, two edge variables touching the same vertex
are the same. This is because $p_r=q_r$ for $r\in\{1,2,\cdots, n\}
\setminus\{s,t\}$, so we replace $q_r, r\in\{1,2,\cdots, n\}
\setminus\{s,t\}$ in Fig.~\ref{fig:mouse1114}(b) with $p_r$.
Moreover, $q_s$ and $q_t$ in Fig.~\ref{fig:mouse1114}(b) are
replaced with $p_s$ and $p_t$, respectively, indicating that
$q_s:=p_s$ and $q_t:=p_t$. In each of
Figs.~\ref{fig:rabbit1115}(a)--(c), the product of $\delta$ and
$R_\psi$ functions in (\ref{eq:mouse0318}) and (\ref{eq:cow0318}),
respectively, are depicted in the form of directed flow(s),
indicated by thick lines traversing edges. The thick line passing
through an edge with variables $p_\gamma$ and $p_\epsilon$ (or
$q_\epsilon$) represents $\delta(p_\gamma
T_c+\tau_{k_\gamma},p_\epsilon T_c+\tau_{k_\epsilon})$ (or
$\delta(p_\gamma T_c+\tau_{k_\gamma},q_\epsilon
T_c+\tau_{k_\epsilon})$) for (\ref{eq:mouse0318}) and
$R_\psi((p_\gamma-p_\epsilon)
T_c+\tau_{k_\gamma}-\tau_{k_\epsilon})$ (or
$R_\psi((p_\gamma-q_\epsilon)
T_c+\tau_{k_\gamma}-\tau_{k_\epsilon})$) for (\ref{eq:cow0318}). For
an edge with variables $p_\gamma+uN$ and $p_\epsilon$, it
corresponds to $\delta((p_\gamma+uN) T_c+\tau_{k_\gamma},p_\epsilon
T_c+\tau_{k_\epsilon})$ for (\ref{eq:mouse0318}).
Conditions i)--iii) are taken into account below.

\begin{itemize}

\item Condition i): Consider the chip-synchronous case. The product
of $\delta$ functions in (\ref{eq:mouse0825_1}) is zero if $p_s\neq
p_t$. Thus, we consider $p_s=p_t$, resulting in $p_s=q_s=p_t=q_t$,
and the expectation in the first line of (\ref{eq:mouse0318}) is
$O(N^{-2})$ because the fourth moment of $c_k^{(p)}$ is $O(N^{-2})$.
It can be taken out from multi-dimensional summation, yielding
\begin{eqnarray}
&& O(N^{-n}) \lim_{M\rightarrow\infty} (2M+1)^{-1}\sum_{{\cal M}\in
{\cal Y}}\sum_{p_1\in {\cal Z}'_1}\sum_{p_2\in {\cal Z}'_2}\cdots
\sum_{p_{n}\in{\cal Z}'_{n}} \delta(p_1 T_c+\tau_{k_1},p_2
T_c+\tau_{k_2})\nn\\
&&\times\delta(p_2 T_c+\tau_{k_2},p_3 T_c+\tau_{k_3})\cdots
\delta(p_{n} T_c+\tau_{k_{n}},p_1
T_c+\tau_{k_1})|_{k_s=k_t}.\label{eq:mouse0319}
\end{eqnarray}
It is readily seen that, in (\ref{eq:mouse0319}), the term behind
$O(N^{-n})$ is equal to $N$. Thus, in this condition, the
contribution of each element in ${\cal X}_{n-1}(e,n-e)$ to
(\ref{eq:mouse0131}) is $O(N^{-n+1})$.

Consider the case of chip-asynchronous. Assume $p_s\neq p_t$. We
replace each $\delta(a,b)$ in (\ref{eq:mouse0825_1}) with
$R_\psi(a-b)$ and plug the resulting product of $R_\psi$ functions
back into (\ref{eq:cow0318}). The expectation of chips in
(\ref{eq:cow0318}) is $N^{-2}$, and it can be taken out from the
multi-sum, resulting in
\begin{eqnarray}
&& N^{-n} \lim_{M\rightarrow\infty} (2M+1)^{-1}\sum_{{\cal M}\in
{\cal Y}}\sum_{p_1\in {\cal Z}'_1}\sum_{p_2\in {\cal Z}'_2}\cdots
\sum_{p_{n}\in{\cal Z}'_{n}}\textrm{E}\left\{R_\psi((p_1-p_2)
T_c+\tau_{k_1}-\tau_{k_2})\right.\nn\\
&&\hspace{1cm}\times  \left.R_\psi((p_2-p_3)
T_c+\tau_{k_2}-\tau_{k_3})\cdots R_\psi((p_{n}-p_1)
T_c+\tau_{k_{n}}-\tau_{k_1})\right\}|_{k_s=k_t}.\label{eq:cow0319}
\end{eqnarray}
By Lemma~\ref{proposition:mouse0515}, the term behind $N^{-n}$ is
equal to $N{\cal W}_\psi^{(n)}$ for bandwidth of $\psi(t)$ either
smaller or wider than $1/(2T_c)$. Thus, (\ref{eq:cow0319}) is equal
to $N^{-n+1}{\cal W}_\psi^{(n)}$. When $p_s=p_t$, the expectation of
chips in (\ref{eq:cow0318}) is $O(N^{-2})$, and (\ref{eq:cow0319})
is revised by changing $N^{-n}$ to $O(N^{-n})$ and imposing a
constraint of $p_s=p_t$. Under these circumstances, the revised
equation is equal to $O(N^{-n+1})$. Thus, with both $p_s=p_t$ and
$p_s\neq p_t$, the contribution of each element of ${\cal
X}_{n-1}(e,n-e)$ to (\ref{eq:cow0131}) is $O(N^{-n+1})$.

\item Condition ii):
Consider the chip-synchronous case. The product of $\delta$
functions is given by (\ref{eq:horse0122}). It can be checked that
the product is zero if either $u\neq 0$ or $p_s\neq q_s$ not
satisfied. Thus, we let $u=0$ and $p_s=q_s$, which yields
$p_s=q_s=p_t=q_t$. By similar arguments as in condition i), the
contribution of each element of ${\cal X}_{n-1}(e,n-e)$ to
(\ref{eq:mouse0131}) is $O(N^{-n+1})$.

In the case of chip-asynchronous, we replace each $\delta(a,b)$ in
(\ref{eq:horse0122}) with $R_\psi(a-b)$ and plug the resulting
product of $R_\psi$ functions back into (\ref{eq:cow0318}). For
either $p_s=q_s$ or $p_s\neq q_s$, the expectation of chips in
(\ref{eq:cow0318}) is $O(N^{-2})$, and we take it out from the
multi-sum. Thus the leading term in (\ref{eq:cow0318}) becomes
$O(N^{-n})$. Tracing the proof of Lemma~\ref{proposition:mouse0515},
we can find that the limiting normalized infinite sum of products of
$R_\psi$ functions in (\ref{eq:cow0318}), i.e., the term behind the
leading term, is equal to
$$
\frac{N}{2\pi T_c^{n-1}}\int
e^{-j2uNT_c\Omega}\left|\Psi(\Omega)\right|^{2n}\textrm{d}\Omega.
$$
When $N$ is large, the integral is nonzero (equal to $2\pi
T_c^{n-1}{\cal W}_\psi^{(n)}$) only if $u=0$. Thus, the contribution
of each element of ${\cal X}_{n-1}(e,n-e)$ to (\ref{eq:cow0131}) is
$O(N^{-n+1})$.

\item Condition iii): In chip-synchronous case,
the product of $\delta$ functions is given by (\ref{eq:sheep0122}).
This product is zero if $u\neq 0$. Thus, we consider $u=0$ and
$q_s=p_t\neq q_t=p_s$, which implies $m_s=m_t$, and
(\ref{eq:mouse0318}) becomes
\begin{eqnarray}
&&N^{-n}\lim_{M\rightarrow\infty}(2M+1)^{-1}
\mathop{\sum_{m_1,\cdots,m_{s-1},m_t,m_{t+1},\cdots,m_{n}}}_{p_1,\cdots,p_{s-1},p_t,p_{t+1},\cdots,p_{n}}
\mathop{\prod_{(\gamma,\epsilon)\in{\cal
I}_3}}_{k_t=k_s}\delta(p_\gamma T_c+\tau_{k_\gamma},p_\epsilon
T_c+\tau_{k_\epsilon})\nn\\
&&\times
\mathop{\mathop{\sum_{m_s,m_{s+1},\cdots,m_{t-1}}}_{p_s,p_{s+1},\cdots,p_{t-1}}}_{m_s=m_t}\prod_{(\eta,\zeta)\in{\cal
I}_4}\delta(p_\eta T_c+\tau_{k_\eta},p_\zeta
T_c+\tau_{k_\zeta}).\label{eq:tiger0825_1}
\end{eqnarray}
Note that, for any particular $m_t$, the constraint of $m_s=m_t$
results in only $N$ choices for $p_s\in [m_t N,(m_t+1)N-1]$ in the
second line of (\ref{eq:tiger0825_1}). It follows that
(\ref{eq:tiger0825_1}) is equal to $N^{-n+2}$.

In the case of chip-asynchronous, (\ref{eq:cow0318}) now becomes
\begin{eqnarray}
&&N^{-n}\lim_{M\rightarrow\infty}(2M+1)^{-1}\textrm{E}\Biggl\{\nn\\
&&\mathop{\sum_{m_1,\cdots,m_{s-1},m_t,m_{t+1},\cdots,m_{n}}}_{p_1,\cdots,p_{s-1},p_t,p_{t+1},\cdots,p_{n}}
R_\psi((p_{s-1}-p_t-uN)T_c+\tau_{k_{s-1}}-\tau_{k_s})\mathop{\prod_{(\gamma,\epsilon)\in{\cal
I}_3}}_{k_t=k_s}R_\psi((p_\gamma-p_\epsilon)
T_c+\tau_{k_\gamma}-\tau_{k_\epsilon})\nn\\
&&\mathop{\mathop{\sum_{m_s,m_{s+1},\cdots,m_{t-1}}}_{p_s,p_{s+1},\cdots,p_{t-1}}}_{m_s=m_t}
R_\psi((p_{t-1}-p_s+uN)T_c+\tau_{k_{t-1}}-\tau_{k_s})
\prod_{(\eta,\zeta)\in{\cal I}_4}R_\psi((p_\eta-p_\zeta)
T_c+\tau_{k_\eta}-\tau_{k_\zeta})\Biggr\}.\label{eq:tiger0319}
\end{eqnarray}
If the sinc chip waveform is employed, the expectation of
(\ref{eq:tiger0319}) can be removed. By similar arguments as in
condition ii), the infinite sums in the second and third lines of
(\ref{eq:tiger0319}) are zero if $u\neq 0$.  By part 1) of
Lemma~\ref{proposition:mouse0515}, when $u=0$, (\ref{eq:tiger0319})
is equal to $N^{-n}\cdot N{\cal W}_\psi^{(e)}\cdot N{\cal
W}_\psi^{(n-e)}=N^{-n+2}{\cal W}_\psi^{(e)}{\cal W}_\psi^{(n-e)}$.
When the bandwidth of $\psi(t)$ is wider than $1/(2T_c)$, the
calculation is more involved. We also let $u=0$. Note that, in
(\ref{eq:tiger0319}), both the product of $R_\psi$ functions in the
second and third lines contain $\tau_{k_s}$. We can rewrite the
equation by
\begin{eqnarray}
&&N^{-n}\lim_{M\rightarrow\infty}(2M+1)^{-1}\nn\\
&&\times\textrm{E}_{\tau_{k_s}}\Biggl\{
\mathop{\sum_{m_1,\cdots,m_{s-1},m_t,m_{t+1},\cdots,m_{n}}}_{p_1,\cdots,p_{s-1},p_t,p_{t+1},\cdots,p_{n}}
\textrm{E}\Biggl\{\mathop{\prod_{(\gamma,\epsilon)\in{\cal
I}'_3}}_{k_t=k_s}R_\psi((p_\gamma-p_\epsilon)
T_c+\tau_{k_\gamma}-\tau_{k_\epsilon})\Bigr|\tau_{k_s}\Biggr\}\nn\\
&&\hspace{1.5cm}\times\mathop{\mathop{\sum_{m_s,m_{s+1},\cdots,m_{t-1}}}_{p_s,p_{s+1},\cdots,p_{t-1}}}_{m_s=m_t}
\textrm{E}\Biggl\{\Biggl.\prod_{(\eta,\zeta)\in{\cal
I}'_4}R_\psi((p_\eta-p_\zeta)
T_c+\tau_{k_\eta}-\tau_{k_\zeta})\Bigr|\tau_{k_s}\Biggr\}\Biggr\},\label{eq:cow0216}
\end{eqnarray}
where ${\cal I}'_3={\cal I}_3\cup \{(s-1,t)\}$, ${\cal I}'_4={\cal
I}_4\cup \{(t-1,s)\}$, the first expectation at the second line is
w.r.t. $\tau_{k_s}$, and the remaining two are w.r.t.
$\{\tau_{k_i}\}_{i=1}^n$ conditioned on $\tau_{k_s}$. By part 2) of
Lemma~\ref{proposition:mouse0515}, the two inner conditional
expectations in the second and third lines of (\ref{eq:cow0216}) are
equal to $N{\cal W}_\psi^{(e)}$ and $N{\cal W}_\psi^{(n-e)}$,
respectively. Thus, identical to the situation when $\psi(t)$ is the
sinc, the contribution of each element in ${\cal X}_{n-1}(e,n-e)$ to
(\ref{eq:cow0131}) is equal to $N^{-n+2}{\cal W}_\psi^{(e)}{\cal
W}_\psi^{(n-e)}$.
\end{itemize}

To sum up, when $j=n-1$, the contributions of each element in ${\cal
X}_j(e,n-e)$ to (\ref{eq:mouse0131}) are $O(N^{-n+1})$,
$O(N^{-n+1})$, and $N^{-n+2}$ for conditions i)--iii), respectively.
Regarding the contributions to (\ref{eq:cow0131}), conditions
i)--iii) have $O(N^{-n+1})$, $O(N^{-n+1})$, and $N^{-n+2}{\cal
W}_\psi^{(e)}{\cal W}_\psi^{(n-e)}$, respectively.
Thus, as $N\rightarrow\infty$, it is sufficient to consider
condition iii) since the corresponding contribution has the highest
order of $N$. Note that we require $u=0$ to get the result.

We reach the conclusion that, when $j=n-1$ with $k_s$ and $k_t$
partitioned in the same class, the equivalence conditions of edge
variables $\{p_r,q_r\}$ are that $p_r=q_r$ for
$r\in\{1,2,\cdots,n\}\setminus\{s,t\}$, $q_s=p_t$, and $p_s=q_t$.
Observing Fig.~\ref{fig:mouse1114}(b), we find that the equivalence
relation can be stated as: \textit{within each of the two cycles in
the $K$-graph, two edge variables touching the same vertex are
equal}.

\begin{figure*}[t]
\begin{center}
\begin{tabular}{c}
\psfig{figure=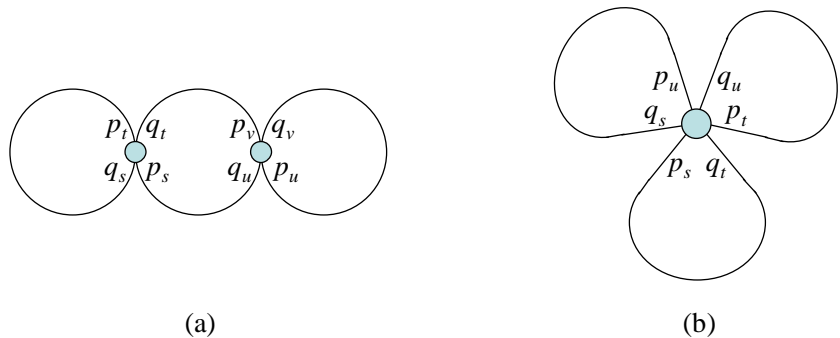,height=5cm}
\end{tabular}
\end{center}
\caption{The $K$-graphs of $(n-2)$-class noncrossing partitions of
the totally ordered set ${\cal K}=\{k_1,k_2,\cdots,k_n\}$: (a)
$k_s,k_t$ and $k_u,k_v$ ($1\leq s<u<v<t\leq n$) are respectively in
the same class and all other elements in $\cal K$ are singletons,
and (b) $k_s,k_t,k_u$ are in the same class and all others are
singletons.}\label{fig:rabbit0319}
\end{figure*}

We consider $j=n-2$ below. Two cases are possible. One is that
$k_s,k_t$ and $k_u,k_v$ are respectively in the same class ($1\leq
s<u<v<t\leq n$) and all others are singletons, whose corresponding
$K$-graph is shown in Fig.~\ref{fig:rabbit0319}(a). The other is
that, except that $k_s,k_t$ and $k_u$ ($s<t<u$) are partitioned in
the same class, all others are singletons, whose $K$-graph is given
in Fig.~\ref{fig:rabbit0319}(b). In Fig.~\ref{fig:rabbit0319}(a),
the edge variables $p_s,q_s,p_t,q_t$ should be paired, and so do
$p_u,q_u,p_v,q_v$. By similar arguments demonstrated above for the
case of $j=n-1$, we can see the equivalence relations of $p_t=q_s$,
$p_s=q_t$, $p_v=q_u$ and $p_u=q_v$ yield a contribution to
(\ref{eq:mouse0131}) and (\ref{eq:cow0131}) with the highest order
of $N$. On the other hand, in Fig.~\ref{fig:rabbit0319}(b), the six
edge variables $p_s,q_s,p_t,q_t,p_u,q_v$ should be in pairs, and the
equivalence relations of $p_s=q_t$, $p_t=q_u$ and $p_u=q_s$ yield a
highest order of $N$ in contribution.

From the discussions about $n-2\leq j\leq n$, the following rule can
be drawn. Let $\{G_j\}$ denote the set of all $K$-graphs
corresponding to $j$-class noncrossing partitions of ${\cal
K}=\{k_1,k_2,\cdots,k_n\}$. From the noncrossing condition, it is
immediate that, for any particular $\bar{G}_j$ in $\{G_j\}$, we can
always find a member $\bar{G}_{j+1}$ in $\{G_{j+1}\}$ such that
$\bar{G}_j$ is obtained from $\bar{G}_{j+1}$ by merging two vertices
in the same cycle of $\bar{G}_{j+1}$. When the two vertices are
merged, the cycle where these two vertices originally locate is torn
into two. Within each of the two newly formed cycles, to yield a
highest order of $N$ in the contribution to (\ref{eq:mouse0131}) and
(\ref{eq:cow0131}), edge variables touching the same vertex should
be set to equal. This observation leads to the following lemma.

\begin{lemma}\label{lemma:1}
Given non-ascending ordered natural numbers
$b_1,b_2,\cdots,b_{n-j+1}$ such that $b_1+b_2+\cdots+b_{n-j+1}=n$.
Suppose that $x\in{\cal X}_j(b_1,b_2,\cdots,b_{n-j+1})$. Let $G(x)$
denote the corresponding $K$-graph of $x$, and $c_1(x)$ and $c_2(x)$
stand for the contributions of $x$ to (\ref{eq:mouse0131}) and
(\ref{eq:cow0131}), respectively. To yield a highest order of $N$ in
$c_1(x)$ and $c_2(x)$, in every cycle of $G(x)$, edge variables
touching the same vertex should take the same value in $[1,N]$.
Moreover, we have
$$
c_1(x)=N^{-j+1}+O(N^{-j})\qquad \mbox{and}\qquad
c_2(x)=N^{-j+1}\prod_{r=1}^{n-j+1}{\cal W}_\psi^{(b_r)}+O(N^{-j}).
$$
\end{lemma}
\begin{proof}
The first part of the lemma can be proved by mathematical induction
on $j$ using the observation stated in the paragraph preceding this
lemma. To prove the highest-order term in $c_1(x)$ is $N^{-j+1}$, we
note that the expectation of $\prod_{r=1}^n
c_{k_r}^{(q_r)}c_{k_r}^{(p_r)}$ in (\ref{eq:mouse0131}) and
(\ref{eq:cow0131}) is equal to $N^{-n}$ when the equivalence
relations of edge variables are satisfied. We take this $N^{-n}$out
from the infinite sum. The limiting infinite sum of products of
$\delta$ functions in (\ref{eq:mouse0131}) can be decomposed into a
product of smaller limiting infinite sum of products with each
corresponding to a cycle in $G(x)$ (see (\ref{eq:tiger0825_1}) as an
example), and each decomposed limiting sum of products has a
contribution of $N$ to (\ref{eq:mouse0131}). Since the number of
cycles in $G(x)$ is $n-j+1$, the term with the highest order of $N$
in $c_1(x)$ is $N^{-n}\prod_{r=1}^{n-j+1}N=N^{-j+1}$. The proof that
the highest-order term in $c_2(x)$ is
$N^{-j+1}\prod_{r=1}^{n-j+1}{\cal W}_\psi^{(b_r)}$ follows trivially
from that for $c_1(x)$.
\end{proof}

By Lemma~\ref{lemma:1}, each element of ${\cal
X}_j(b_1,b_2,\cdots,b_{n-j+1})$ has the same contribution of
$N^{-j+1}+O(N^{-j})$ to (\ref{eq:mouse0131}). Thus, the total
contribution of ${\cal X}_j(b_1,b_2,\cdots,b_{n-j+1})$ is equal to
\begin{eqnarray*}
&&\#{\cal X}_j(b_1,b_2,\cdots,b_{n-j+1})\cdot
(N^{-j+1}+O(N^{-j}))\nn\\
&=&\frac{n(n-1)\cdots(j+1)}{f(b_1,b_2,\cdots,b_{n-j+1})}\prod_{r=0}^{j-1}
(K-r)\cdot (N^{-j+1}+O(N^{-j})).
\end{eqnarray*}
Hence, by (\ref{eq:cow0120}), the total contribution of noncrossing
partitions of $\cal K$ to $\mu(\bb R_\textrm{cs}^n)$ is
\begin{eqnarray}
&&\mathop{\lim_{K,N\rightarrow\infty}}_{K/N\rightarrow\beta}K^{-1}
\sum_{j=1}^n\hspace{2mm}
\mathop{\sum_{b_1+b_2+\cdots+b_{n-j+1}=n}}_{b_1\geq
b_2\geq\cdots\geq b_{n-j+1}\geq 1}
\frac{n(n-1)\cdots(j+1)}{f(b_1,b_2,\cdots,b_{n-j+1})}\prod_{r=0}^{j-1}
(K-r)\cdot (N^{-j+1}+O(N^{-j}))\nn\\
&=&\dfrac{1}{n}\sum_{j=1}^n {n\choose j}{n\choose
j-1}\beta^{j-1},\label{eq:mouse0831}
\end{eqnarray}
where the equality holds by applying change of variables to
(\ref{eq:mouse0804}). Similarly, the contribution of noncrossing
partitions of $\cal K$ to $\mu(\bb R_\textrm{ca}^n)$ is given by
\begin{eqnarray}
&&\mathop{\lim_{K,N\rightarrow\infty}}_{K/N\rightarrow\beta}K^{-1}
\sum_{j=1}^n\hspace{2mm}
\mathop{\sum_{b_1+b_2+\cdots+b_{n-j+1}=n}}_{b_1\geq
b_2\geq\cdots\geq b_{n-j+1}\geq 1}
\frac{n(n-1)\cdots(j+1)}{f(b_1,b_2,\cdots,b_{n-j+1})}\prod_{r=0}^{j-1}
(K-r)\cdot N^{-j+1}\prod_{s=1}^{n-j+1}{\cal W}_\psi^{(b_s)}\nn\\
&=&\sum_{j=1}^n\beta^{j-1}
\mathop{\sum_{b_1+b_2+\cdots+b_{n-j+1}=n}}_{b_1\geq
b_2\geq\cdots\geq b_{n-j+1}\geq 1}
\frac{n(n-1)\cdots(j+1)}{f(b_1,b_2,\cdots,b_{n-j+1})}\prod_{r=1}^{n-j+1}{\cal
W}_\psi^{(b_r)}.\label{eq:tiger0216}
\end{eqnarray}

\subsection{Contributions of Crossing Partitions to $\mu(\bb R_\textrm{cs}^n)$ and $\mu(\bb R_\textrm{ca}^n)$}

Let $G$ be a $K$-graph resulting from a crossing partition of $\cal
K$ into $j$ classes. The graph $G$ can be decomposed into at most $n-j$ cycles.
For example, Fig.~\ref{fig:7}(c) can be decomposed into at most five cycles. Thus, the contribution of $G$ to
(\ref{eq:mouse0131}) is $O(N^{-n}\cdot N^{n-j})\prod_{r=0}^{j-1}
(K-r)=O(1)$. By (\ref{eq:cow0120}), the contribution of crossing
partitions of $\cal K$ to $\mu(\bb R_\textrm{cs}^n)$ is
\begin{equation}\label{eq:mouse0123}
\mathop{\lim_{K,N\rightarrow\infty}}_{K/N\rightarrow\beta}K^{-1}
\sum_{j=1}^n \#{\cal X}_{cro}(j)\cdot O(1)=0,
\end{equation}
where $\#{\cal X}_{cro}(j)$ denote the number of $j$-class crossing
partitions. By (\ref{eq:mouse0831}) and (\ref{eq:mouse0123}), we
complete the proof of Lemma~\ref{lemma:rabbit0120}. Similarly, it
can be shown that the contribution of crossing partitions of $\cal
K$ to $\mu(\bb R_\textrm{ca}^n)$ vanishes in the large system limit.
Thus, we have shown $\mu(\bb R_\textrm{ca}^n)$ is given by
(\ref{eq:tiger0216}), which completes the proof for
Lemma~\ref{theorem:cow0902}.

\section{Proof of Lemma~\ref{lemma:cow0123}}\label{appendix:dragon0831}

By (\ref{eq:tiger0120}), the $n$-th moment of the ESD of $\bb
R_\textrm{cs}^{(K)}$ when $M\rightarrow\infty$ can be represented by
a normalized trace operator, given as
$$
\overline{\textrm{tr}}((\pmb
R_\textrm{cs}^{(K)})^n)=\lim_{M\rightarrow\infty}
(2M+1)^{-1}K^{-1}\tr((\pmb R_{\textrm{cs}}^{(K)})^n).
$$
Define
$$
v_K=\overline{\textrm{tr}}((\pmb R_\textrm{cs}^{(K)})^n)-\mu(\bb
R_\textrm{cs}^n).
$$
By Borell-Cantelli lemma \cite{feller68}, if
$$
\sum_{K=1}^\infty \textrm{Prob}(|v_K|>\epsilon)<\infty,\qquad
\forall \epsilon>0,
$$
then $v_K\rightarrow 0$ in a.s. sense. Using Markov inequality that
$$
\textrm{Prob}(|v_K|>\epsilon)=\textrm{Prob}(v_K^2>\epsilon^2)\leq
\textrm{E}\{v_K^2\}/\epsilon^{2},\qquad\forall\epsilon>0,
$$
we can show the a.s. convergence of $v_K$ to 0 by proving
$\sum_{K=1}^\infty\textrm{E}\left\{ v_K^2 \right\}<\infty$. In the
following, we will show that
\begin{equation}\label{eq:rabbit0831}
\sum_{K=1}^\infty\textrm{E}\left\{  \left[
\overline{\textrm{tr}}((\pmb R_\textrm{cs}^{(K)})^n)-\mu(\pmb
R_\textrm{cs}^n) \right]^2 \right\}<\infty
\end{equation}
for all $n\in \mathbb{N}$, and the superscript of $\bb
R_\textrm{cs}^{(K)}$ will be omitted for simplicity. In the proof,
for simplicity, we suppose that the spreading sequences are
independent from users to users, and for a particular user, the
sequence is independent across symbols. That is, a long-code system
is assumed. However, by following similar arguments of the proofs
in Appendix~\ref{appendix:dragon0120}, it is straightforward to
extend the proof here to a short-code system.

We have
\begin{eqnarray}
&&\textrm{E}\left\{ \left[ \overline{\textrm{tr}}(\pmb
R_\textrm{cs}^n)-\mu(\pmb R_\textrm{cs}^n) \right]^2 \right\}\nn\\
&=&\textrm{E}\left\{ \left[ \overline{\textrm{tr}}(\pmb
R_\textrm{cs}^n)\right]^2\right\} -[\mu(\pmb R_\textrm{cs}^n)]^2
\nn\\
&=&\lim_{M\rightarrow\infty}(2M+1)^{-2}K^{-2}\sum
Q(m_1,\cdots,m_n,m_{n+1},\cdots,m_{2n};k_1,\cdots,k_n,k_{n+1},\cdots,k_{2n}),
\label{eq:cow0818}
\end{eqnarray}
where
\begin{eqnarray}
&&Q(m_1,\cdots,m_n,m_{n+1},\cdots,m_{2n};k_1,\cdots,k_n,k_{n+1},\cdots,k_{2n})\\
&=&\textrm{E}\{ [\pmb R_\textrm{cs}]_{m_1 m_2,k_1 k_2} [\pmb
R_\textrm{cs}]_{m_2 m_3,k_2 k_3} \cdots [\pmb
R_\textrm{cs}]_{m_{n}m_1, k_{n} k_1}\nn\\
&&\times [\pmb R_\textrm{cs}]_{m_{n+1} m_{n+2},k_{n+1} k_{n+2}}
[\pmb R_\textrm{cs}]_{m_{n+2} m_{n+3},k_{n+2} k_{n+3}} \cdots [\pmb
R_\textrm{cs}]_{m_{2n}m_{n+1},
k_{2n} k_{n+1}}\}\nn\\
&&-\textrm{E}\{ [\pmb R_\textrm{cs}]_{m_1 m_2,k_1 k_2} [\pmb
R_\textrm{cs}]_{m_2 m_3,k_2 k_3} \cdots [\pmb
R_\textrm{cs}]_{m_{n}m_1, k_{n} k_1} \}\nn\\
&&\times\textrm{E}\{[\pmb R_\textrm{cs}]_{m_{n+1} m_{n+2},k_{n+1}
k_{n+2}} [\pmb R_\textrm{cs}]_{m_{n+2} m_{n+3},k_{n+2} k_{n+3}}
\cdots [\pmb R_\textrm{cs}]_{m_{2n}m_{n+1}, k_{2n}
k_{n+1}}\},\label{eq:mouse0818}
\end{eqnarray}
and the summation is over all $-M\leq m_1,\cdots,m_{2n}\leq M$ and
$1\leq k_1,\cdots,k_{2n}\leq K$. Moreover, we have
$|m_j-m_{j+1}|\leq 1$ for $\{m_j\}_{j=1}^n$, and so do
$\{m_j\}_{j=n+1}^{2n}$.

We consider two $n$-element noncrossing partitions. One is for
$\{k_1,\cdots,k_n\}$, and the other is for
$\{k_{n+1},\cdots,k_{2n}\}$. Suppose that there are $j$ and $l$
classes in the former and latter partitions, respectively. Assume
$j$ classes of noncrossing partitions of $\{k_1,\cdots,k_n\}$ take
distinct values $\{u_1,\cdots,u_j\}$ in $[1,K]$, and they have sizes
$(a_1,\cdots,a_j)$, respectively. Similarly,
$\{k_{n+1},\cdots,k_{2n}\}$ take values $\{v_1,\cdots,v_l\}$, which
have sizes $(b_1,\cdots,b_l)$, respectively.

First, consider the case that $\{u_1,\cdots,u_j\}$ and
$\{v_1,\cdots,v_l\}$ have no common element, i.e., all of
$u_1,\cdots,u_j,v_1,\cdots,v_l$ are distinct. Due to independence of
spreading codes $c_k^{(p)}$'s, the first term of
(\ref{eq:mouse0818}) (expectation of a product of $2n$ elements) is
equal to the second term. Thus, $Q(m_1,\cdots,m_n,$
$m_{n+1},\cdots,m_{2n};k_1,\cdots,k_n,k_{n+1},\cdots,k_{2n})$ is
equal to zero, and (\ref{eq:rabbit0831}) follows trivially.

Secondly, consider the situation that $\{u_1,\cdots,u_j\}$ and
$\{v_1,\cdots,v_l\}$ have only one element in common. Without loss
of generality, we let $u_1=v_1=w$. In this case,
(\ref{eq:mouse0818}) is equal to
\begin{eqnarray}
&&\sum\left\{ \textrm{E}\left(\prod_{\alpha(1)=1}^{a_1}
c_{w}^{(p_{1,\alpha(1)})}c_{w}^{(q_{1,\alpha(1)})}\prod_{\gamma(1)=1}^{b_1}
c_{w}^{(r_{1,\gamma(1)})}c_{w}^{(s_{1,\gamma(1)})}\right)\right.\label{eq:mouse0821}\\
&&\hspace{1cm}- \left.\textrm{E}\left(\prod_{\alpha(1)=1}^{a_1}
c_{w}^{(p_{1,\alpha(1)})}c_{w}^{(q_{1,\alpha(1)})}\right)
\textrm{E}\left(\prod_{\gamma(1)=1}^{b_1}
c_{w}^{(r_{1,\gamma(1)})}c_{w}^{(s_{1,\gamma(1)})}\right) \right\}\nn\\
&&\times\prod_{i=2}^j \textrm{E}\left(\prod_{\alpha(i)=1}^{a_i}
c_{u_i}^{(p_{i,\alpha(i)})}c_{u_i}^{(q_{i,\alpha(i)})}\right)\cdot
\prod_{i=2}^l \textrm{E}\left(\prod_{\gamma(i)=1}^{b_i}
c_{v_i}^{(r_{i,\gamma(i)})}c_{v_i}^{(s_{i,\gamma(i)})}\right)\times
\mbox{product of $\delta$ functions}, \nn
\end{eqnarray}
where, for given $i$ and $\alpha(i)$, $p_{i,\alpha(i)}$ and
$q_{i,\alpha(i)}$ are edge variables touching the same vertex within
a cycle of the $K$-graph, and so do $r_{i,\gamma(i)}$ and
$s_{i,\gamma(i)}$ with given $i$ and $\gamma(i)$. The summation in
(\ref{eq:mouse0821}) is over all $p_{i,\alpha(i)}$,
$q_{i,\alpha(i)}$, $i=1,\cdots,j$, $\alpha(i)=1,\cdots,a_i$ and
$r_{i,\gamma(i)}$, $s_{i,\gamma(i)}$, $i=1,\cdots,l$,
$\gamma(i)=1,\cdots,b_i$.
Equation (\ref{eq:mouse0821}) is nonzero if and only if all the
following conditions are met:

\begin{itemize}
\item[1)] for each $2\leq i\leq j$, elements in $\{p_{i,1},\cdots,p_{i,a_i},
q_{i,1},\cdots,q_{i,a_i}\}$ are in pairs,

\item[2)] for each $2\leq i\leq l$, elements in $\{r_{i,1},\cdots,r_{i,b_i},
s_{i,1},\cdots,s_{i,b_i}\}$ are in pairs,

\item[3)] elements of $\{p_{1,1},\cdots,p_{1,a_1},q_{1,1},\cdots,q_{1,a_1},r_{1,1},
\cdots,r_{1,b_1},s_{1,1},\cdots,s_{1,b_1}\}$ are in pair, and some
of $p_{1,1},$ $\cdots,p_{1,a_1},$ $q_{1,1},\cdots,q_{1,a_1}$ are in
pair with elements of $r_{1,1},\cdots,r_{1,b_1},
s_{1,1},\cdots,s_{1,b_1}$,
\end{itemize}
where we say members of a set are in pairs, if each element of the
set can find odd number of other elements that take the same value.
To have the largest cardinality of summation variables
$m_1,\cdots,m_{2n}$, $p_{i,\alpha(i)}$'s, $q_{i,\alpha(i)}$'s,
$r_{i,\gamma(i)}$'s and $s_{i,\gamma(i)}$'s in (\ref{eq:mouse0821}),
the pairing constraints listed in conditions 1)--3) above should be
as least as possible, which yields
\begin{itemize}
\item[4)] for each $2\leq i\leq j$, $\{p_{i,1},\cdots,p_{i,a_i},
q_{i,1},\cdots,q_{i,a_i}\}$ is paired by, for each $1\leq
\alpha(i)\leq a_i$, $p_{i,\alpha(i)}$ is only paired with
$q_{i,\alpha(i)}$,

\item[5)] for each $2\leq i\leq l$, $\{r_{i,1},\cdots,r_{i,b_i},
s_{i,1},\cdots,s_{i,b_i}\}$ is paired by, for each $1\leq
\gamma(i)\leq b_i$, $r_{i,\gamma(i)}$ is only paired with
$s_{i,\gamma(i)}$,

\item[6)] there exists a unique $(\theta,\nu)\in[1,a_1]\times[1,b_1]$, denoted by $(\theta_0,\nu_0)$,
such that $p_{1,\theta_0}$ is paired with $r_{1,\nu_0}$, and
$q_{1,\theta_0}$ is paired with $s_{1,\nu_0}$ (or $p_{1,\theta_0}$
paired with $s_{1,\nu_0}$ and $q_{1,\theta_0}$ paired with
$r_{1,\nu_0}$),

\item[7)] for $\{p_{1,1},\cdots,p_{1,a_1},
q_{1,1},\cdots,q_{1,a_1}\}$, $p_{1,\alpha(1)}$ is only paired with
$q_{1,\alpha(1)}$ when $\alpha(1)\neq\theta_0$,

\item[8)] for $\{r_{1,1},\cdots,r_{1,b_1},
s_{1,1},\cdots,s_{1,b_1}\}$, $r_{1,\gamma(1)}$ is only paired with
$s_{1,\gamma(1)}$ when $\gamma(1)\neq\nu_0$.
\end{itemize}
Under these circumstances, the summand of (\ref{eq:mouse0821})
without the product of $\delta$ functions is $O(N^{-2n})$. On the
other hand, for each $m_1$ and $m_{n+1}$, the product of $\delta$
functions in (\ref{eq:mouse0821}) summed over dummy variables
$m_1,\cdots,m_{n},$ $m_{n+1},\cdots,m_{2n}$, $p_{i,\alpha(i)}$'s,
$q_{i,\alpha(i)}$'s, $r_{i,\gamma(i)}$'s and $s_{i,\gamma(i)}$'s is
$$
O(N^{n-j+1}\cdot N^{n-l+1}\cdot N^{-1})=O(N^{2n-j-l+1}),
$$
where $N^{n-j+1}$ and $N^{n-l+1}$ are because $\{k_1,\cdots,k_n\}$
and $\{k_{n+1},\cdots,k_{2n}\}$ form $j$- and $l$-class noncrossing
partitions, respectively, and $N^{-1}$ is because conditions 6)--8)
causes the cardinality reduced by one\footnote{This is the same as
conditions i) and ii) of Appendix~\ref{sub_appendix:mouse0124},
where we compute the contribution of an element of ${\cal
X}_{n-1}(e,n-e)$ to (\ref{eq:mouse0131}). In these two conditions,
edge variables touching the same vertex in a cycle of a $K$-graph do
not always take the same value.}. Thus, (\ref{eq:cow0818}) is qual
to
\begin{equation}
K^{-2}\cdot O(N^{-2n})\cdot O(K^{j+l-1}) \cdot
O(N^{2n-j-l+1})=O(K^{-2}),\label{eq:tiger0821}
\end{equation}
where $O(K^{j+l-1})$ is because $u_1,\cdots,u_j,v_1,\cdots,v_l$ are
all distinct except for $u_1=v_1$. Consider the infinite sum over
$K$ of (\ref{eq:rabbit0831}). It is finite when (\ref{eq:tiger0821})
is summed over $K$ from $1$ to $\infty$.

Next, we consider the situation that $\{u_1,\cdots,u_j\}$ and
$\{v_1,\cdots,v_l\}$ have $t$ elements in common, where $2\leq
t\leq\min(j,l)$. Without loss of generality, we assume
$u_i=v_i=w_i$, $1\leq i\leq t$. In this case, (\ref{eq:mouse0818})
is equal to
\begin{eqnarray}
&&\sum\left\{\prod_{i=1}^t \textrm{E}\left(\prod_{\alpha(i)=1}^{a_i}
c_{w_i}^{(p_{i,\alpha(i)})}c_{w_i}^{(q_{i,\alpha(i)})}\prod_{\gamma(i)=1}^{b_i}
c_{w_i}^{(r_{i,\gamma(i)})}c_{w_i}^{(s_{i,\gamma(i)})}\right)\right.\label{eq:mouse0901}\\
&& \left.\hspace{1cm}- \prod_{i=1}^t
\textrm{E}\left(\prod_{\alpha(i)=1}^{a_i}
c_{w_i}^{(p_{i,\alpha(i)})}c_{w_i}^{(q_{i,\alpha(i)})}\right)
\textrm{E}\left(\prod_{\gamma(i)=1}^{b_i}
c_{w_i}^{(r_{i,\gamma(i)})}c_{w_i}^{(s_{i,\gamma(i)})}\right) \right\}\nn\\
&&\times\prod_{i=t+1}^j \textrm{E}\left(\prod_{\alpha(i)=1}^{a_i}
c_{u_i}^{(p_{i,\alpha(i)})}c_{u_i}^{(q_{i,\alpha(i)})}\right)\cdot
\prod_{i=t+1}^l \textrm{E}\left(\prod_{\gamma(i)=1}^{b_i}
c_{v_i}^{(r_{i,\gamma(i)})}c_{v_i}^{(s_{i,\gamma(i)})}\right)\times\mbox{product
of $\delta$ functions}.\nn
\end{eqnarray}
The pairing constraints listed in 4)--8) is one of the conditions
that yield (\ref{eq:mouse0901}) nonzero and have the largest
cardinality of summation variables. Thus, the contribution of the
sum of products of $\delta$ functions is the same as that of when
$\{u_1,\cdots,u_j\}$ and $\{v_1,\cdots,v_l\}$ have only one common
element. That is, it is equal to $O(N^{2n-j-l+1})$. It is not
difficult to see, when $\{u_1,\cdots,u_j\}$ and $\{v_1,\cdots,v_l\}$
have $t$ common elements, the contribution to (\ref{eq:cow0818}) is
\begin{equation}
K^{-2}\cdot O(N^{-2n})\cdot O(K^{j+l-t}) \cdot
O(N^{2n-j-l+1})=O(K^{-t-1}).\label{eq:rabbit0821}
\end{equation}
When (\ref{eq:rabbit0821}) is summed over $K$ from 1 to $\infty$, it
is finite.

Thus, we have proved
$$
\sum_{K=1}^\infty\textrm{E}\left\{  \left[
\overline{\textrm{tr}}(\pmb R_\textrm{cs}^n)-\mu(\pmb
R_\textrm{cs}^n) \right]^2 \right\}<\infty
$$
for all natural numbers $n$, meaning that
$\overline{\textrm{tr}}(\pmb R_\textrm{cs}^n)$ converges a.s. to
$\mu(\pmb R_\textrm{cs}^n)$ when $K\rightarrow\infty$. Moreover, it
is seen $\mu(\bb R_\textrm{cs}^n)$ given in (\ref{eq:dragon0124}) is
equal to the $n$-th moment of the Mar\u{c}enko-Pastur distribution.
It has been proved in \cite{jonsson82} that the moment sequence of
the distribution satisfies the Carleman's criterion.

\section{Proof of Lemma~\ref{corollary:cow0807}}\label{appendix:cow0128}

Let $\pmb P=\pmb A\pmb A^\dag$. Matrix $\pmb P$ has the same
structure as $\pmb A$. The $k$-th diagonal entry in the $m$-th block
diagonal component of $\pmb P$ is $P_k(m)=|A_k(m)|^2$. We have
$$
\tr\{(\pmb A^\dag \pmb R_\textrm{cs}\pmb A)^n\} = \tr\{(\pmb
R_\textrm{cs}\pmb P)^n\},
$$
and
\begin{eqnarray}\label{eq:rabbit0728}
&&\mathop{\lim_{K,N,M\rightarrow\infty}}_{K/N\rightarrow\beta}(2M+1)^{-1}K^{-1}\textrm{E}\{
\tr((\pmb
R_\textrm{cs}\pmb P)^n)\}\nn\\
&=&\mathop{\lim_{K,N,M\rightarrow\infty}}_{K/N\rightarrow\beta}(2M+1)^{-1}K^{-1}
\sum_{j=1}^n\mathop{\sum_{b_1+b_2+\cdots+b_{n-j+1}=n}}_{b_1\geq
b_2\geq\cdots b_{n-j+1}\geq 1}\sum_{{\cal K}\in {\cal
X}_j(b_1,b_2,\cdots,b_{n-j+1})} \sum_{{\cal M}\in {\cal Y}}
\sum_{{\cal P}_1\in {\cal Z}_1}\cdots \sum_{{\cal P}_{n}\in{\cal
Z}_{n}}\nn\\
&& \textrm{E}\left\{\left(c_{k_1}^{(p_1)} c_{k_2}^{(q_2)}\right)
\left(c_{k_2}^{(p_2)} c_{k_3}^{(q_3)}\right) \cdots
\left(c_{k_{n}}^{(p_{n})} c_{k_1}^{(q_1)}\right)
\right\}\textrm{E}\{P_{k_1}(m_1)P_{k_2}(m_2)\cdots
P_{k_{n}}(m_{n})\}\nn\\
&&\times\delta(p_1 T_c+\tau_{k_1},q_2 T_c+\tau_{k_2}) \delta(p_2
T_c+\tau_{k_2},q_3 T_c+\tau_{k_3})\cdots \delta(p_{n}
T_c+\tau_{k_{n}},q_1 T_c+\tau_{k_1}),
\end{eqnarray}
with $|m_j-m_{j+1}|\leq 1$. Consider the contribution of an element
$x$ in ${\cal X}_j(b_1,b_2,\cdots,b_{n-j+1})$, which has a $K$-graph
composed of $n-j+1$ concatenated cycles with the number of edges
$b_1,b_2,\cdots,b_{n-j+1}$ in a non-ascending order. Suppose that
this $K$-graph is yielded by a $j$-class noncrossing partition of
$\cal K$ with non-ascending class sizes $(c_1,c_2,\cdots,c_j)$. As
shown in (\ref{eq:cow0804}), the number of elements in ${\cal
X}_j(b_1,b_2,\cdots,b_{n-j+1})$ satisfying these conditions is equal
to
$$
\dfrac{n(n-j)!(j-1)!}
{f(b_1,b_2,\cdots,b_{n-j+1})f(c_1,c_2,\cdots,c_j)}.
$$
For $x$, the limit of the corresponding
$\textrm{E}\{P_{k_1}(m_1)P_{k_2}(m_2)\cdots P_{k_{n}}(m_{n})\}$ in
(\ref{eq:rabbit0728}) becomes $\prod_{r=1}^j {\cal P}^{(c_r)}$,
where note that if $k_s$ and $k_t$ of $\cal K$ are partitioned in
the same class, $m_s$ is equal to $m_t$.\footnote{An illustration
for the note is in case iii) of
Appendix~\ref{sub_appendix:mouse0124}, where we compute the
contribution of an element of ${\cal X}_{n-1}(e,n-e)$ to
(\ref{eq:mouse0131}).} Thus, for either short-code or long-code
system, (\ref{eq:rabbit0728}) can be expressed as
\begin{eqnarray}
&&
\mathop{\lim_{K,N,M\rightarrow\infty}}_{K/N\rightarrow\beta}K^{-1}
\sum_{j=1}^n\mathop{\sum_{b_1+b_2+\cdots+b_{n-j+1}=n}}_{b_1\geq
b_2\geq\cdots b_{n-j+1}\geq
1}\mathop{\sum_{c_1+c_2+\cdots+c_j=n}}_{c_1\geq c_2\geq\cdots
c_j\geq 1} \dfrac{n(n-j)!(j-1)!}
{f(b_1,b_2,\cdots,b_{n-j+1})f(c_1,c_2,\cdots,c_j)}
\nn\\
&&\times  \prod_{s=0}^{j-1}(K-s)\cdot N^{-j+1}\cdot \prod_{r=1}^j
{\cal
P}^{(c_r)},\label{eq:mouse0807}\\
&&=\sum_{j=1}^n
\beta^{j-1}\mathop{\sum_{c_1+c_2+\cdots+c_j=n}}_{c_1\geq
c_2\geq\cdots\geq c_j\geq
1}\dfrac{n(n-1)\cdots(n-j+2)}{f(c_1,c_2,\cdots,c_j)}\prod_{r=1}^j
{\cal P}^{(c_r)},
\end{eqnarray}
where we make use of the equality
\begin{equation}\label{eq:mouse0218}
\mathop{\sum_{b_1+b_2+\cdots+b_{n-j+1}=n}}_{b_1\geq b_2\geq\cdots
b_{n-j+1}\geq 1}\dfrac{n(n-j)!(j-1)!} {f(b_1,b_2,\cdots,b_{n-j+1})}
=n(n-1)\cdots(n-j+2).
\end{equation}

\section{Proof of
Lemma~\ref{proposition:mouse0515}}\label{appendix:mouse0826}

We prove parts 1) and 2) of this lemma in
Appendices~\ref{subsection:tiger0128} and
\ref{subsection:rabbit0128}, respectively.
\subsection{Bandwidth of $\psi(t)$ Less Than $1/(2T_c)$}\label{subsection:tiger0128}

Define $\theta_1=n_0 T_c+\eta_0-\eta_1$, $\theta_2=n_2
T_c+\eta_2-\eta_1$, $\tilde{R}_{\psi,1}(x)=R_\psi(-x+\theta_1)$, and
$\tilde{R}_{\psi,2}(x)=R_\psi(x-\theta_2)$. Then
$R_\psi((n_0-n_1)T_c+\eta_0-\eta_1)$ is the sample of
$\tilde{R}_{\psi,1}(x)$ at $x=n_1 T_c$. The discrete-time signal
$\tilde{R}_{\psi,1}(n)$ obtained from the continuous-time
$\tilde{R}_{\psi,1}(x)$ by a sampling period of $T_c$ is denoted by
the same notation, but the argument is an integer indicating the
sample index.

By Parseval's theorem, a partial sum w.r.t. $n_1$ in
(\ref{eq:tiger0826}) can be given as
\begin{eqnarray}
&&\sum_{n_1=-\infty}^\infty
R_\psi((n_0-n_1)T_c+\eta_0-\eta_1)R_\psi((n_1-n_2)T_c+\eta_1-\eta_2)\label{eq:tiger0804}\\
&=& \sum_{n_1=-\infty}^\infty \tilde{R}_{\psi,1}(n_1)
\tilde{R}_{\psi,2}(n_1)\nn\\
 &=&\frac{1}{2\pi}\int_{-\pi}^\pi
\textrm{DTFT}\{\tilde{R}_{\psi,1}(n_1)\}
\textrm{DFTF}^*\{\tilde{R}_{\psi,2}(n_1)\}\textrm{d}\omega,\label{eq:mouse0512}
\end{eqnarray}
where $\textrm{DTFT}\{\cdot\}$ is the operator of discrete-time
Fourier transform (DTFT) with
$$
\textrm{DTFT}\{x(n)\}=\sum_{n=-\infty}^\infty x(n) e^{-j\omega n}.
$$
As $\tilde{R}_{\psi,1}(n_1)$ is the sample of $R_\psi(-x+\theta_1)$
at time $x=n_1 T_c$, and the Fourier transform of
$R_\psi(x)=\psi(x)\ast\psi(-x)$ is $|\Psi(\Omega)|^2$, we have
$$
\textrm{DFTF}\{\tilde{R}_{\psi,1}(n_1)\}=\dfrac{1}{T_c}
\sum_{k=-\infty}^\infty e^{-j\frac{\omega-2\pi k}{T_c}\theta_1}
\left|\Psi\left(\dfrac{\omega-2\pi k}{T_c}\right)\right|^2,
$$
where $\omega=\Omega T_c$. Consequently, (\ref{eq:mouse0512}) is
equal to
\begin{equation}\label{eq:monkey0804}
\frac{1}{2\pi T_c^2}\sum_{k,l=-\infty}^\infty \int_{-\pi}^\pi
e^{-j\frac{\omega-2\pi k}{T_c}\theta_1+j\frac{\omega-2\pi
l}{T_c}\theta_2} \left|\Psi\left(\frac{\omega-2\pi
k}{T_c}\right)\right|^2 \left|\Psi\left(\frac{\omega-2\pi
l}{T_c}\right)\right|^2\textrm{d}\omega,
\end{equation}
where, since $\Psi(\Omega)$ is bandlimited to $\pi/T_c$, only
$k=l=0$ has nonzero integral. Thus, (\ref{eq:tiger0804}) is equal to
\begin{equation}\label{eq:cow0512}
\frac{1}{2\pi T_c^2}\int_{-\pi}^\pi
e^{-j\frac{\omega}{T_c}((n_0-n_2)T_c+\eta_0-\eta_2)}
\left|\Psi\left(\frac{\omega}{T_c}\right)\right|^4\textrm{d}\omega.
\end{equation}

We consider the summation w.r.t. $n_2$ in (\ref{eq:tiger0826}).
Define $\theta_3=n_3 T_c+\eta_3-\eta_2$ and
$\tilde{R}_{\psi,3}(x)=R_\psi(x-\theta_3)$. We have
\begin{eqnarray}
&&\sum_{n_2=-\infty}^\infty \left( \frac{1}{2\pi
T_c^2}\int_{-\pi}^\pi
e^{-j\frac{\omega}{T_c}((n_0-n_2)T_c+\eta_0-\eta_2)}
\left|\Psi\left(\frac{\omega}{T_c}\right)\right|^4\textrm{d}\omega
\right)
R_\psi((n_2-n_3)T_c+\eta_2-\eta_3)\nn\\
&=& \frac{1}{2\pi T_c^2}\int_{-\pi}^\pi e^{-j\frac{\omega}{T_c}(n_0
T_c+\eta_0-\eta_2)}
\left|\Psi\left(\frac{\omega}{T_c}\right)\right|^4 \left(
\sum_{n_2=-\infty}^\infty e^{j\omega n_2} \tilde{R}_{\psi,3}(n_2)
\right) \textrm{d}\omega,\label{eq:rabbit0804}
\end{eqnarray}
where the summation inside the brackets of (\ref{eq:rabbit0804}) is
the complex conjugate of the DTFT of $\tilde{R}_{\psi,3}(n_2)$,
given by
\begin{equation}\label{eq:dragon0804}
\frac{1}{T_c}\sum_{k=-\infty}^\infty e^{j\frac{\omega-2\pi
k}{T_c}\theta_3}\left|\Psi\left(\frac{\omega-2\pi
k}{T_c}\right)\right|^2.
\end{equation}
Plugging (\ref{eq:dragon0804}) back to (\ref{eq:rabbit0804}), we can
see that the integral is nonzero only when $k=0$, which results in
\begin{equation}\label{eq:snake0804}
\frac{1}{2\pi T_c^3} \int_{-\pi}^\pi
e^{-j\frac{\omega}{T_c}((n_0-n_3)T_c+\eta_0-\eta_3)}
\left|\Psi\left(\frac{\omega}{T_c}\right)\right|^6 \textrm{d}\omega.
\end{equation}
In consequence, when the summations w.r.t. $n_1$ and $n_2$ are taken
into account, the result is given in (\ref{eq:snake0804}).
Continuing this process, we obtain the final result as
$$
\frac{1}{2\pi T_c^m}\int_{-\pi}^\pi
\left|\Psi\left(\frac{\omega}{T_c}\right)\right|^{2m}\textrm{d}\omega,
$$
which is equal to (\ref{eq:horse0804}) by setting
$\Omega=\omega/T_c$.

\subsection{Bandwidth of $\psi(t)$ Greater Than $1/(2T_c)$}\label{subsection:rabbit0128}

Suppose that the bandwidth of $\psi(t)$ is greater than
$\alpha/(2T_c)$ but less than $(\alpha+1)/(2T_c)$ for $\alpha\in
\mathbb{N}$. Using the equality
$\textrm{E}_{u,v}\{g(u,v)\}=\textrm{E}_v\{\textrm{E}_u\{g(u,v)|v\}\}$
with $\textrm{E}_{u}\{\cdot|\cdot\}$ denoting the conditional
expectation w.r.t. $u$, we can see that
\begin{equation}\label{eq:dog0804}
\sum_{n_1=-\infty}^\infty
\textrm{E}_{\eta_1}\{R_\psi((n_0-n_1)T_c+\eta_0-\eta_1)
R_\psi((n_1-n_2)T_c+\eta_1-\eta_2)|\eta_2\}
\end{equation}
is nested in the multi-dimensional summation of
(\ref{eq:chicken0804}), where note that $\eta_0$ is deterministic.
By employing the same procedures of getting (\ref{eq:monkey0804}),
it is immediate (\ref{eq:dog0804}) becomes
\begin{equation}
\frac{1}{2\pi T_c^2}\sum_{k,l=-\infty}^\infty \int_{-\pi}^\pi
\textrm{E}_{\eta_1}\left\{\left.e^{-j\frac{\omega-2\pi
k}{T_c}\theta_1+j\frac{\omega-2\pi
l}{T_c}\theta_2}\right|\eta_2\right\}
\left|\Psi\left(\frac{\omega-2\pi k}{T_c}\right)\right|^2
\left|\Psi\left(\frac{\omega-2\pi
l}{T_c}\right)\right|^2\textrm{d}\omega,\nn
\end{equation}
whose imaginary part is definitely zero. Since
$\textrm{E}_{\eta_1}\left\{ \cos(2\pi k\eta_1/T_c)\right\}=0$ for
any nonzero integer $k$, it is readily seen that we only need to
consider $k=l$ in the above equation, which is given by
\begin{equation}
\frac{1}{2\pi
T_c^2}\sum_{k=-\lceil\alpha/2\rceil}^{\lceil\alpha/2\rceil}
\int_{-\pi}^\pi e^{-j\frac{\omega-2\pi
k}{T_c}((n_0-n_2)T_c+\eta_0-\eta_2)}
\left|\Psi\left(\frac{\omega-2\pi
k}{T_c}\right)\right|^4\textrm{d}\omega.
\end{equation}

Consider one more layer of summations in (\ref{eq:chicken0804}),
i.e., w.r.t. $n_2$, which yields
\begin{equation}\label{eq:dragon0512}
\frac{1}{2\pi
T_c^2}\sum_{k=-\lceil\alpha/2\rceil}^{\lceil\alpha/2\rceil}
\int_{-\pi}^\pi \left|\Psi\left(\frac{\omega-2\pi
k}{T_c}\right)\right|^4 \left( \sum_{n_2=-\infty}^\infty
\textrm{E}_{\eta_2}\left\{\left. e^{-j\frac{\omega-2\pi
k}{T_c}((n_0-n_2)T_c+\eta_0-\eta_2)}\tilde{R}_{\psi,3}(n_2)
\right|\eta_3\right\} \right) \textrm{d}\omega,
\end{equation}
where the term inside the brackets can be written as
\begin{eqnarray}
&&\textrm{E}_{\eta_2}\left\{ \left. e^{-j\frac{\omega-2\pi
k}{T_c}(n_0 T_c+\eta_0-\eta_2)} \sum_{n_2=-\infty}^\infty e^{j\omega
n_2} \tilde{R}_{\psi,3}(n_2) \right|\eta_3\right\}\nn\\
&=&\frac{1}{T_c}\sum_{l=-\infty}^\infty \textrm{E}_{\eta_2}\left\{
\left. e^{-j\frac{\omega-2\pi k}{T_c}(n_0 T_c+\eta_0-\eta_2)}
e^{j\frac{\omega-2\pi l}{T_c}(n_3 T_c+\eta_3-\eta_2)}\right|\eta_3
\right\} \left|\Psi\left( \frac{\omega-2\pi l}{T_c}
\right)\right|^2\label{eq:mouse0806}
\end{eqnarray}
with the expectation in (\ref{eq:mouse0806}) being nonzero only when
$l=k$. Thus, the summations w.r.t. $n_1$ and $n_2$ of
(\ref{eq:chicken0804}), i.e., (\ref{eq:dragon0512}), become
$$
\frac{1}{2\pi
T_c^3}\sum_{k=-\lceil\alpha/2\rceil}^{\lceil\alpha/2\rceil}\int_{-\pi}^\pi
e^{-j\frac{\omega-2\pi k}{T_c}((n_0-n_3)T_c+\eta_0-\eta_3)}
\left|\Psi\left(\frac{\omega-2\pi k}{T_c}\right)\right|^6
\textrm{d}\omega.
$$
Continuing this process, we obtain the final result as
$$
\frac{1}{2\pi
T_c^m}\sum_{k=-\lceil\alpha/2\rceil}^{\lceil\alpha/2\rceil}\int_{-\pi}^\pi
\left|\Psi\left(\frac{\omega-2\pi k}{T_c}\right)\right|^{2m}
\textrm{d}\omega=\frac{1}{2\pi
T_c^m}\int_{-(2\lceil\alpha/2\rceil+1)\pi}^{(2\lceil\alpha/2\rceil+1)\pi}
\left|\Psi\left(\frac{\omega}{T_c}\right)\right|^{2m}
\textrm{d}\omega,
$$
which is equal to (\ref{eq:sheep0804}) by changing variable from
$\omega$ to $\Omega=\omega/T_c$.

\section{Proof of Theorem~\ref{theorem:mouse0810}}\label{appendix:rabbit0826}

The following two lemmas are helpful to prove
Theorem~\ref{theorem:mouse0810}.

\begin{lemma}\label{lemma:mouse0605}
Suppose that $\varpi$ is a noncrossing partition of a finite totally
ordered set $S$, where every class in $\varpi$ has at least two
elements. Then, there exist some classes in $\varpi$ that contain
adjacent elements of $S$, where the adjacency is cyclic ordering,
i.e., the first and last elements of $S$ are adjacent.
\end{lemma}

\begin{proof}
Denote the $K$-graph corresponding to $\varpi$ by $G(\varpi)$. From
the properties of a $K$-graph given in
Appendix~\ref{appendix:dragon0120}, there is a bijective
correspondence between the class set of $\varpi$ and the vertex set
of $G(\varpi)$. We have the following three observations. First, the
size of the $r$-th class of $\varpi$ is equal to $d(v_r)/2$, where
$v_r$ is the vertex in $G(\varpi)$ that corresponds to the $r$-th
class, and $d(v_r)$ is the degree\footnote{The degree of a vertex is
the number of edges that connect to that vertex. The singly vertex
in a self-loop has the degree equal to two.} of $v_r$. Secondly,
whenever two adjacent elements of $S$ locate in the same class of
$\varpi$, there is a self-loop in $G(\varpi)$. Thirdly, for all
noncrossing partitions $\varpi$ of $S$, we cannot find any
$G(\varpi)$ that contains no self-loops and whose every vertex has
the degree equal to or greater than four. Based on the first two
observations, the third one can be interpreted as the statement of
the lemma. Thus, we have completed the proof.
\end{proof}

\begin{lemma}\label{lemma:mouse0811}
Let $\bb D$ be the diagonal random matrix described in
Theorem~\ref{theorem:mouse0810}. We have, for
$\textrm{x}\in\{\textrm{cs,ca}\}$,
\begin{eqnarray}
&&\mathop{\lim_{K,N,M\rightarrow\infty}}_{K/N\rightarrow\beta}
(2M+1)^{-1}K^{-1} \mathop{\sum_{{\cal K}\in{\cal X}}}_{k_t=k_{t+1}}
\sum_{{\cal M}\in {\cal Y}}\textrm{E}\{ [\pmb
R_\textrm{x}^{r_1}]_{m_1 m_2,k_1 k_2} [\pmb D^{s_1}]_{m_2 m_2,k_2
k_2}\nn\\
&&\hspace{1cm}[\pmb R_\textrm{x}^{r_2}]_{m_2 m_3,k_2 k_3} [\pmb
D^{s_2}]_{m_3 m_3,k_3 k_3}\cdots [\pmb
R_\textrm{x}^{r_n}]_{m_{n}m_1, k_{n} k_1}[\pmb D^{s_n}]_{m_1 m_1,k_1
k_1}\},\label{eq:mouse0326}\\
&=&\mu(\pmb R_\textrm{x}^{r_t})
\mathop{\lim_{K,N,M\rightarrow\infty}}_{K/N\rightarrow\beta}
(2M+1)^{-1}K^{-1} \sum_{{\cal K}\setminus\{k_{t+1}\}} \sum_{{\cal
M}\setminus\{m_{t+1}\}}\nn\\
&&\hspace{.5cm}\textrm{E}\{ (d_{k_2}(m_2))^{s_1}\cdots
(d_{k_{t-1}}(m_{t-1}))^{s_{t-2}}(d_{k_t}(m_t))^{s_{t-1}+s_t}
(d_{k_{t+2}}(m_{t+2}))^{s_{t+1}}\cdots
(d_{k_1}(m_1))^{s_n}\}\nn\\
&&\hspace{.5cm}\textrm{E}\{ [\pmb R_\textrm{x}^{r_1}]_{m_1 m_2,k_1
k_2} \cdots[\pmb R_\textrm{x}^{r_{t-1}}]_{m_{t-1} m_t,k_{t-1} k_t}
[\pmb R_\textrm{x}^{r_{t+1}}]_{m_t m_{t+2}, k_t k_{t+2}}\cdots [\pmb
R_\textrm{x}^{r_n}]_{m_n m_1, k_n k_1}\},\label{eq:mouse0827}
\end{eqnarray}
where $r_i$'s and $s_i$'s are non-negative integers, notations $\cal
K$, $\cal X$, $\cal M$ and $\cal Y$ have the same definitions as in
(\ref{eq:mouse0131}) and (\ref{eq:cow0131}). When $\bb D$ is set as
the identity matrix and the constraint $k_{t}=k_{t+1}$ in
(\ref{eq:mouse0326}) is replaced with $k_{t}=k_u$ for any $t<u\leq
n$, (\ref{eq:mouse0326}) is equal to $\mu(\bb
R_\textrm{x}^{r_t+\cdots+r_{u-1}})\mu(\bb
R_\textrm{x}^{r_1+\cdots+r_{t-1}+r_u+\cdots+r_n})$.
\end{lemma}

\begin{proof}
We expand the multi-sum of matrix product in (\ref{eq:mouse0326}) as
\begin{eqnarray}
&& \mathop{\mathop{\sum_{1\leq u_{j,l(j)}\leq K,-M\leq
v_{j,l(j)}\leq M}}_{1\leq j\leq n,1\leq l(j)\leq
r_j}}_{u_{t,1}=u_{t+1}}
\textrm{E}\{(d_{u_{2,1}}(v_{2,1}))^{s_1}(d_{u_{3,1}}
(v_{3,1}))^{s_2}\cdots
(d_{u_{1,1}}(v_{1,1}))^{s_n} \}\nn\\
&&\times \textrm{E}\{\underbrace{[\pmb
R_\textrm{x}]_{v_{1,1}v_{1,2},u_{1,1}u_{1,2}} [\pmb
R_\textrm{x}]_{v_{1,2}v_{1,3},u_{1,2}u_{1,3}}\cdots [\pmb
R_\textrm{x}]_{v_{1,r_1}v_{2,1},u_{1,r_1}u_{2,1}}}_{[\pmb
R_\textrm{x}^{r_1}]_{m_1 m_2, k_1 k_2}}\cdots\nn\\
&&\cdot \underbrace{[\pmb
R_\textrm{x}]_{v_{n,1}v_{n,2},u_{n,1}u_{n,2}} [\pmb
R_\textrm{x}]_{v_{n,2}v_{n,3},u_{n,2}u_{n,3}}\cdots [\pmb
R_\textrm{x}]_{v_{n,r_{n}}v_{1,1},u_{n,r_n}u_{1,1}}} _{[\pmb
R_\textrm{x}^{r_n}]_{m_{n}m_1, k_{n}k_1}}\},\label{eq:cow0326}
\end{eqnarray}
where we let $u_{j,1}:=k_j$ and $v_{j,1}:=m_j$ for $1\leq j\leq n$.
To compute (\ref{eq:cow0326}), we consider noncrossing partitions of
the ordered set $\{u_{j,l(j)}: 1\leq j\leq n, 1\leq l(j)\leq r_j\}$
with a constraint that $u_{t,1}$ and $u_{t+1,1}$ are in the same
class. The ordering of the set is clear from the equation. The
$K$-graphs corresponding to these noncrossing partitions can be
generated by vertex mergence of a $K$-graph, called \textit{original
$K$-graph}, similar to the one in Fig.~\ref{fig:mouse1114}(b). In
specific, consider a cycle with $r_1+\cdots+r_n$ vertices. In a
counter-clockwise direction, the vertices on this cycle are ${\cal
U}=\{u_{1,1},\cdots,u_{1,r_1},u_{2,1},\cdots,u_{2,r_2},\cdots,u_{n,1},\cdots,u_{n,r_n}\}$.
The original $K$-graph is obtained by merging vertices $u_{t,1}$ and
$u_{{t+1},1}$ into one vertex. Let us call the cycles at the left-
and right-hand-side as $L$-cycle and $R$-cycle, respectively.

By the property of noncrossing, it is sufficient to consider only
$K$-graphs obtained by executing vertex mergence individually on
$L$- and $R$-cycles. That is, noncrossing partitions are performed
respectively on the two ordered sets $\{u_{t,1},\cdots,u_{t,r_t}\}$
and ${\cal U}\setminus\{u_{t,2},\cdots,u_{t,r_t},u_{t+1,1}\}$. Note
that the two sets have a common element $u_{t,1}$. The $K$-graphs
yielded in this way give non-vanishing contributions to
(\ref{eq:mouse0827}) in the large-system regime. It follows that
(\ref{eq:cow0326}) can be written as
\begin{eqnarray}
&&\mathop{\sum_{{\cal
U}\setminus\{u_{t,2},\cdots,u_{t,r_t},u_{t+1,1}\}}}_{{\cal
V}\setminus\{v_{t,2},\cdots,v_{t,r_t},v_{t+1,1}\}} \textrm{E}\{
(d_{u_{2,1}}(v_{2,1}))^{s_1}\cdots
(d_{u_{{t-1},1}}(v_{{t-1},1}))^{s_{t-2}}(d_{u_{t,1}}(v_{t,1}))^{s_{t-1}+s_t}
(d_{u_{{t+2},1}}(v_{{t+2},1}))^{s_{t+1}}\nn\\
&&\cdots(d_{u_{1,1}}(v_{1,1}))^{s_n}\} \textrm{E}\{ [\bb
R_\textrm{x}^{r_1}]_{v_{1,1}v_{2,1},u_{1,1}u_{2,1}}\cdots [\bb
R_\textrm{x}^{r_{t-1}}]_{v_{{t-1},1}v_{t,1},u_{{t-1},1}u_{t,1}} [\bb
R_\textrm{x}^{r_{t+1}}]_{v_{{t},1}v_{t+2,1},u_{{t},1}u_{t+2,1}}\nn\\
&&\cdots [\bb R_\textrm{x}^{r_n}]_{v_{n,1}v_{1,1},u_{n,1}u_{1,1}}\}
\mathop{\sum_{u_{t,2},\cdots,u_{t,r_t}}}_{v_{t,2},\cdots,v_{t,r_t}}
\textrm{E}\{ [\bb
R_\textrm{x}]_{v_{t,1}v_{t,2},u_{t,1}u_{t,2}}\cdots [\bb
R_\textrm{x}]_{v_{t,r_t}v_{t,1},u_{t,r_t}u_{t,1}} \},
\label{eq:tiger0326}
\end{eqnarray}
where ${\cal
V}=\{v_{1,1},\cdots,v_{1,r_1},v_{2,1},\cdots,v_{2,r_2},\cdots,v_{n,1},\cdots,v_{n,r_n}\}$,
each $[\bb R_\textrm{x}^{r_i}]_{v_{i,1}v_{i+1,1},u_{i,1}u_{i+1,1}}$
in the second and third lines is expanded as the product $[\bb
R_\textrm{x}]_{v_{i,1}v_{i,2},u_{i,1}u_{i,2}}[\bb
R_\textrm{x}]_{v_{i,2}v_{i,3},u_{i,2}u_{i,3}}\cdots[\bb
R_\textrm{x}]_{v_{i,r_i}v_{i+1,1},u_{i,r_1}u_{i+1,1}}$. The integers
in $[1,K]$ chosen by elements in ${\cal
U}\setminus\{u_{t,1},u_{t,2},\cdots,u_{t,r_t},u_{t+1,1}\}$ are all
distinct from those chosen by elements of
$u_{t,2},\cdots,u_{t,r_t}$. That is, besides $u_{t,1}$, elements in
sets $\{u_{t,1},\cdots,u_{t,r_t}\}$ and ${\cal
U}\setminus\{u_{t,2},\cdots,u_{t,r_t},u_{t+1,1}\}$ choose no common
integers. So do sets ${\cal
V}\setminus\{v_{t,1},v_{t,2},\cdots,v_{t,r_t},v_{t+1,1}\}$ and
$v_{t,2},\cdots,v_{t,r_t}$. Note that, although the second and third
expectations of (\ref{eq:tiger0326}) are both concerned with common
summation variables $u_{t,1}$ and $v_{t,1}$, the random variables
indexed by $u_{t,1}$ and $v_{t,1}$ can still be put in two different
expectations. For details, see the discussion of condition iii) in
Appendix~\ref{sub_appendix:mouse0124}. Since the limit of the
multi-sum in the third line of (\ref{eq:tiger0326}) is equal to
$\mu(\bb R_\textrm{x}^{r_t})$ for any integers $u_{t,1}\in[1,K]$ and
$v_{t,1}\in[-M,M]$, it can be factored out to the head of the
equation, and we obtain (\ref{eq:mouse0827}).

When $\bb D$ is set as the identity matrix and the constraint
$k_{t}=k_{t+1}$ is replaced with $k_{t}=k_u$, (\ref{eq:mouse0827})
can be revised accordingly. We can see the revised equation is equal
to $\mu(\bb R_\textrm{x}^{r_t+\cdots+r_{u-1}})\mu(\bb
R_\textrm{x}^{r_1+\cdots+r_{t-1}+r_u+\cdots+r_n})$.

\end{proof}

\begin{proof}
\textbf{[Theorem~\ref{theorem:mouse0810}]}

Suppose that, for $1\leq j\leq n$, polynomials
$$
p_j(x)=\sum_{r_j\geq 0} a_{j,r_j} x^{r_j}\qquad\mbox{and}\qquad
q_j(x)=\sum_{s_j\geq 0} b_{j,s_j} x^{s_j}
$$
give
$$
\sum_{r_j\geq 0} a_{j,r_j}\mu(\pmb R_\textrm{x}^{r_j})=\sum_{s_j\geq
0} b_{j,s_j}\mu(\pmb D^{s_j})=0.
$$
We have
\begin{eqnarray}\label{eq:tiger0613}
&&\mu(p_1(\pmb R_\textrm{x})q_1(\pmb D)\cdots p_{n}(\pmb
R_\textrm{x})q_{n}(\pmb D))\nn\\
&=& \mathop{\sum_{r_1,\cdots,r_{n}}}_{s_1,\cdots,s_{n}}
a_{1,r_1}b_{1,s_1}\cdots
a_{n,r_{n}}b_{n,s_{n}}\mathop{\lim_{M,N,K\rightarrow\infty}}_{K/N\rightarrow\beta}
(2M+1)^{-1}K^{-1}\textrm{E}\{\tr( \pmb R_\textrm{x}^{r_1}\pmb
D^{s_1}\cdots \pmb R_\textrm{x}^{r_{n}}\pmb D^{s_{n}})\},
\end{eqnarray}
where
\begin{eqnarray}
&&\textrm{E}\{\tr( \pmb R_\textrm{x}^{r_1}\pmb D^{s_1}\cdots \pmb
R_\textrm{x}^{r_n}\pmb D^{s_n})\}\nn\\
&=&\mathop{\sum_{m_1,\cdots,m_{n}}}_{k_1,\cdots,k_{n}} \textrm{E}\{
[\pmb R_\textrm{x}^{r_1}]_{m_1 m_2,k_1 k_2} [\pmb D^{s_1}]_{m_2
m_2,k_2 k_2}
\cdots [\pmb R_\textrm{x}^{r_n}]_{m_{n} m_1,k_{n}
k_1} [\pmb D^{s_n}]_{m_1 m_1,k_1 k_1}\}\label{eq:mouse0328}\\
&=&\mathop{\sum_{m_1,\cdots,m_{n}}}_{k_1,\cdots,k_{n}}
\textrm{E}\{(d_{k_2}(m_2))^{s_1}(d_{k_3}(m_3))^{s_2}\cdots
(d_{k_1}(m_1))^{s_n} \}\nn\\
&&\times\textrm{E}\{ [\pmb R_\textrm{x}^{r_1}]_{m_1 m_2,k_1 k_2}
[\pmb R_\textrm{x}^{r_2}]_{m_2 m_3,k_2 k_3} \cdots [\pmb
R_\textrm{x}^{r_n}]_{m_{n} m_1,k_{n} k_1} \}\\
&=& \mathop{\sum_{1\leq u_{j,l(j)}\leq K,-M\leq v_{j,l(j)}\leq
M}}_{1\leq j\leq n,1\leq l(j)\leq r_j}
\textrm{E}\{(d_{u_{2,1}}(v_{2,1}))^{s_1}(d_{u_{3,1}}
(v_{3,1}))^{s_2}\cdots
(d_{u_{1,1}}(v_{1,1}))^{s_n} \}\nn\\
&&\times \textrm{E}\{\underbrace{[\pmb
R_\textrm{x}]_{v_{1,1}v_{1,2},u_{1,1}u_{1,2}} [\pmb
R_\textrm{x}]_{v_{1,2}v_{1,3},u_{1,2}u_{1,3}}\cdots [\pmb
R_\textrm{x}]_{v_{1,r_1}v_{2,1},u_{1,r_1}u_{2,1}}}_{[\pmb
R_\textrm{x}^{r_1}]_{m_1 m_2, k_1 k_2}}\cdots\nn\\
&&\cdot \underbrace{[\pmb
R_\textrm{x}]_{v_{n,1}v_{n,2},u_{n,1}u_{n,2}} [\pmb
R_\textrm{x}]_{v_{n,2}v_{n,3},u_{n,2}u_{n,3}}\cdots [\pmb
R_\textrm{x}]_{v_{n,r_{n}}v_{1,1},u_{n,r_n}u_{1,1}}} _{[\pmb
R_\textrm{x}^{r_n}]_{m_{n}m_1, k_{n}k_1}}\}.\label{eq:tiger0603}
\end{eqnarray}
In (\ref{eq:tiger0603}), we use $v_{j,1}:= m_j$ and $u_{j,1}:= k_j$
for $1\leq j\leq n$. Our goal is to show $\mu(p_1(\pmb
R_\textrm{x})q_1(\pmb D)\cdots $ $p_{n}(\pmb R_\textrm{x})q_{n}(\pmb
D))=0$.

To compute (\ref{eq:tiger0603}), noncrossing partitions of the
ordered set $\{u_{j,l(j)}: 1\leq j\leq n, 1\leq l(j)\leq r_j\}$ (the
ordering is as shown in (\ref{eq:tiger0603})) are considered. The
partitioning can be decomposed into two stages. In the first stage,
we perform noncrossing partitions on the ordered set $\{u_{j,1}:
1\leq j\leq n\}$; in the second stage, elements in $\{u_{j,l(j)}:
1\leq j\leq n, 1\leq l(j)\leq r_j\}\setminus\{u_{j,1}: 1\leq j\leq
n\}$ are partitioned into classes of the noncrossing partitions
performed in the first stage according to the noncrossing condition.
We divide the noncrossing partitions performed in the first stage
into two groups as follows.

\textit{Group 1}: At least one of the classes contain only one
element.

\textit{Group 2}: Every class contains at least two elements.

For Group 1, without loss of generality, we suppose that $u_{1,1}$
is a singleton. Then, the expectation of $d_{u_{j,1}}(v_{j,1})$'s in
(\ref{eq:tiger0603}) can be written as
\begin{equation}\label{eq:mouse0613}
\textrm{E}\{(d_{u_{1,1}}(v_{1,1}))^{s_{n}}\}
\textrm{E}\{(d_{u_{2,1}}(v_{2,1}))^{s_1}
(d_{u_{3,1}}(v_{3,1}))^{s_2} \cdots (d_{u_{n,1}}(v_{n,1}))^{s_{n-1}}
\},
\end{equation}
since $d_k(m_1)$ and $d_l(m_2)$ are independent if $k\neq l$. Then,
(\ref{eq:tiger0603}) can be written as
\begin{eqnarray}\label{eq:cow0613}
&&\sum_{u_{1,1},v_{1,1}}
\textrm{E}\{(d_{u_{1,1}}(v_{1,1}))^{s_{n}}\}\times\nn\\
&& \mathop{\sum}_{ \{u_{j,l(j)},v_{j,l(j)}:1\leq j\leq n,1\leq
l(j)\leq
r_j\}\setminus\{u_{1,1},v_{1,1}\}}\textrm{E}\{(d_{u_{2,1}}(v_{2,1}))^{s_1}
(d_{u_{3,1}}(v_{3,1}))^{s_2} \cdots (d_{u_{n,1}}(v_{n,1}))^{s_{n-1}}
\}\nn\\
&&\hspace{1cm}\cdot\textrm{E}\{\underbrace{[\pmb
R_\textrm{x}]_{v_{1,1}v_{1,2},u_{1,1}u_{1,2}} [\pmb
R_\textrm{x}]_{v_{1,2}v_{1,3},u_{1,2}u_{1,3}}\cdots [\pmb
R_\textrm{x}]_{v_{1,r_1}v_{2,1},u_{1,r_1}u_{2,1}}}_{[\pmb
R_\textrm{x}^{r_1}]_{m_1 m_2, k_1 k_2}}\nn\\
&&\hspace{2cm}\cdots \underbrace{[\pmb
R_\textrm{x}]_{v_{n,1}v_{n,2},u_{n,1}u_{n,2}} [\pmb
R_\textrm{x}]_{v_{n,2}v_{n,3},u_{n,2}u_{n,3}}\cdots [\pmb
R_\textrm{x}]_{v_{n,r_{n}}v_{1,1},u_{n,r_n}u_{1,1}}} _{[\pmb
R_\textrm{x}^{r_n}]_{m_{n}m_1, k_{n}k_1}}\}.
\end{eqnarray}
Denote the summation comprised of the second to fourth lines of
(\ref{eq:cow0613}) as ${\cal
A}(u_{1,1},v_{1,1};\{s_i\}_{i=1}^{n-1};\{r_i\}_{i=1}^{n})$. We are
going to show the limiting value of ${\cal
A}(u_{1,1},v_{1,1};\{s_i\}_{i=1}^{n-1};\{r_i\}_{i=1}^{n})$ is
$O(1)$. Let
$$
B(\{u_{j,1},v_{j,1}\}_{j=2}^n;\{s_i\}_{i=1}^{n-1})=\textrm{E}\{(d_{u_{2,1}}(v_{2,1}))^{s_1}
(d_{u_{3,1}}(v_{3,1}))^{s_2} \cdots (d_{u_{n,1}}(v_{n,1}))^{s_{n-1}}
\}
$$
and
\begin{eqnarray}
&&C(\{u_{j,l(j)},v_{j,l(j)}\}_{1\leq j\leq n,1\leq l(j)\leq
r_j};\{r_i\}_{i=1}^n)\nn\\
&=&\textrm{E}\{[\pmb R_\textrm{x}]_{v_{1,1}v_{1,2},u_{1,1}u_{1,2}}
[\pmb R_\textrm{x}]_{v_{1,2}v_{1,3},u_{1,2}u_{1,3}}\cdots [\pmb
R_\textrm{x}]_{v_{1,r_1}v_{2,1},u_{1,r_1}u_{2,1}}\nn\\
&&\hspace{2cm}\cdots [\pmb
R_\textrm{x}]_{v_{n,1}v_{n,2},u_{n,1}u_{n,2}} [\pmb
R_\textrm{x}]_{v_{n,2}v_{n,3},u_{n,2}u_{n,3}}\cdots [\pmb
R_\textrm{x}]_{v_{n,r_{n}}v_{1,1},u_{n,r_n}u_{1,1}}\}.\nn
\end{eqnarray}
That is, $B(\{u_{j,1},v_{j,1}\};\{s_i\})$ and
$C(\{u_{j,l(j)},v_{j,l(j)}\};\{r_i\})$ stand for the first and
second expectations in ${\cal A}(u_{1,1},v_{1,1};\{s_i\};\{r_i\})$,
respectively. We have
\begin{eqnarray}
&&\left|{\cal A}(u_{1,1},v_{1,1};\{s_i\};\{r_i\})\right|\nn\\
&\leq& \mathop{\sum}_{ \{u_{j,l(j)},v_{j,l(j)}:1\leq j\leq n,1\leq
l(j)\leq r_j\}\setminus\{u_{1,1},v_{1,1}\}}
\left|B(\{u_{j,1},v_{j,1}\};\{s_i\})\right|\cdot\left|C(\{u_{j,l(j)},v_{j,l(j)}\};\{r_i\})\right|.\label{eq:mouse0226}
\end{eqnarray}
Due to H\"{o}lder's inequality and bounded moments of $d_k(m)$'s, we
have
\begin{eqnarray}
\left|B(\{u_{j,1},v_{j,1}\};\{s_i\})\right|&\leq& \textrm{E}\{|
(d_{u_{2,1}}(v_{2,1}))|^{s_1 (n-1)}\}^{1/(n-1)} \cdots \textrm{E}\{
|(d_{u_{n,1}}(v_{n,1}))|^{s_{n-1} (n-1)}\}^{1/(n-1)}\nn\\
&=& O(1).\label{eq:cow0222}
\end{eqnarray}
When $\textrm{x}=\textrm{cs}$, suppose that
$\left|B(\{u_{j,1},v_{j,1}\};\{s_i\})\right|$ is less than a
constant $c_1$. Thus, by (\ref{eq:mouse0226}),
$$
\left|{\cal A}(u_{1,1},v_{1,1};\{s_i\};\{r_i\})\right|\leq c_1
\mathop{\sum}_{ \{u_{j,l(j)},v_{j,l(j)}:1\leq j\leq n,1\leq l(j)\leq
r_j\}\setminus\{u_{1,1},v_{1,1}\}}
\left|C(\{u_{j,l(j)},v_{j,l(j)}\};\{r_i\})\right|.
$$
Following the proof of Lemma~\ref{lemma:rabbit0120} given in
Appendix~\ref{appendix:dragon0120}, we have
\begin{eqnarray*}
&&\mathop{\sum}_{ \{u_{j,l(j)},v_{j,l(j)}:1\leq j\leq n,1\leq
l(j)\leq
r_j\}\setminus\{u_{1,1},v_{1,1}\}}\left|C(\{u_{j,l(j)},v_{j,l(j)}\};\{r_i\})\right|=\textrm{E}\{[\bb
R_\textrm{cs}^{r_1+\cdots+r_n}]_{v_{1,1}v_{1,1},u_{1,1}u_{1,1}}\},
\end{eqnarray*}
which is equal to $\mu(\bb R_\textrm{cs}^{r_1+\cdots+r_n})$ in the
limit of $K,M,N\rightarrow\infty$ and $K/N\rightarrow\beta$ for any
$v_{1,1}$ and $u_{1,1}$. Thus, we have shown the limiting value of
${\cal A}(u_{1,1},v_{1,1};\{s_i\};\{r_i\})$ is $O(1)$ when
$\textrm{x}=\textrm{cs}$.

When $\textrm{x}=\textrm{ca}$ and condition 2) in
Theorem~\ref{theorem:mouse0810} is true, we can show
\begin{equation}\label{eq:tiger0222}
\mathop{\sum}_{ \{u_{j,l(j)},v_{j,l(j)}:1\leq j\leq n,1\leq l(j)\leq
r_j\}\setminus\{u_{1,1},v_{1,1}\}}\left|C(\{u_{j,l(j)},v_{j,l(j)}\};\{r_i\})\right|=O(1).
\end{equation}
Thus, we can use (\ref{eq:cow0222}) and (\ref{eq:tiger0222}) to
demonstrate that $\lim_{N,K=\beta N\rightarrow\infty}{\cal
A}(u_{1,1},v_{1,1};\{s_i\};\{r_i\})=O(1)$. Consider the case that
$\textrm{x}=\textrm{ca}$ and condition 1) of
Theorem~\ref{theorem:mouse0810} holds. Since $d_k(m)$'s are
non-negative random variables, $B(\{u_{j,1},v_{j,1}\};\{s_i\})$ is
non-negative, and we suppose it is upper-bounded by constant $c_2$.
Then
\begin{eqnarray*}
{\cal A}(u_{1,1},v_{1,1};\{s_i\};\{r_i\})&\leq& c_2\mathop{\sum}_{
\{u_{j,l(j)},v_{j,l(j)}:1\leq j\leq n,1\leq l(j)\leq
r_j\}\setminus\{u_{1,1},v_{1,1}\}}
C(\{u_{j,l(j)},v_{j,l(j)}\};\{r_i\})\nn\\
&=&c_2\textrm{E}\{[\bb
R_\textrm{ca}^{r_1+\cdots+r_n}]_{v_{1,1}v_{1,1},u_{1,1}u_{1,1}}\},
\end{eqnarray*}
which is equal to $c_2\mu(\bb R_\textrm{ca}^{r_1+\cdots+r_n})$
asymptotically. Thus, ${\cal A}(u_{1,1},v_{1,1};\{s_i\};\{r_i\})$ is
asymptotically equal to $O(1)$ as well.

It follows that, for Group 1 of noncrossing partitions of the
ordered set $\{u_{j,1}: 1\leq j\leq n\}$,
\begin{eqnarray}
&& \mu(p_1(\pmb R_\textrm{x})q_1(\pmb D)\cdots p_n(\pmb
R_\textrm{x})q_n(\pmb D))\nn\\
&=&\mathop{\lim_{M,N,K\rightarrow\infty}}_{K/N\rightarrow\beta}(2M+1)^{-1}K^{-1}
\sum_{s_n}b_{n,s_n} \sum_{u_{1,1},v_{1,1}}
\textrm{E}\{(d_{u_{1,1}}(v_{1,1}))^{s_n}\}\nn\\
&& \times\mathop{\sum_{r_1,\cdots,r_n}}_{s_1,
\cdots,s_{n-1}}a_{1,r_1}b_{1,s_1}\cdots
a_{n,r_n}b_{n-1,s_{n-1}}{\cal A}(u_{1,1},v_{1,1};\{s_i\};\{r_i\})\nn\\
&=& 0.\label{eq:mouse0225}
\end{eqnarray}
Note that the second line of (\ref{eq:mouse0225}) is equal to
$\sum_{s_n}b_{n,s_n}\mu(\pmb D^{s_n})=0$, while the third line is
$O(1)$. Thus, in this case, $\mu(p_1(\pmb R_\textrm{x})q_1(\pmb
D)\cdots p_n(\pmb R_\textrm{x})q_n(\pmb D))=0$.

Now, we consider Group 2 of noncrossing partitions of the ordered
set $\{u_{j,1}: 1\leq j\leq n\}$. By Lemma~\ref{lemma:mouse0605}, we
can suppose that $u_{t,1}$ and $u_{t+1,1}$ are partitioned in the
same class. Under this circumstance, we can use
Lemma~\ref{lemma:mouse0811} to show $\mu(p_1(\pmb
R_\textrm{x})q_1(\pmb D)\cdots p_{n}(\pmb R_\textrm{x})q_{n}(\pmb
D))$ given by (\ref{eq:tiger0613}) and (\ref{eq:mouse0328}) can be
written as
\begin{eqnarray}
&&\mu(p_1(\pmb R_\textrm{x})q_1(\pmb D)\cdots p_{n}(\pmb
R_\textrm{x})q_{n}(\pmb D))\nn\\
&=& \sum_{r_t} a_{t,r_t}\mu(\bb
R_\textrm{x}^{r_t})\mathop{\lim_{K,N,M\rightarrow\infty}}_{K/N\rightarrow\beta}
(2M+1)^{-1}K^{-1}\mathop{\sum_{r_1,\cdots,r_{t-1},r_{t+1},\cdots,r_{n}}}_{s_1,\cdots,s_{n}}
\underbrace{a_{1,r_1}b_{1,s_1}\cdots
a_{n,r_{n}}b_{n,s_{n}}}_{\textrm{without } a_{t,r_t}}
\times\label{eq:cow0328}\\
&& \sum_{{\cal K}\setminus\{k_{t+1}\}} \sum_{{\cal
M}\setminus\{m_{t+1}\}}\textrm{E}\{ (d_{k_2}(m_2))^{s_1}\cdots
(d_{k_{t-1}}(m_{t-1}))^{s_{t-2}}(d_{k_t}(m_t))^{s_{t-1}+s_t}
(d_{k_{t+2}}(m_{t+2}))^{s_{t+1}}\cdots\nn\\
&&(d_{k_1}(m_1))^{s_n}\}\textrm{E}\{ [\pmb R_\textrm{x}^{r_1}]_{m_1
m_2,k_1 k_2} \cdots[\pmb R_\textrm{x}^{r_{t-1}}]_{m_{t-1}
m_t,k_{t-1} k_t} [\pmb R_\textrm{x}^{r_{t+1}}]_{m_t m_{t+2}, k_t
k_{t+2}}\cdots [\pmb R_\textrm{x}^{r_n}]_{m_n m_1, k_n k_1}\}.\nn
\end{eqnarray}
Using similar arguments as in the discussion of Group 1, we can show
the limiting sum behind $\sum_{r_t} a_{t,r_t}\mu(\bb
R_\textrm{x}^{r_t})$ in (\ref{eq:cow0328}) is $O(1)$ for
$\textrm{x}\in\{\textrm{cs},\textrm{ca}\}$. Since $\sum_{r_t}
a_{t,r_t}\mu(\bb R_\textrm{x}^{r_t})=0$, for Group 2 of noncrossing
partitions of the ordered set $\{u_{j,1}: 1\leq j\leq n\}$, we have
$\mu(p_1(\pmb R_\textrm{x})q_1(\pmb D)\cdots p_{n}(\pmb
R_\textrm{x})q_{n}(\pmb D))=0$.

As we have shown both Groups 1 and 2 have contributions to
$\mu(p_1(\pmb R_\textrm{x})q_1(\pmb D)\cdots p_{n}(\pmb
R_\textrm{x})q_{n}(\pmb D))$ equal to zero, the proof is completed.

\end{proof}

\bibliography{Hwang}
\bibliographystyle{Hwang}

\newpage

\end{document}